\documentclass[11pt, twoside, sort, a4paper]{article}

\newcommand{\dom}{{\scalebox{0.7}{$d$}}}
\newcommand{\f}{{\scalebox{0.7}{$f$}}}
\usepackage{url}
\usepackage{wrapfig}
\usepackage{enumitem}
\usepackage{graphicx}
\usepackage{subcaption}

\DeclareCaptionLabelFormat{lowercaseparens}{(#2)}
\usepackage{hyperref, bm}
\hypersetup{
  hidelinks,
  pdftitle={Spouse-Protected Tontines: Household Decumulation via Neural-Network Optimization},
  pdfauthor={Duy-Minh Dang; Yukan Perumal},
  pdfsubject={Household decumulation, spouse protection, and book-level diversification in tontines},
  pdfkeywords={spouse-protected tontines, household decumulation, longevity pooling, mortality credits, spouse-continuation costs, neural-network policies, conditional value-at-risk}
}
\usepackage{mathtools}
\usepackage{float}
\usepackage{xcolor}

\usepackage{footnote}
\makesavenoteenv{tabular}
\makesavenoteenv{table}
\usepackage{geometry}
\usepackage{empheq}
 \usepackage{mathrsfs}
 \usepackage{tikz}

\AtBeginDocument{
\addtolength{\abovedisplayskip}{-0.5ex}
\addtolength{\abovedisplayshortskip}{-0.5ex}
\addtolength{\belowdisplayskip}{-0.5ex}
\addtolength{\belowdisplayshortskip}{-0.5ex}
}
\usepackage{cases}
\makeatletter
\newcommand{\subalign}[1]{%
  \vcenter{%
    \Let@ \restore@math@cr \default@tag
    \baselineskip\fontdimen10 \scriptfont\tw@
    \advance\baselineskip\fontdimen12 \scriptfont\tw@
    \lineskip\thr@@\fontdimen8 \scriptfont\thr@@
    \lineskiplimit\lineskip
    \ialign{\hfil$\m@th\scriptstyle##$&$\m@th\scriptstyle{}##$\crcr
      #1\crcr
    }%
  }
}

\usepackage[numbers]{natbib}
\usepackage{amsmath, amsthm, amssymb}
\usepackage{tikz}
\usetikzlibrary{matrix, positioning, calc}
\usepackage{multirow, nccmath}
\usepackage{algorithm}
\usepackage{algorithmic}
\usepackage{tikz}
\usetikzlibrary{positioning,arrows.meta}

\usepackage{comment}

\usepackage[titletoc,title]{appendix}

\usepackage[running, mathlines]{lineno}

\newcommand{\EQ}{\begin{equation}}
\newcommand{\EN}{\end{equation}}
\newcommand{\EQS}{\begin{equation*}}
\newcommand{\ENS}{\end{equation*}}
\newcommand{\EQA}{\begin{eqnarray}}
\newcommand{\ENA}{\end{eqnarray}}
\newcommand{\EQAS}{\begin{eqnarray*}}
\newcommand{\ENAS}{\end{eqnarray*}}

\usepackage{accents}
\newlength{\dhatheight}

\newtheorem{remark}{Remark}

\numberwithin{equation}{section}
\numberwithin{table}{section}
\numberwithin{figure}{section}

\numberwithin{definition}{section}
\numberwithin{theorem}{section}
\numberwithin{lemma}{section}
\numberwithin{remark}{section}
\numberwithin{assumption}{section}
\numberwithin{condition}{section}
\numberwithin{property}{section}
\numberwithin{proposition}{section}
\numberwithin{corollary}{section}
\numberwithin{algorithm}{section}

\makeatletter
\renewcommand*\env@matrix[1][c]{\hskip -\arraycolsep
  \let\@ifnextchar\new@ifnextchar
  \array{*\c@MaxMatrixCols #1}}
\makeatother

\makeatletter
\def\thm@space@setup{\thm@preskip=3pt
\thm@postskip=3pt}
\makeatother
\usepackage{titlesec}
\titlespacing*{\section}
  {0pt}{0.6\baselineskip}{0.2\baselineskip}
\titlespacing*{\subsection}
  {0pt}{0.6\baselineskip}{0.2\baselineskip}
\titlespacing*{\subsubsection}
  {0pt}{0.6\baselineskip}{0.2\baselineskip}

\usepackage[normalem]{ulem}
\usepackage{cancel}

\newcounter{savedcontloadline}


\usepackage{caption}
\begin{document}

\title{Spouse-Protected Tontines: Household Decumulation via Neural-Network Optimization}

\author{
Duy-Minh Dang\thanks{School of Mathematics and Physics, The University of Queensland, St Lucia, Brisbane 4072, Australia,
email: \texttt{duyminh.dang@uq.edu.au}
}
\and
Yukan Perumal\thanks{School of Mathematics and Physics, The University of Queensland, St Lucia, Brisbane 4072, Australia,
email: \texttt{y.perumal@student.uq.edu.au}
}
}
\date{\today}
\maketitle
\begin{abstract}
We develop a spouse-protected tontine in which a first death changes
the household state but generates no pool transfer. The same account remains
attached to the household contract until extinction and funds a spouse-only
continuation phase if the retiree dies first.
We derive contract-level actuarial-fairness conditions and a
finite-pool mortality-credit allocation rule with exact ex post budget
balance.
Under a homogeneous large-pool approximation, we formulate a
multidimensional decumulation problem for a representative household,
with withdrawal and rebalancing controls while the retiree is alive.
The objective balances expected cumulative real household payments
against the Conditional Value-at-Risk (CVaR) of terminal shortfalls relative to
household reserve targets, without conditioning on survival to the horizon. We develop a numerical solution method for this problem based on a
global-in-time neural-network parameterization of admissible control policies.
We quantify policy-induced spouse-continuation costs using expected-cost and
risk-loaded payment-scale loads at both representative-contract and
book levels. We characterize the large-book limit of average per-contract
continuation cost as the expected representative-contract cost conditional
on common market information.

Numerical experiments calibrated to Australia show that the
spouse-only phase is a first-order and persistent household event.
In these experiments, book-level diversification removes
most of the representative-contract excess upper-tail cost of spouse
continuation in the large-book limit, without changing expected
per-contract continuation cost.
A cross-country mortality comparison indicates that the prevalence and
persistence of the spouse-only phase are not specific to Australia.

\vspace{.1in}

\vspace{.1in}
\end{abstract}

\section{Introduction}
\label{sec:intro}
Tontines pool idiosyncratic longevity risk by redistributing the account
balances of contracts that exit the pool to contracts that remain active.
The resulting mortality credits can enhance retirement income while retaining
greater investment and withdrawal flexibility than a conventional life annuity
\cite{donnelly2014bringing,donnelly2015actuarial,
milevsky2015optimal,Fullmer2019Tontines,forsyth2024optimal}.
Many modern tontine optimization models nevertheless take the pooled unit to be
a single retiree who purchases the contract: the contract exits when the
retiree dies, and no household payment is made thereafter.

Spouse protection addresses a different retirement-income need.
Commercial lifetime-income products in Australia and other markets, such as
Canada, the UK, and the US, can offer joint-life options under which payments
continue for a surviving spouse or partner
\cite{qsuper2025pds,MyNorth2025PDS,AllianzRetirePlus2026PDS,
RBCJointLifeAnnuity,LegalGeneralJointLifeAnnuity,TIAATwoLifeAnnuity}.
QSuper's Lifetime Pension provides a pooled example: its spouse-protection
option continues payments for the life of the last survivor, bases the initial
payment rate on the younger spouse's age, and applies a lower payment rate than
the single option \cite{qsuper2025pds}. These product terms specify the
last-survivor income benefit but leave open, for a tontine design, how a
spouse-protected contract should participate in the pool, when its balance
should be forfeited, and how mortality-credit gains should depend on household
survival status.

This paper draws on two related but largely separate literatures.
On the one hand, modern tontine and pooled-annuity research develops
actuarially fair pooling rules and optimal retirement-income designs,
predominantly for single-life participants
\cite{donnelly2015actuarial,milevsky2015optimal,bernhardt2019modern,
forsyth2024optimal,orozco2026money}. On the other hand, research on retired
couples and joint-life products recognizes that consumption, portfolio choice,
survivor income, and annuity values depend on both lifetimes
\cite{hubener2014retiredcouples,sanders2016jointsurvival,
ventura2023jointlife}.
Related work develops longevity-credit mechanisms for joint-life
benefits \cite{donnelly2023jointlife} and extends longevity pooling to
multi-life benefits, with large-pool convergence results for survivor-only
credits \cite{donnelly2026multilife}.
To our knowledge, existing studies do not jointly address
last-survivor tontine pooling, controlled household withdrawals and portfolio
choice under investment risk, and the resulting policy-induced
spouse-continuation cost and its book-level tail risk.

We develop a spouse-protected tontine in which each pooled unit is a household
contract that remains active until household extinction. Under this design,
pool-exit probabilities and mortality-credit gain rates are household-state
dependent. While both members are alive, the contract exits during a period
only if both members die during that period; if the retiree dies first and the
spouse survives, the contract enters a spouse-only continuation phase and
remains active in the pool. Crucially, a first death changes the
household state but generates no forfeiture or other pool transfer; the existing
account remains attached to the household contract until extinction. To
isolate this household-state mechanism, we consider a homogeneous pool in which
every contract includes spouse protection; mixed pools containing both
single-life and spouse-protected contracts are outside the present scope.
Throughout, \emph{spouse protection} denotes the overall product-design
feature, whereas \emph{spouse continuation} denotes the spouse-only payment
phase or stream that may follow the retiree's death.

The household contract has four survival states: both members alive, retiree
only, spouse only, and neither alive. We derive the state-dependent pool-exit
probabilities, conditions for individual-contract actuarial fairness, exact
ex post budget balance at the finite-pool level, and the corresponding
large-pool tontine gain rates. Withdrawals and portfolio rebalancing are
controlled while the retiree remains alive; if the retiree dies first, the
surviving spouse receives a capped, rule-based spouse-continuation payment and
the portfolio follows a passive rebalancing rule. Household reward is expected
cumulative real household payments, while risk is the CVaR of a terminal
shortfall measured against household-state-dependent reserve targets. Both
criteria are evaluated without conditioning on survival to the horizon.

Under the homogeneous large-pool approximation, the resulting
decumulation problem for a representative household is a multidimensional
stochastic control problem. It combines a four-asset investment universe
comprising domestic and foreign equity and bond indices with discrete
household mortality states, household-state-dependent mortality-credit gain
rates and payment rules, insolvency, and a nonlinear tail-risk objective.
We develop a numerical solution method for this problem based on a
global-in-time neural-network (NN) parameterization of admissible withdrawal
and asset-allocation control policies. Admissibility is enforced through the
network output transformations.
This approach follows the NN policy literature for multi-period portfolio
and decumulation problems
\cite{li2019data,chen2023benchmark,hu2024recent,orozco2026money};
recent convergence theory provides convergence-in-probability results for NN
policy approximations in constrained risk--reward control problems
\cite{chang2026convergence}.

We quantify the spouse-continuation payment obligation induced by
the learned household policy through expected-cost and risk-loaded
payment-scale loads. We then distinguish representative-contract tail risk
from contract-book risk. In a finite book, contracts share market returns but
have conditionally independent household mortality histories
given common market information. We characterize the large-book
limit of average per-contract continuation cost as the expected
representative-contract cost conditional on this information.
Household-specific mortality risk therefore diversifies across contracts,
while common-market risk remains.
Expected per-contract continuation cost is unchanged by pooling, whereas the
prudential tail buffer must reflect book-level risk.

During the preparation of this paper, we became aware of the
contemporaneous work of \cite{borac2026coupletontines}, who develop a fully
funded mutual couple tontine with
dependent partner lifetimes and actuarially fair mortality and
widowing credits. Withdrawals are prescribed to achieve a target average
income together with random credits, rather than optimized jointly with
portfolio choice under investment risk. First death releases reserves for
redistribution as widowing credits to couples joint-alive at the beginning
of the period.

We study a different same-account continuation design: first death
generates no forfeiture or other pool transfer, and the existing account remains
attached to the household contract until extinction. This convention preserves
the link between retiree-alive withdrawal and investment decisions and the
subsequent spouse-continuation stream.
Our focus is therefore on how a household should decumulate and invest when
the same account may later support a surviving spouse, how large the resulting
policy-induced spouse-continuation cost is,
and how its tail risk changes with book size.
This book-level analysis concerns cumulative continuation payments
under common market risk, rather than the large-pool behavior of individual
longevity credits analyzed in \cite{donnelly2026multilife}.

The contributions of the paper are threefold.
\begin{itemize}[noitemsep, topsep=2pt, leftmargin=*]

\item \emph{Household-contract tontine design.} We develop a
spouse-protected tontine in which a first death changes the household state but
generates no forfeiture or other pool transfer, and the same account remains
attached to the household until household extinction. We derive the associated
household-state pool-exit probabilities, contract-level actuarial-fairness
conditions for mortality-credit allocation,
and a finite-pool allocation rule with exact ex post budget balance.

\item \emph{Household reward--risk optimization.} We formulate a decumulation
problem under investment risk that distinguishes controlled retiree-alive
withdrawals from rule-based spouse-continuation payments in the spouse-only
state and solve it using admissible global-in-time
NN policies.

\item \emph{Continuation-cost measurement and book-level
diversification.} We define representative-contract and book-level
expected-cost and risk-loaded spouse-continuation loads, characterize the
conditional large-book limit, and quantify the
reduction in excess upper-tail cost through household-specific
mortality diversification.

\end{itemize}
The numerical experiments use the 2021 Australian male and female period life
tables from the Human Mortality Database \cite{HMD}. The baseline household
consists at inception of a male retiree aged 65 and a female spouse aged 60 and
is followed for 35 years, until the spouse reaches age 95. The initial contract
balance is AUD $1$ million. The Australian calibration yields three main
findings:
\begin{itemize}[noitemsep, topsep=2pt, leftmargin=*]

\item The spouse-only continuation phase is a first-order event:
approximately $71\%$ of contracts enter this state, which lasts about
$12.25$ years on average conditional on entry.

\item Under capped continuation, expected representative-contract
continuation payments remain substantial and relatively stable as the
evaluated policies vary along the reward--risk frontier. The expected-cost
spouse-continuation load nevertheless increases as policies become more
conservative.

\item Contract-book diversification leaves expected continuation
cost unchanged but sharply reduces its excess upper tail.
Under the baseline capped design, the conditional large-book
limit reduces the representative-contract excess upper-tail cost by
approximately $97\%$ for the policies evaluated along the reward--risk
frontier.

\end{itemize}
The Australian household-state pattern is not exceptional. Under the same
household ages, 35-year horizon, and independent-mortality specification, the
corresponding 2021 male and female period life tables for Canada, the United
States, Japan, and the Netherlands imply that, across the five countries, the
probability of entering the spouse-only phase ranges from approximately
$68.1\%$ to $77.0\%$. Its unconditional mean duration ranges from $8.70$ to
$10.23$ years, while its mean duration conditional on entry ranges from
$12.25$ to $13.34$ years. The spouse-only continuation phase is therefore a
first-order and persistent household event across these mortality environments.

The remainder of the paper is organized as follows.
Section~\ref{sc:tontine_modeling} develops the household-state tontine and
mortality-credit mechanics. Section~\ref{sc:hh_modeling} specifies the
household payment, portfolio, and insolvency framework.
Section~\ref{sc:household_reward_risk} formulates the household reward--risk
problem, and Section~\ref{sec:NNs} presents the neural-network policy
approximation. Section~\ref{sc:spouse_continuation_load} defines the
representative-contract and book-level spouse-continuation loads and develops
the conditional large-book diversification framework.
Numerical experiments and their implications for spouse-protected tontine
design are reported in Section~\ref{sec:numerics}.
Section~\ref{sec:conclusion} concludes the paper and outlines some possible
research directions.
Appendix~\ref{app:international_household_states} reports the
cross-country household-state comparison.
Appendix~\ref{app:num_continuation_cap_sensitivity} examines the effects
of removing the continuation cap on expected continuation costs and
tail risk. 
\section{Tontine modeling with spouse-protection}
\label{sc:tontine_modeling}
We build on and generalize the individual-account tontine framework of
\cite{forsyth2024optimal} to a spouse-protected household-contract setting.
At inception, each pooled unit is a \emph{spouse-protected household contract} comprising a retiree and spouse. There is no mixed entry pool: all pooled units enter as
spouse-protected household contracts, rather than as a mixture of
spouse-protected and single-life contracts. As retiree and spouse deaths occur,
a contract may move from the two-member state to a one-survivor state, but it
remains the same contract until it exits the pool. Hereafter, these pooled units
are simply referred to as contracts unless otherwise noted.

The central modeling distinction is that mortality pooling is based on
contract exit from the pool, not on the retiree's death alone.
If the retiree
dies while the spouse remains alive, the contract continues in the pool and
household payments may continue under the spouse-protection rule.
Mortality
credits are therefore determined by the household survival state of each contract and
are generated only when a contract exits the pool. In this section,
we focus on the mortality-pooling mechanics of the spouse-protected tontine.

\subsection{Preliminaries}
\label{ssc:tontine_preliminaries}
Unlike the ``plan-to-live'' convention in retirement planning
\cite{pfau2018overview,orozco2026money,forsyth2024optimal}, we do not condition
on household survival to the horizon. Mortality is incorporated through the
household-state process introduced below. The finite horizon is therefore a
computational truncation horizon rather than an assumption that either the
retiree or the spouse survives to that date. Once neither household member
remains alive, household payments and mortality credits cease.

Let $a_R$ and $a_S$ denote the retiree's and spouse's ages at inception. Since
spouse continuation may remain relevant after the retiree's death, choosing the
horizon solely from the retiree's age can mechanically truncate the spouse-only
phase when the spouse is younger. To avoid this, we select a terminal reference
age $x^\star$ for the younger of the retiree and spouse and set
\[
T=\left\lceil x^\star-\min\{a_R,a_S\}\right\rceil.
\]
This convention does not condition on survival of either household member to
the reference age $x^\star$.

We consider the finite planning horizon $[0,T]$ with annual decision times
\[
\mathcal{T}=\{t_m \mid t_m=m,\; m=0,\ldots,M\},
\qquad
M=T,
\]
at which mortality credits are allocated, fees are deducted, household payments
are made, and portfolio positions are rebalanced.\footnote{Annual spacing is
assumed for simplicity. In practice, review and rebalancing dates are typically
fixed in advance, e.g.\ annually or semi-annually.}
Here, $t_0=0$ is the inception date. Household payments and rebalancing
decisions are made at $t_m$, $m=0,\ldots,M-1$, while mortality credits are
allocated at $t_m^-$, $m=1,\ldots,M$.

For notational convenience, let $t^- = t-\varepsilon$ and $t^+ = t+\varepsilon$
denote the instants immediately before and after time $t$, with
$\varepsilon \to 0^+$. For a generic time-dependent quantity $f(t)$ and any
$t_m\in\mathcal{T}$, we write
\[
f_{m^-} := \lim_{\varepsilon\to 0^+} f(t_m-\varepsilon),
\qquad
f_{m^+} := \lim_{\varepsilon\to 0^+} f(t_m+\varepsilon),
\]
as shorthand for $f(t_m^-)$ and $f(t_m^+)$, respectively.

The tontine pool consists of $L$ contracts, indexed by $\ell=1,\ldots,L$. A contract is called \emph{active} at time $t$ if at least one household member is
alive at $t$.

During the inter-decision period $[t_{m-1}^+,t_m^-]$, asset prices evolve and mortality outcomes are realized. At $t_m^-$, these outcomes are observed, after which the period-$m$ decision and implementation sequence is
\[
t_m^- \;\longrightarrow\; t_m \;\longrightarrow\; t_m^+.
\]
For contracts active at the start of the period, the events are ordered as
follows.
\begin{itemize}[noitemsep, topsep=1pt, leftmargin=*]
\item At $t_m^-$, each contract's portfolio balance is observed and
its household survival state is updated. Any contract for which neither member
remains alive exits the pool and forfeits its balance.

The forfeited balances are redistributed immediately as mortality credits
to the contracts that remain active after the mortality update,
subject to the pool-closing convention in
Subsection~\ref{ssc:household_large_pool}.

\item At $t_m$, $m=0,\ldots,M-1$, the period-$m$ household payment is made. When the
retiree is alive, this payment is the controlled withdrawal; when only the
spouse is alive, it is the spouse-protection continuation payment, determined
by the continuation rule rather than by a new withdrawal decision.

\smallskip
At $t_m^+$, the post-payment portfolio is then updated: contracts for which the retiree remains alive and post-payment wealth is strictly positive may rebalance; if
post-payment wealth is nonpositive, insolvency is triggered; and spouse-only
contracts follow the passive rebalancing rule.

\end{itemize}
At the terminal date $t_M=T$, the portfolio is liquidated and no further
household payment or rebalancing decision is made.

To formalize mortality credits, we denote by $\{H_t^\ell\}_{0\le t\le T}$  the
household survival-state process for contract $\ell$, where $H_t^\ell$ records at
time $t$ whether the retiree and spouse are alive.
In the period-by-period recursion, the relevant quantities are the
pre- and post-decision states
\[
H_{m^-}^\ell:=H_{t_m^-}^\ell,
\qquad
H_{m^+}^\ell:=H_{t_m^+}^\ell,
\qquad
\ell=1,\ldots,L,
\]
with values in
\[
\mathcal H:=\{11,10,01,00\}.
\]
The two digits indicate the survival status of the retiree and spouse,
respectively:
\[
11:\text{ both alive},\qquad
10:\text{ retiree only},\qquad
01:\text{ spouse only},\qquad
00:\text{ neither alive}.
\]
Since mortality is updated at $t_m^-$ and no further mortality event occurs
between $t_m^-$ and $t_m^+$, we have $H_{m^+}^\ell=H_{m^-}^\ell$.
Contract $\ell$ remains active in the pool after the mortality update at $t_m^-$
if and only if $H_{m^-}^\ell\neq 00$; if $H_{m^-}^\ell=00$, it exits the pool at
$t_m^-$.

\subsection{Spouse protection}
\label{ssc:spouse_continuation}

Spouse protection changes the pooled unit from a single retiree to a
household contract. The retiree's death alone does not cause pool exit: the
contract remains active in states $11$, $10$, and $01$, and exits only when
$H_{m^-}^\ell=00$. Consequently, spouse protection changes the
household-state-dependent pool-exit probabilities and hence the corresponding
mortality-credit gain rates. In particular, while both household members are
alive, pool exit is governed by their joint-death probability.

The payment convention distinguishes the retiree-alive control states
from the spouse-only continuation state. In states $11$ and $10$, the household
payment is a controlled withdrawal and the portfolio may be rebalanced. In
state $01$, no new withdrawal or rebalancing control is chosen; the surviving
spouse instead receives a rule-based \emph{spouse-continuation payment}, and
the portfolio follows the passive update convention. The precise payment,
continuation-payment cap, admissibility, and insolvency rules are specified in
Section~\ref{sc:hh_modeling}.

The amount of the spouse-continuation payment does not enter the
current-period mortality-credit gain rate directly; it affects post-payment
wealth and therefore the subsequent contract evolution. The resulting
spouse-only payment stream is quantified later through the policy-induced
\mbox{spouse-continuation load.}

\subsection{Household-state transitions and contract exit}
\label{ssc:household_tontine_accounts}
Let $\delta_{R,m-1}^\ell$ and $\delta_{S,m-1}^\ell$ denote the conditional one-year
death probabilities of the retiree and spouse in contract $\ell$ over the period
$[t_{m-1}^+,t_m^-]$, given survival at $t_{m-1}^+$. Throughout this subsection,
we assume conditional independence of retiree and spouse mortality within each
period. The resulting one-period household-state transitions are
\begin{align}
\mathbb P\!\left(H_{m^-}^\ell = 11 \mid H_{(m-1)^+}^\ell = 11\right)
&= (1-\delta_{R,m-1}^\ell)(1-\delta_{S,m-1}^\ell), \nonumber \\
\mathbb P\!\left(H_{m^-}^\ell = 10 \mid H_{(m-1)^+}^\ell = 11\right)
&= (1-\delta_{R,m-1}^\ell)\delta_{S,m-1}^\ell, \nonumber \\
\mathbb P\!\left(H_{m^-}^\ell = 01 \mid H_{(m-1)^+}^\ell = 11\right)
&= \delta_{R,m-1}^\ell(1-\delta_{S,m-1}^\ell), \nonumber \\
\mathbb P\!\left(H_{m^-}^\ell = 00 \mid H_{(m-1)^+}^\ell = 11\right)
&= \delta_{R,m-1}^\ell\delta_{S,m-1}^\ell, \label{eq:hh_transition_11}
\end{align}
together with
\begin{align}
\mathbb P\!\left(H_{m^-}^\ell = 10 \mid H_{(m-1)^+}^\ell = 10\right)
&= 1-\delta_{R,m-1}^\ell, &
\mathbb P\!\left(H_{m^-}^\ell = 00 \mid H_{(m-1)^+}^\ell = 10\right)
&= \delta_{R,m-1}^\ell, \label{eq:hh_transition_10}\\
\mathbb P\!\left(H_{m^-}^\ell = 01 \mid H_{(m-1)^+}^\ell = 01\right)
&= 1-\delta_{S,m-1}^\ell, &
\mathbb P\!\left(H_{m^-}^\ell = 00 \mid H_{(m-1)^+}^\ell = 01\right)
&= \delta_{S,m-1}^\ell. \label{eq:hh_transition_01}
\end{align}
State $00$ is absorbing and corresponds to permanent exit of the
contract from the pool.

Let
\[
\mathcal{L}_{m-1}:=\left\{\ell\in\{1,\ldots,L\}:\,H_{(m-1)^+}^\ell\neq 00\right\}
\]
denote the set of contracts active at time $t_{m-1}^+$.
For each $\ell\in\mathcal{L}_{m-1}$, the one-period pool-exit probability over $[t_{m-1}^+,t_m^-]$ is given by
\begin{equation}
\bar\delta_{m-1}^\ell
:=
\mathbb P\!\left(H_{m^-}^\ell = 00 \,\middle|\, H_{(m-1)^+}^\ell\right)
=
\begin{cases}
\delta_{R,m-1}^\ell\,\delta_{S,m-1}^\ell, & H_{(m-1)^+}^\ell = 11, \\[6pt]
\delta_{R,m-1}^\ell, & H_{(m-1)^+}^\ell = 10, \\[6pt]
\delta_{S,m-1}^\ell, & H_{(m-1)^+}^\ell = 01.
\end{cases}
\label{eq:hh_exit_prob}
\end{equation}
Thus, a contract in state $11$ exits the pool during the period only if both members die before $t_m^-$. A contract already in state $10$ or $01$ exits during the period
if the surviving household member dies before~$t_m^-$.

For each $\ell\in\mathcal{L}_{m-1}$, let $\mathcal{W}_{m^-}^\ell$ denote the real
(inflation-adjusted) account balance of contract $\ell$ at time $t_m^-$, before
mortality credits are distributed; hereafter, we also refer to this account balance
as the wealth of contract $\ell$.
The mortality-credit allocation in this section is formulated for
solvent tontine accounts, so we assume $\mathcal{W}_{m^-}^\ell>0$ for all
$\ell\in\mathcal{L}_{m-1}$. Once an account is depleted, it receives no
further mortality credits, and subsequent minimum household payments are
financed from other assets or sources of finance under the insolvency
convention in Subsection~\ref{ssc:hh_insolvency}.
We define the post-mortality-update contract-active indicator at $t_m^-$
and its conditional expectation by
\begin{equation}
\mathbf{1}_m^\ell
=
\begin{cases}
1, & H_{m^-}^\ell \neq 00, \\[4pt]
0, & H_{m^-}^\ell = 00,
\end{cases}
\qquad
\mathbb E_{m-1}\!\left[\mathbf 1_m^\ell\right]
=
1-\bar\delta_{m-1}^\ell,
\qquad \ell\in\mathcal{L}_{m-1}.
\label{eq:hh_indicator}
\end{equation}
Here, $\mathbb E_{m-1}[\cdot]$  denotes expectation conditional on all information available at $t_{m-1}^+$ (the start of  $[t_{m-1}^+, t_m^-]$).
If $\mathbf 1_m^\ell=0$, the contract exits the pool and its account
balance is forfeited. If $\mathbf 1_m^\ell=1$, the contract remains active in
the pool and receives a mortality credit $c_m^\ell$ at time $t_m^-$.

\begin{remark}[Pool exit and mortality credits]
\label{rm:hh_contract_exit}
For spouse-protected contracts, the relevant event for mortality pooling is
contract exit from the pool, not the retiree's death alone. The cases in
\eqref{eq:hh_exit_prob} therefore encode the relevant one-period pool-exit
probability for each contract state. Consequently, the conditional probability
entering the actuarially fair mortality credit formulas is the pool-exit
probability $\bar\delta_{m-1}^\ell$ for contract $\ell$ in \eqref{eq:hh_exit_prob}, rather than the retiree death probability alone.
\end{remark}
We next derive the mortality-credit allocation
associated with this pool-exit probability, proceeding from
individual-contract fairness, through the pool-level budget constraint and
large-pool approximation, to the contract-level tontine gain rate used in the wealth
recursion. Throughout this derivation, we abstract from administration,
investment-management, and transaction costs; fees are introduced separately
in the next section.
\subsection{Individual-contract fairness condition}
\label{ssc:household_fairness}
The fairness condition is imposed at the contract level. During
$[t_{m-1}^+,t_m^-]$, a contract that exits the pool forfeits its account
balance, while a contract that remains active is eligible to receive a mortality
credit.
Let $\Omega_m$ be the sigma-field generated by the information
available at $t_{m-1}^+$ and the realized balances
$\{\mathcal W_{m^-}^k\}_{k\in\mathcal L_{m-1}}$, before forfeiture and
mortality-credit allocation. We assume that current-period mortality outcomes
and asset returns are conditionally independent given the start-of-period
information, so
$\mathbb P(\mathbf 1_m^\ell=1\mid\Omega_m)=1-\bar\delta_{m-1}^\ell$.
For $\bar\delta_{m-1}^\ell<1$, individual-contract fairness balances expected
forfeitures against expected credit receipts conditional on $\Omega_m$:
\begin{equation}
\bar\delta_{m-1}^\ell\,\mathcal{W}_{m^-}^\ell
=
(1-\bar\delta_{m-1}^\ell)\,
\mathbb E\!\left[c_m^\ell \,\middle|\, \Omega_m,\;\mathbf 1_m^\ell=1\right],
\label{eq:hh_fairness}
\end{equation}
where the expectation conditions on contract $\ell$ remaining active,
but not on the current-period mortality outcomes of the other contracts.
Solving \eqref{eq:hh_fairness} gives the actuarial-fairness benchmark
\begin{equation}
\mathbb E\!\left[c_m^\ell \,\middle|\, \Omega_m,\;\mathbf 1_m^\ell=1\right]
=
\frac{\bar\delta_{m-1}^\ell}{1-\bar\delta_{m-1}^\ell}\,\mathcal{W}_{m^-}^\ell.
\label{eq:hh_fair_credit}
\end{equation}
Thus, the expected mortality credit is proportional to
the contract's own wealth, with the proportionality factor determined by
the contract's pool-exit probability.
Substituting the pool-exit probabilities in
\eqref{eq:hh_exit_prob} into the general fairness condition
\eqref{eq:hh_fair_credit} gives the state-specific fair mortality-credit
factor using
$\delta_{R,m-1}^\ell\delta_{S,m-1}^\ell$ in state $11$,
$\delta_{R,m-1}^\ell$ in state $10$, and
$\delta_{S,m-1}^\ell$ in state $01$. In particular, since the fair-credit
multiplier $p/(1-p)$ increases with the pool-exit probability $p$, the
state-$11$ factor is no greater than either corresponding one-survivor
factor.

\begin{remark}[Finite-pool budget balance and actuarial fairness]
\label{rm:hh_finite_pool_bias}
Equation~\eqref{eq:hh_fair_credit} is a contract-level ex ante
actuarial-fairness condition, not an ex post budget identity. In a finite pool,
actual mortality credits must be funded by realized forfeitures and therefore
depend on the realized pool-exit outcomes and account balances of all
contracts. This distinction is especially important under spouse protection:
a first death produces a one-survivor state without a forfeiture, while
spouse-continuation payments may reduce the account balance before household
extinction. Hence realized forfeitures may be insufficient to pay all unscaled
fair mortality-credit amounts in \eqref{eq:hh_fair_credit}.
\end{remark}
The group-gain factor introduced below scales these unscaled fair
mortality-credit amounts to match aggregate realized forfeitures, thereby
enforcing exact ex post pool-level budget balance
while surviving recipients remain.

\subsection{Pool-level budget constraint and large-pool approximation}
\label{ssc:household_large_pool}
For the finite-pool allocation below, assume
$0<\bar\delta_{m-1}^\ell<1$ for all $\ell\in\mathcal L_{m-1}$.
When at least one contract remains active after the mortality update,
realized mortality credits must match realized forfeitures:
\begin{equation}
\sum_{\ell \in \mathcal{L}_{m-1}} \mathbf 1_m^\ell\,c_m^\ell
=
\sum_{\ell \in \mathcal{L}_{m-1}} (1-\mathbf 1_m^\ell)\,\mathcal{W}_{m^-}^\ell.
\label{eq:hh_budget}
\end{equation}
The left-hand side is the total mortality credit paid to contracts that remain
active after the mortality update, and the right-hand side is the total
forfeited account balance of contracts that exit the pool during the period.
This identity is an ex post pool-level budget constraint.

As in \cite{sabin2016analytics}, introduce the realized group-gain factor
\begin{equation}
\Gamma_m
=
\frac{\displaystyle \sum_{k \in \mathcal{L}_{m-1}} (1-\mathbf 1_m^k)\,\mathcal{W}_{m^-}^k}
     {\displaystyle \sum_{k \in \mathcal{L}_{m-1}} \mathbf 1_m^k\,
      \frac{\bar\delta_{m-1}^k}{1-\bar\delta_{m-1}^k}\,\mathcal{W}_{m^-}^k}.
\label{eq:hh_group_gain}
\end{equation}
The factor $\Gamma_m$ is a random variable ex ante: it is determined by the
realized pool-exit outcomes and account balances observed at $t_m^-$.
Its numerator is the total realized forfeited account balance available
for redistribution, whereas its denominator is the aggregate actuarially fair
mortality-credit amount for contracts that remain active after the mortality
update. Thus, on the event that the denominator is positive, $\Gamma_m>1$ means
realized forfeitures exceed this aggregate amount, $\Gamma_m<1$ means realized
forfeitures are insufficient, and $\Gamma_m=0$ occurs when no account balance is
forfeited.

Scaling the fair-credit benchmark on the right-hand side of
\eqref{eq:hh_fair_credit} by $\Gamma_m$ defines the finite-pool mortality-credit
allocation rule
\begin{equation}
c_m^\ell
=
\frac{\bar\delta_{m-1}^\ell}{1-\bar\delta_{m-1}^\ell}\,\mathcal{W}_{m^-}^\ell\,\Gamma_m,
\qquad \text{for all } \ell \in \mathcal{L}_{m-1}\ \text{with}\ \mathbf 1_m^\ell=1.
\label{eq:hh_credit_finite_pool}
\end{equation}
This allocation satisfies the pool-level budget identity
\eqref{eq:hh_budget} exactly whenever at least one contract
remains active: the total realized mortality credits paid to
contracts that remain active equal the total realized forfeited account balance
collected from contracts that exit the pool. Since $\Gamma_m\ge 0$ whenever
account balances are nonnegative, the mortality credits paid under
\eqref{eq:hh_credit_finite_pool} are also nonnegative.

Following \cite[Remark~3.2]{forsyth2024optimal}, if all contracts
in $\mathcal L_{m-1}$ exit during the period, we set $\Gamma_m=0$,
close the pool, and pay the remaining account balances to the respective
estates. This exceptional settlement replaces the mortality-credit
allocation in \eqref{eq:hh_budget}--\eqref{eq:hh_credit_finite_pool}.

The group-gain factor enforces exact ex post pool-level budget balance,
but in a finite heterogeneous pool the allocation rule can introduce a bias in
expected mortality credits relative to the individual-contract
actuarial-fairness condition \eqref{eq:hh_fair_credit}. Following the
finite-pool analysis in \cite{sabin2016analytics} and Condition~3.1 in
\cite{forsyth2024optimal}, this bias is negligible when the pool is
sufficiently large and each contract's unscaled actuarially fair
mortality-credit amount is small relative to aggregate expected
forfeitures:
\begin{equation}
\frac{\bar\delta_{m-1}^\ell}{1-\bar\delta_{m-1}^\ell}\,\mathcal{W}_{m^-}^\ell
\ll
\sum_{k \in \mathcal{L}_{m-1}} \bar\delta_{m-1}^k\,\mathcal{W}_{m^-}^k,
\qquad \text{for all } \ell \in \mathcal{L}_{m-1}.
\label{eq:hh_small_bias}
\end{equation}

\begin{remark}[Spouse protection and the small-bias condition]
\label{rm:hh_small_bias_spouse_protection}
In a spouse-protected pool, a large number of contracts does not by
itself ensure \eqref{eq:hh_small_bias}. All contracts begin in state $11$, where
pool exit requires both household members to die; a first death instead creates
a one-survivor state without forfeiture, and spouse-continuation payments may
reduce the account balance before eventual exit. The condition therefore also
requires sufficiently large aggregate expected forfeitures to support the
mortality-credit allocation.
\end{remark}
When \eqref{eq:hh_small_bias} holds, and the pool is large enough that
realized pool exits and realized forfeited account balances are close to their
expected aggregate levels, the random group-gain factor satisfies
$\mathbb E\!\left[\Gamma_m\right]\simeq 1$ with small variance; see
\cite{sabin2016analytics,forsyth2024optimal}. Accordingly, for the remainder of
the paper, we adopt the large-pool approximation $\Gamma_m\equiv 1$.
{With this approximation, the mortality credit for contract $\ell$ that
remains active after the mortality update at time $t_m^-$ becomes}
\begin{equation}
c_m^\ell
=
\frac{\bar\delta_{m-1}^\ell}{1-\bar\delta_{m-1}^\ell}\,\mathcal{W}_{m^-}^\ell,
\qquad \text{for all } \ell \in \mathcal{L}_{m-1}\ \text{with}\ \mathbf 1_m^\ell=1.
\label{eq:hh_credit_large_pool}
\end{equation}
Under the large-pool and small-bias assumptions in
\eqref{eq:hh_small_bias}, the rule \eqref{eq:hh_credit_large_pool} preserves
the key properties of the finite-pool allocation with high accuracy:
(i) individual-contract actuarial fairness, as expressed by
\eqref{eq:hh_fair_credit}; (ii) the pool-level budget constraint
\eqref{eq:hh_budget}, exactly when the realized group-gain factor is retained in
\eqref{eq:hh_credit_finite_pool} while surviving recipients remain
and approximately under the large-pool
approximation $\Gamma_m\equiv 1$; and (iii) non-negativity of mortality credits
whenever account balances are nonnegative.

\subsection{Household-state tontine gain rate}
\label{ssc:household_tontine_gain_rate}

{For the homogeneous spouse-protected pool studied in this paper, we
suppress the superscript $\ell$ and consider a representative contract conditional
on the contract remaining active at $t_m^-$ and having a positive account
balance.} Combining the large-pool
mortality-credit rule \eqref{eq:hh_credit_large_pool} with the household-state
pool-exit probabilities in \eqref{eq:hh_exit_prob}, the mortality credit can be
written in tontine-gain form as
\begin{equation}
c_m = g_m(H_{(m-1)^+})\,\mathcal{W}_{m^-},
\label{eq:hh_tontine_gain_rule}
\end{equation}
where the state-dependent tontine gain rate is
\begin{equation}
g_m(H_{(m-1)^+})
=
\begin{cases}
\dfrac{\delta_{R,m-1}\,\delta_{S,m-1}}
      {1-\delta_{R,m-1}\,\delta_{S,m-1}},
& \qquad H_{(m-1)^+}=11, \\[8pt]
\dfrac{\delta_{R,m-1}}{1-\delta_{R,m-1}},
& \qquad H_{(m-1)^+}=10, \\[8pt]
\dfrac{\delta_{S,m-1}}{1-\delta_{S,m-1}},
& \qquad H_{(m-1)^+}=01.
\end{cases}
\label{eq:hh_tontine_gain_piecewise}
\end{equation}
For each start-of-period household state, the gain rate has the usual
tontine-uplift form: the numerator is the corresponding one-period pool-exit
probability, and the denominator is the corresponding probability of remaining
active in the pool. Thus, the mortality credit remains proportional to the
contract's own account balance, while the proportionality factor depends on the
household state. Equations~\eqref{eq:hh_tontine_gain_rule}--%
\eqref{eq:hh_tontine_gain_piecewise} define the mortality-credit mechanism used
in the household control framework below.

\section{Household control framework}
\label{sc:hh_modeling}
Following the decision-time chronology and household-state conventions
specified in Section~\ref{sc:tontine_modeling}, we formulate the controlled
wealth recursion for a representative spouse-protected contract. Mortality
credits enter through the household-state tontine gain rate
$g_m(H_{(m-1)^+})$ in \eqref{eq:hh_tontine_gain_piecewise}. We denote the
retiree-alive states by
\begin{equation*}
\mathcal H_R:=\{11,10\}.
\end{equation*}
Withdrawal and rebalancing controls are available only in
$\mathcal H_R$; state $01$ follows the spouse-continuation payment and passive
portfolio-update conventions, while state $00$ is absorbing.

\subsection{Index positions and contract wealth}
\label{ssc:hh_index_dynamics}
We consider the investment portfolio held in a spouse-protected contract,
with access to four real (inflation-adjusted) asset classes:
(i) a domestic stock index fund,
(ii) a domestic bond index fund,
(iii) a foreign stock index fund, converted to domestic currency, and
(iv) a foreign bond index fund, converted to domestic currency.
A subscript $\iota\in\{\dom,\f\}$ distinguishes domestic and foreign asset
classes, and a superscript $s$ or $b$ denotes stock and bond index positions,
respectively.

Let $S^{\dom}(t)$, $B^{\dom}(t)$, $S^{\f}(t)$, and $B^{\f}(t)$
respectively denote the real (inflation-adjusted) \emph{amounts}
invested in the stock and bond indices of the domestic and foreign markets, respectively, at time $t\in[0,T]$. To avoid notational clutter, we occasionally write
$S_t^{\dom}\equiv S^{\dom}(t)$, $B_t^{\dom}\equiv B^{\dom}(t)$,
$S_t^{\f}\equiv S^{\f}(t)$, and $B_t^{\f}\equiv B^{\f}(t)$.

We denote by $\{X_t\}_{0\le t\le T}$ the controlled index-position process,
where
\begin{equation}
\label{eq:hh_Xt}
{
X_t
=
\bigl(S_t^{\dom},\,B_t^{\dom},\,S_t^{\f},\,B_t^{\f}\bigr)^\top
\in\mathbb R^4.
}
\end{equation}
Between decision times $t_m\in\mathcal T$, the process evolves passively
according to the index-return dynamics. For any
$t\in[t_m^+,t_{m+1}^-]$, we write
\begin{equation}
\label{eq:hh_dynamics_generic}
X_t
=
\mathcal M_{t_m,t}\bigl(X_{m^+},\,\varepsilon_{m+1}\bigr),
\qquad
t\in[t_m^+,t_{m+1}^-],
\end{equation}
where $\mathcal M_{t_m,t}$ is a transition operator and
$\varepsilon_{m+1}$ collects the exogenous return drivers over the interval
$[t_m^+,t_{m+1}^-]$. These drivers may include Brownian or jump shocks in
parametric models, or resampled blocks of historical returns in bootstrapped
settings.

We define by $\{\mathcal W_t\}_{0\le t\le T}$ the contract wealth process  before applying mortality credits and the tontine management fee, where
\begin{equation}
\label{eq:hh_financial_wealth}
\mathcal W_t
=
S_t^{\dom}+B_t^{\dom}+S_t^{\f}+B_t^{\f}
\end{equation}
In particular, at decision time $t_m^-$ we write
$\mathcal W_{m^-}:=\mathcal W_{t_m^-}$. The quantity $\mathcal W_t$ is allowed to become non-positive; under the insolvency convention, negative wealth is interpreted as accumulated debt.

\subsection{Mortality credits and tontine fee}
\label{ssc:hh_mortality_credit_fee}
{To rule out mortality-credit distribution at $t_0$ and after contract
termination or insolvency, we set the mortality credit to zero at $t_0$ and
apply it for $m\ge 1$ only when the contract remains active and satisfies the
solvency condition $\mathcal W_{m^-}>0$. Accordingly, define the modified
state-dependent tontine gain rate by}
\begin{equation}
\label{eq:hh_mod_tontine_gain}
\widehat g_m=
\begin{cases}
g_m\bigl(H_{(m-1)^+}\bigr),
& m=1,\ldots,M,\ H_{m^-}\neq 00,\ \mathcal W_{m^-}>0,\\[4pt]
0, & \text{otherwise}.
\end{cases}
\end{equation}
Here, $g_m(\cdot)$ is defined in \eqref{eq:hh_tontine_gain_piecewise}. The
resulting contract wealth after mortality-credit distribution at time
$t_m^-$, but before any fee, household payment, or portfolio allocation, is
\begin{equation}
\label{eq:hh_Wtilde}
\widetilde W_{m^-}=(1+\widehat g_m)\,\mathcal W_{m^-}.
\end{equation}
We allow for a baseline tontine management fee, deducted once per year at
 times $t_m$, $m=1,\ldots,M$, and modeled as a proportional charge on
the account balance. Let $\varrho\in(0,1)$ denote the yearly fee rate, and define
\begin{equation}
\label{eq:hh_varrho}
\varrho_m
=
\varrho\,\mathbf 1_{\{\,m\ge 1,\ \widetilde W_{m^-}>0,\ H_{m^-}\neq 00\,\}},
\qquad m=0,\ldots,M.
\end{equation}
The contract wealth after fee deduction is
\begin{equation}
\label{eq:hh_Wm_minus}
W_{m^-}=(1-\varrho_m)\,\widetilde W_{m^-},
\qquad
\text{$\widetilde W_{m^-}$ given in \eqref{eq:hh_Wtilde}}.
\end{equation}
In the remainder of the paper, $W_{m^-}$ denotes contract wealth at time
$t_m^-$ after mortality credits and tontine management fee, but before the
period-$m$ household payment and any portfolio allocation. If $W_{m^-}\le 0$,
the contract is insolvent at the pre-payment stage; scheduled household
payments may still be made, but no further trading or mortality credits are
applied (see Subsection~\ref{ssc:hh_insolvency}).

\subsection{Household payments}
\label{ssc:hh_controls}
This subsection defines the household payment process used in the control
problem. The withdrawal control is applied only in retiree-alive states $11$ and
$10$; in the spouse-only state $01$,  the automatic spouse-protection continuation payment applies. We first specify the
withdrawal bands and admissible withdrawal control, then combine controlled
withdrawals and spouse-protection continuation payments into a unified payment rule.

\subsubsection{Household withdrawal bands}
\label{ssc:hh_withdrawal_bands}
We specify withdrawal bounds directly at the household level. Let
$[\underline q_{11},\,\overline q_{11}]$ denote the admissible real (inflation-adjusted) withdrawal band when both household members are alive,
i.e.\ when $H_{m^-}=11$, and let $[\underline q_{1},\,\overline q_{1}]$ denote the
corresponding one-survivor real (inflation-adjusted) withdrawal band. We impose
\begin{equation}
\label{eq:hh_band_ordering}
0<\underline q_1\le \underline q_{11},
\qquad
\underline q_{11}\le \overline q_{11},
\qquad
\underline q_1\le \overline q_1,
\qquad
\overline q_1\le \overline q_{11}.
\end{equation}
The band $[\underline q_{11},\overline q_{11}]$ applies in state $11$ and
corresponds to a two-person household.
The band
$[\underline q_{1},\overline q_{1}]$ is the one-survivor withdrawal band and
applies when the retiree is the sole survivor, i.e.\ in state $10$.
We emphasize that state $01$ is
also a one-survivor household, but it is not a control state: once only the
spouse remains alive, payments are determined by the automatic
spouse-protection continuation rule rather than by a controlled withdrawal.
The ordering in \eqref{eq:hh_band_ordering} reflects the assumption that a
one-survivor household requires no more real (inflation-adjusted) spending than a two-person household.

For subsequent use, define the state-dependent lower and upper controlled
withdrawal bounds by
\begin{equation}
\label{eq:hh_state_bounds}
\underline q(H)
:=
\begin{cases}
\underline q_{11}, & H=11,\\[4pt]
\underline q_1, & H=10,
\end{cases}
\qquad
\overline q(H)
:=
\begin{cases}
\overline q_{11}, & H=11,\\[4pt]
\overline q_1, & H=10.
\end{cases}
\end{equation}

\subsubsection{Withdrawal control}
\label{ssc:hh_withdrawal_control}
For $m=0,\ldots,M-1$, we represent the withdrawal control by a feedback function
that maps pre-payment wealth, time, and household state into a withdrawal amount:
\[
q_m(\cdot) : (W_{m^-},t_m,H_{m^-}) \mapsto q_m=q(W_{m^-},t_m,H_{m^-}),
\]
where $W_{m^-}$ is given in \eqref{eq:hh_Wm_minus}. The withdrawal control is
available only in retiree-alive states, i.e.\ when
$H_{m^-}\in\mathcal H_R$. Its admissible set is household-state dependent and
is given by $\mathcal Z_q\bigl(W_{m^-},t_m,H_{m^-}\bigr)$, where
\begin{subnumcases}{
\label{eq:hh_Zq}
\mathcal Z_q(\cdot)=
}
[\underline q(H_{m^-}),\,\overline q(H_{m^-})],
& \(H_{m^-}\in\mathcal H_{R},
\ W_{m^-}\ge \overline q(H_{m^-}),\)
\label{eq:hh_Zq_a}
\\[5pt]
[\underline q(H_{m^-}),\,\max\{\underline q(H_{m^-}),W_{m^-}\}],
& $H_{m^-}\in\mathcal H_{R},
\ 0<W_{m^-}<\overline q(H_{m^-}),$
\label{eq:hh_Zq_b}
\\[5pt]
\{\underline q(H_{m^-})\},
& $H_{m^-}\in\mathcal H_{R},
\ W_{m^-}\le 0$,
\label{eq:hh_Zq_c}
\\[5pt]
\{0\},
& $H_{m^-}\notin\mathcal H_{R}.$
\label{eq:hh_Zq_d}
\end{subnumcases}
Here, $\underline q(\cdot)$ and $\overline q(\cdot)$ are defined in
\eqref{eq:hh_state_bounds}.
Cases~\eqref{eq:hh_Zq_a} and \eqref{eq:hh_Zq_b} apply in retiree-alive states
when the contract is solvent at $t_m^-$. In case~\eqref{eq:hh_Zq_a},
pre-payment wealth is large enough that the full household-state-dependent
withdrawal band is admissible. In case~\eqref{eq:hh_Zq_b}, the contract remains
solvent but wealth has fallen below the relevant upper bound, so the upper
endpoint is tightened while the relevant lower bound is preserved.
Case~\eqref{eq:hh_Zq_c} applies when the contract is already insolvent at
$t_m^-$; the admissible set then collapses to the relevant lower bound, so no
withdrawal control is applied. Case~\eqref{eq:hh_Zq_d} records that the
withdrawal control is inactive outside retiree-alive states.
No withdrawal control is introduced at $t_M$.

This formulation is also consistent with evidence that retired households
adjust spending in response to cash-flow changes, including categories often
perceived as relatively fixed; see the related discussion in
\cite{orozco2026money}. It is therefore natural to model withdrawals as a
flexible control within state-dependent bounds rather than as a rigid
deterministic rule.

This constraint \eqref{eq:hh_Zq} admits a natural economic interpretation. In state $11$, the
upper bound $\overline q_{11}$ represents the household's desired annual real
spending level while both members are alive, while $\underline q_{11}$ is a
contingency floor the household is willing to adopt to reduce depletion risk. In
state $10$, the corresponding quantities $\overline q_1$ and $\underline q_1$
represent the desired and minimum real spending levels for a one-survivor
retiree household. The optimal control uses the flexibility in
\eqref{eq:hh_Zq} to preserve higher withdrawals when the account is well funded,
but cuts withdrawals toward the relevant lower bound in adverse market or wealth
states.

\subsubsection{Spouse-protection continuation payment}
\label{ssc:hh_continuation_cashflow}
We now define the automatic spouse-protection continuation payment. Let
\begin{equation}
\label{eq:hh_tau_R}
\tau_R
:=
\inf\{m\in\{1,\ldots,M\}: H_{m^-}=01\},
\end{equation}
with the convention $\inf\varnothing=\infty$. The event
$\{\tau_R\le m\}$ is determined by the observed household-state history up to
time $t_m^-$. On the event $\{\tau_R<\infty\}$, the retiree dies during
$[t_{\tau_R-1}^+,t_{\tau_R}^-]$ and the spouse is alive at $t_{\tau_R}^-$.

Since the last controlled withdrawal before the retiree's death may have been
chosen under the larger two-person band, we cap the spouse-protection
continuation payment at the one-survivor upper bound. Define the
continuation-base variable recursively by
\begin{equation}
\label{eq:hh_continuation_base}
B_{-1}:=0,
\qquad
B_m=
\begin{cases}
\min\{q_m,\overline q_1\}, & H_{m^-}\in\mathcal H_{R},\\[4pt]
B_{m-1}, & H_{m^-}\in\{01,00\}.
\end{cases}
\end{equation}
The process $\{B_m\}_{m=-1}^{M-1}$ stores the capped one-survivor continuation
base associated with the most recent controlled withdrawal while the retiree was
alive. If the household first enters the spouse-only state at time $\tau_R$,
then
$B_{\tau_R-1}=\min\{q_{\tau_R-1},\overline q_1\}$ is the automatic
continuation amount inherited by the surviving spouse.

\subsubsection{Unified household payment rule}
\label{ssc:hh_household_payment_rule}
We now unify the withdrawal control in retiree-alive states 11 and 10 with automatic
spouse-protection continuation payments in the spouse-only state 01.
Specifically, for
$m=0,\ldots,M-1$, the household payment at time $t_m$ is
\begin{subnumcases}{\label{eq:hh_household_cashflow} C_m=}
q_m,
& \(H_{m^-}\in\{11,10\},\ W_{m^-}>0,\)
\label{eq:hh_household_cashflow_a}
\\[4pt]
\underline q(H_{m^-}),
& \(H_{m^-}\in\{11,10\},\ W_{m^-}\le 0,\)
\label{eq:hh_household_cashflow_b}
\\[4pt]
B_{m-1},
& \(H_{m^-}=01,\ W_{m^-}>0,\)
\label{eq:hh_household_cashflow_c}
\\[4pt]
\underline q_1,
& \(H_{m^-}=01,\ W_{m^-}\le 0,\)
\label{eq:hh_household_cashflow_d}
\\[4pt]
0,
& \(H_{m^-}=00.\)
\label{eq:hh_household_cashflow_e}
\end{subnumcases}
Cases~\eqref{eq:hh_household_cashflow_a} and \eqref{eq:hh_household_cashflow_c}
cover positive pre-payment wealth. In retiree-alive states $11$ and $10$, the
payment is the controlled withdrawal $q_m$; in the spouse-only state $01$, it is
the automatic spouse-protection continuation amount $B_{m-1}$. Cases
\eqref{eq:hh_household_cashflow_b} and~\eqref{eq:hh_household_cashflow_d}
apply when the contract is already insolvent,
with the relevant minimum payment imposed mechanically. Case~\eqref{eq:hh_household_cashflow_e} records that no household
payment is made after pool exit.

After the household payment $C_m$ is made, post-payment wealth is
\begin{equation}
\label{eq:hh_Wm_plus}
W_{m^+}
=
\begin{cases}
W_{m^-}-C_m, & H_{m^-}\in\{11,10,01\},\\[4pt]
0, & H_{m^-}=00,
\end{cases}
\qquad m=0,\ldots,M-1.
\end{equation}
At the terminal date $t_M=T$, no further household payment or rebalancing is
made. We define terminal wealth by
\begin{equation}
\label{eq:hh_terminal_wealth}
W_T
=
\begin{cases}
W_{M^-}, & H_{M^-}\in\{11,10,01\},\\[4pt]
0, & H_{M^-}=00.
\end{cases}
\end{equation}
Thus, terminal wealth is recorded only if at least one household member remains
alive at the horizon.

\begin{remark}[Minimum payments and insolvency]
\label{rm:hh_insolvency_minimum}
At first sight it may seem natural to cease household payments once the account
is depleted, i.e.\ when $W_{m^-}\le 0$. However, the lower bounds $\underline q_{11}$ and $\underline q_1$
represent minimum real spending levels for the two-person state $11$
and the one-survivor states $10$ and $01$.
Accordingly,  in retiree-alive states $11$ and $10$,
\eqref{eq:hh_household_cashflow_b} sets
$C_m=\underline q(H_{m^-})$ even when the contract is already insolvent at
$t_m^-$. Hence $W_{m^+}=W_{m^-}-C_m<0$,
which is equivalent to borrowing the shortfall.

The same principle applies to the spouse-only state $01$. When $W_{m^-}>0$,
\eqref{eq:hh_household_cashflow_c} pays the spouse-protection continuation
amount $C_m=B_{m-1}$; if this amount exceeds available wealth, the shortfall is
recorded as debt. When the contract is already insolvent at $t_m^-$,
\eqref{eq:hh_household_cashflow_d} applies the one-survivor minimum
$C_m=\underline q_1$ mechanically. Any resulting debt then evolves according to
the insolvency convention in Subsection~\ref{ssc:hh_insolvency}. Accumulated debt increases terminal shortfall if at least one
household member remains alive at the horizon.
\end{remark}

\subsection{Rebalancing control}
\label{ssc:hh_rebalancing_control}
For $m=0,\ldots,M-1$ and household states $H_{m^-}\in\mathcal H_R$, we
denote the rebalancing control by $\boldsymbol p_m(\cdot)$, a vector of
proportions of post-payment wealth allocated to the asset indices. The control
depends on the post-payment wealth $W_{m^+}$ (defined in \eqref{eq:hh_Wm_plus}),
the decision time $t_m$, and the household state $H_{m^-}$:
\[
\boldsymbol p_m(\cdot):(W_{m^+},t_m,H_{m^-})
\mapsto
\boldsymbol p_m=\boldsymbol p(W_{m^+},t_m,H_{m^-}),
\qquad H_{m^-}\in\mathcal H_R,
\]
where
\begin{equation}
\label{eq:hh_reb_control}
{
\boldsymbol p_m
=
\bigl(p_{s,m}^{\dom},\,p_{b,m}^{\dom},\,p_{s,m}^{\f}\bigr).
}
\end{equation}
We enforce no leverage and no short-selling whenever the contract is solvent.
{To define the admissible set of rebalancing controls, introduce}
\[
{
\Delta^{(3)}
:=
\bigg\{(p_1,p_2,p_3)\in[0,1]^3 \,\bigm|\,
\sum_{i=1}^{3}p_i\le 1\bigg\},
\qquad
\mathbf b^{(3)}:=(0,1,0).
}
\]
{The three explicitly parameterized proportions correspond to domestic
stock, domestic bond, and foreign stock, while the residual proportion is
allocated to the foreign bond.}

For retiree-alive states $H_{m^-}\in\mathcal H_R$, the admissible
rebalancing control set is
\begin{equation}
\label{eq:hh_Zp}
{
\mathcal Z_p(W_{m^+},t_m)
=
\begin{cases}
\Delta^{(3)}, & W_{m^+}>0,\\[4pt]
\{\mathbf b^{(3)}\}, & W_{m^+}\le 0.
\end{cases}
}
\end{equation}
{The singleton $\{\mathbf b^{(3)}\}$ is a bookkeeping placeholder for
the control variable after insolvency and represents full allocation to the
domestic bond/debt component; it is not a genuine rebalancing decision.}
Outside retiree-alive states, the admissible set \eqref{eq:hh_Zp} is not used.
No rebalancing control is introduced at $t_M$.

For $m=0,\ldots,M-1$, if $H_{m^-}\in\mathcal H_R$ and
$W_{m^+}>0$, the post-rebalancing state $X_{m^+}$ is
\begin{equation}
\label{eq:hh_Xmplus_active}
X_{m^+}
=
W_{m^+}
\bigl(
 p_{s,m}^{\dom},
 p_{b,m}^{\dom},
 p_{s,m}^{\f},
 p_{b,m}^{\f}
\bigr)^\top,
\quad\text{where}\quad
p_{b,m}^{\f}
:=
1-p_{s,m}^{\dom}-p_{b,m}^{\dom}-p_{s,m}^{\f}.
\end{equation}

If the household payment results in $W_{m^+}\le 0$, no rebalancing control is applied; instead, the insolvency convention in
Subsection~\ref{ssc:hh_insolvency} applies. If $H_{m^-}=01$ and $W_{m^+}> 0$, no rebalancing control is applied; the portfolio is updated according to the passive
rebalancing rule in Subsection~\ref{ssc:hh_spouse_only_terminated}.
If $H_{m^-}=00$, the state is absorbing and no further portfolio update occurs.

\subsection{Passive rebalancing and contract termination}
\label{ssc:hh_spouse_only_terminated}
If $H_{m^-}=01$, no new withdrawal or rebalancing control is chosen.
The spouse-protection continuation payment is determined by \eqref{eq:hh_household_cashflow}. To describe the passive
portfolio update in this state, recall that $\mathcal W_{m^-}$, defined in
\eqref{eq:hh_financial_wealth}, is the contract wealth before mortality credits
and fees, while $W_{m^-}$, defined in \eqref{eq:hh_Wm_minus}, is the wealth
after mortality credits and the tontine management fee, but before the household
payment.

When $\mathcal W_{m^-}>0$ and $W_{m^-}>0$, define the pre-payment asset-position vector by
\begin{equation}
\label{eq:hh_X_prepay}
X_{m^-}^{\mathrm{pay}}
:= \frac{W_{m^-}}{\mathcal W_{m^-}}\,X_{m^-}.
\end{equation}
The scaling in \eqref{eq:hh_X_prepay} updates the asset positions to
reflect mortality credits and the tontine fee while preserving the pre-payment
portfolio proportions.
If either $\mathcal W_{m^-}\le 0$ or $W_{m^-}\le 0$, the contract is already insolvent at the pre-payment stage, and \eqref{eq:hh_X_prepay} is not used.

If $H_{m^-}=01$ and the spouse-protection continuation payment leaves the contract solvent, \mbox{i.e.\ $W_{m^+}>0$,} the passive rebalancing rule is
\begin{equation}
\label{eq:hh_spouse_passive_update}
X_{m^+}
=
\frac{W_{m^+}}{W_{m^-}}\,X_{m^-}^{\mathrm{pay}},
\qquad
\text{when } H_{m^-}=01,\ \mathcal W_{m^-}>0,\ W_{m^-}>0,\ W_{m^+}>0.
\end{equation}
Equation \eqref{eq:hh_spouse_passive_update} preserves the pre-payment
portfolio proportions after the spouse-protection continuation payment is made.
Equivalently, the payment is funded by reducing each asset position
proportionally; no new portfolio allocation is chosen in state $01$. If
$W_{m^+}\le 0$, the passive rebalancing rule is not used and the insolvency
convention applies.

If $H_{m^-}=00$, the contract has exited the pool and neither household member remains alive.
We then set
\begin{equation}
\label{eq:hh_terminated}
C_m=0,
\qquad
W_{m^+}=0,
\qquad
X_{m^+} = (0,0,0,0)^\top.
\end{equation}
State $00$ is absorbing: once the contract has exited the pool, no further
household payments, mortality credits, or portfolio updates occur.
The reset in \eqref{eq:hh_terminated} ends the household wealth
recursion; settlement of external debt after extinction is outside the model.

\subsection{Insolvency}
\label{ssc:hh_insolvency}

For an active contract, insolvency occurs at $t_m^+$ whenever the
post-payment wealth in \eqref{eq:hh_Wm_plus} satisfies
\[
H_{m^-}\neq 00,
\qquad
W_{m^+}\le 0.
\]
Thus, insolvency is triggered after the household payment has been made and
before any rebalancing control or passive rebalancing rule is applied. If
$W_{m^-}\le 0$, the contract has carried debt forward from an earlier period.

Once the contract is insolvent, no further discretionary withdrawal or
rebalancing control is applied and no further mortality credits are
distributed. The applicable scheduled minimum payments under
\eqref{eq:hh_household_cashflow} continue while the contract remains active,
with any resulting shortfall carried as debt in the domestic bond/debt
component. Accordingly, the portfolio of an insolvent active contract is set to
\begin{equation}
\label{eq:hh_insolvent_X}
X_{m^+}
=
W_{m^+}(0,1,0,0)^\top,
\qquad
\text{when } H_{m^-}\neq 00 \text{ and } W_{m^+}\le 0.
\end{equation}
In the baseline implementation, the debt position accrues at the domestic
bond-index return plus a borrowing spread of $2\%$ per annum.

\subsection{Admissible policy set}
\label{ssc:hh_admissible_policy}
A {household control policy} is a sequence of non-anticipative feedback rules
\[
\nu=\{(q_m,\boldsymbol p_m)\}_{m=0}^{M-1},
\]
where $q_m$ is the withdrawal control and $\boldsymbol p_m$ is the
rebalancing control. The withdrawal control is applied using
$(W_{m^-},t_m,H_{m^-})$ and is constrained by
$\mathcal Z_q\bigl(W_{m^-},t_m,H_{m^-}\bigr)$ defined in \eqref{eq:hh_Zq}. The rebalancing control is defined only on retiree-alive states $H_{m^-}\in\mathcal H_R$, and is
constrained by $\mathcal Z_p(W_{m^+},t_m)$ defined in \eqref{eq:hh_Zp}.
{The admissible household control policy set is}
\begin{equation}
\label{eq:hh_A}
\mathcal A
=
\left\{
\nu=\{(q_m,\boldsymbol p_m)\}_{m=0}^{M-1}
\ \middle|\
\begin{array}{l}
q_m(W_{m^-},t_m,H_{m^-})
\in
\mathcal Z_q\bigl(W_{m^-},t_m,H_{m^-}\bigr),\\[3pt]
\boldsymbol p_m(W_{m^+},t_m,H_{m^-})
\in
\mathcal Z_p(W_{m^+},t_m),
\quad \text{whenever } H_{m^-}\in\mathcal H_R
\end{array}
\right\}.
\end{equation}
Outside retiree-alive states, no rebalancing control is defined: state $01$
uses the passive rebalancing rule, while state $00$ is absorbing. The
solvency restrictions are already embedded in
$\mathcal Z_q$ and $\mathcal Z_p$, defined in
\eqref{eq:hh_Zq} and \eqref{eq:hh_Zp}, respectively.

\section{Household reward-risk optimization}
\label{sc:household_reward_risk}
We now specify the household-level reward and risk criteria used in the
household optimization problem. Throughout, all household payments and contract
wealth are measured in real (inflation-adjusted) terms. The formulation reflects
two central features of the spouse-protection setting: household payments depend
on the household state, and the terminal reserve targets
depend on the terminal household state.

Let $x_0=X_{0^-}$ and $h_0=H_{0^-}$ denote the initial index-position vector
and household state, respectively. For any admissible policy $\nu$, we write
$\mathbb E_{x_0,h_0}^{\nu}[\cdot]
:=
\mathbb E^{\nu}\!\left[\cdot \,\middle|\,
X_{0^-}=x_0,\ H_{0^-}=h_0\right]$,
where the expectation is taken over the joint law of the exogenous asset-return
and household-state processes, with contract wealth and household payments
induced by the policy $\nu$.

\subsection{Reward}
\label{ssc:household_reward}
{The reward criterion is the expected cumulative real (inflation-adjusted) household payments over the planning horizon.
Given $\nu\in\mathcal A$, let $C_m^{\nu}$ denote the
period-$m$ household payment generated by the unified payment rule
\eqref{eq:hh_household_cashflow}. This rule already distinguishes controlled
withdrawals in retiree-alive states from automatic spouse-protection
continuation payments in the spouse-only state, and incorporates the lower-bound payment convention when the contract is already insolvent. We therefore define the household reward functional by}
\begin{equation}
\label{eq:household_reward}
\mathcal R(\nu)
:=
\mathbb E_{x_0,h_0}^{\nu}
\!\left[
\sum_{m=0}^{M-1} C_m^{\nu}
\right].
\end{equation}
\begin{remark}[Discounting and household-state averaging]
\label{rm:hh_discounting_mortality}
All payments and wealth are expressed in real dollars, and the baseline reward
functional \eqref{eq:household_reward} sets the real discount factor equal to
one at each decision time. More generally, prescribed real discount factors
$\{d_m\}_{m=0}^{M-1}$ could be introduced by replacing
$\sum_{m=0}^{M-1} C_m^\nu$ with
$\sum_{m=0}^{M-1} d_m C_m^\nu$.

Mortality is incorporated through the stochastic household-state process
$\{H_t\}_{0\le t\le T}$. Accordingly, the expectation in
\eqref{eq:household_reward} averages the state-contingent household payments
over the joint law of asset returns and household mortality, so no additional
survival-probability weighting is required.
\end{remark}

\subsection{Risk}
\label{ssc:household_risk}
{On the risk side, we do not use terminal contract wealth itself as the risk variable, nor do we penalize low terminal wealth uniformly across all
terminal household states. In a single-retiree formulation with a fixed planning horizon, terminal wealth can often be used directly as a proxy for
terminal adequacy. In a spouse-protected contract, however, the same terminal
wealth has different implications depending on the household state:
it may be inadequate for a two-person household, sufficient for a one-survivor household,
or irrelevant for continuing household reserve needs once neither household
member remains alive. We therefore measure terminal risk through a
state-contingent shortfall variable, applying CVaR to the shortfall from
state-dependent terminal reserve targets rather than directly to terminal
wealth.}

Let $W_T^\nu$ denote the {contract wealth} at time $T$ under policy
$\nu\in\mathcal A$, and write $H_T:=H_{M^-}$ for the terminal household state
after the final mortality update. To measure downside risk under
policy $\nu$, we define the associated state-contingent terminal shortfall, denoted by $L_T^{\nu}$.

Specifically, let $w_{11}^\ast$, $w_{10}^\ast$, and $w_{01}^\ast$ be
nonnegative {terminal reserve targets}, corresponding respectively to
the terminal household states $11$, $10$, and $01$. These reserve targets are
model inputs and may be {specified externally using} household
retirement-spending benchmarks. A natural ordering is
\[
w_{11}^\ast \ge \max\{w_{10}^\ast,w_{01}^\ast\},
\]
since state $11$ corresponds to a two-person household. The one-survivor
targets $w_{10}^\ast$ and $w_{01}^\ast$ may be chosen equal or specified
differently, depending on whether retiree-only and spouse-only terminal reserve
needs are treated as symmetric.

The {state-contingent terminal shortfall} random variable $L_T^{\nu}$ is
\begin{equation}
\label{eq:household_shortfall}
L_T^{\nu}
:=
\mathbf 1_{\{H_T=11\}}(w_{11}^\ast-W_T^\nu)_+
+
\mathbf 1_{\{H_T=10\}}(w_{10}^\ast-W_T^\nu)_+
+
\mathbf 1_{\{H_T=01\}}(w_{01}^\ast-W_T^\nu)_+,
\end{equation}
where $(x)_+ := \max\{x,0\}$. No shortfall term is included when
$H_T=00$, since no household member remains alive at the horizon.

To focus on adverse household outcomes, we apply CVaR directly to the
state-contingent terminal shortfall. For a confidence level
$\alpha\in(0,1)$, define the household risk functional under policy $\nu$ by
\begin{equation}
\label{eq:household_cvar_explicit}
\mathcal Q_\alpha(\nu)
:=
\operatorname{CVaR}_\alpha\!\left(L_T^{\nu}\right)
:=
\inf_{\eta\in\mathbb R}
\left\{
\eta+\frac{1}{1-\alpha}
\mathbb E_{x_0,h_0}^{\nu}
\!\left[(L_T^{\nu}-\eta)_+\right]
\right\}.
\end{equation}
For example, $\operatorname{CVaR}_{0.95}(L_T^\nu)$ corresponds to the upper $5\%$
tail of the household terminal shortfall. This specification penalizes
inadequate terminal reserves only in household states where at least one
household member remains alive, with the shortfall target determined by the
terminal household state.

\subsection{Household value function}
\label{ssc:household_scalarized_objective}
The household optimization problem trades off two competing criteria: increasing expected cumulative household payments and reducing terminal reserve shortfall risk. We combine these criteria using a scalarization parameter $\lambda>0$, which converts the reward--risk trade-off into a single scalar objective. Since $\mathcal Q_\alpha(\nu)$ is a risk functional applied to a loss variable, it enters the objective with a negative sign.

{The household value function associated with this scalarized
reward--risk criterion is}
\begin{equation}
\label{eq:household_scalarized_objective}
V(x_0,h_0,t_0^-)
:=
\sup_{\nu\in\mathcal A}
\left\{
\mathcal R(\nu)-\lambda\,\mathcal Q_\alpha(\nu)
\right\}.
\end{equation}
{For any admissible household control policy
$\nu\in\mathcal A$ and auxiliary CVaR threshold $\eta\in\mathbb R$, define the
household objective associated with $(\nu,\eta)$ by}
\begin{equation}
\label{eq:Vhh_eta}
{
\begin{aligned}
V(x_0,h_0,t_0^-;\nu,\eta)
:=
\mathbb E_{x_0,h_0}^{\nu}\Bigg[
\sum_{m=0}^{M-1} C_m^{\nu}
-\lambda\left(
\eta+\frac{1}{1-\alpha}(L_T^{\nu}-\eta)_+
\right)
\Bigg],
\end{aligned}
}
\end{equation}
{subject to the following system evolution and control-admissibility
conditions:}
\begin{equation}
\label{eq:hh_control_problem_constraints}
{
\begin{cases}
\text{(initial state)}
&
X_{0^-}=x_0,
\qquad
H_{0^-}=h_0,
\\[4pt]
\text{(index-position dynamics)}
&
\{X_t\}\text{ evolves via }\eqref{eq:hh_dynamics_generic},
\qquad t\notin\mathcal T,
\\[4pt]
\text{(household state)}
&
H_{m^-}\text{ evolves according to }
\eqref{eq:hh_transition_11}\text{--}\eqref{eq:hh_transition_01},
\\[4pt]
\text{(tontine and fee)}
&
\mathcal W_{m^-},\widetilde W_{m^-},W_{m^-}
\text{ are computed via }
\eqref{eq:hh_mod_tontine_gain}\text{--}\eqref{eq:hh_Wm_minus},
\\[4pt]
\text{(withdrawal control)}
&
q_m\in
\mathcal Z_q(W_{m^-},t_m,H_{m^-})
\text{ as defined in }\eqref{eq:hh_Zq},
\\[4pt]
\text{(household payment)}
&
B_m,C_m,W_{m^+}
\text{ are computed via }
\eqref{eq:hh_continuation_base}\text{--}\eqref{eq:hh_Wm_plus},
\\[4pt]
\text{(rebalancing control)}
&
\boldsymbol p_m\in
\mathcal Z_p(W_{m^+},t_m)
\text{ when }H_{m^-}\in\mathcal H_R,
\text{ as defined in }\eqref{eq:hh_Zp},
\\[4pt]
\text{(asset update)}
&
X_{m^+}\text{ is computed via }
\eqref{eq:hh_Xmplus_active},
\eqref{eq:hh_X_prepay}\text{--}\eqref{eq:hh_spouse_passive_update},
\eqref{eq:hh_terminated},
\text{ or }\eqref{eq:hh_insolvent_X},
\end{cases}
}
\end{equation}
{Using the explicit CVaR representation
\eqref{eq:household_cvar_explicit}, and since $\nu$ and $\eta$ range over the
independent feasible sets $\mathcal A$ and $\mathbb R$, respectively, the
iterated suprema commute. Hence, subject to the system evolution and
control-admissibility conditions in
\eqref{eq:hh_control_problem_constraints}, the household value function
$V(x_0,h_0,t_0^-)$ can equivalently be written as}
\begin{equation}
\label{eq:Vhh_value_function}
{
V(x_0,h_0,t_0^-)
=
\sup_{\nu\in\mathcal A}\ \sup_{\eta\in\mathbb R}
V(x_0,h_0,t_0^-;\nu,\eta)
=
\sup_{\eta\in\mathbb R}\ \sup_{\nu\in\mathcal A}
V(x_0,h_0,t_0^-;\nu,\eta).
}
\end{equation}
\begin{remark}[Pre-commitment and induced time-consistent interpretation]
\label{rm:hh_precommitment_time_consistency}
The scalarized problem \eqref{eq:household_scalarized_objective} is evaluated
at inception and therefore defines a pre-commitment objective. Since CVaR is
generally not time-consistent in multi-period settings, the reward-risk problem
is formally time-inconsistent: after observing future wealth and household
states, the household may prefer to solve a different continuation problem.
Nevertheless, the pre-commitment formulation is useful for identifying the
inception-time reward-risk trade-off. {If
$(\nu_\lambda^\ast,\eta_\lambda^\ast)$ attains the suprema in
\eqref{eq:Vhh_value_function}, then fixing $\eta_\lambda^\ast$ gives the
induced expected-value control problem}
\[
\sup_{\nu\in\mathcal A}
\mathbb E_{x_0,h_0}^{\nu}
\!\left[
\sum_{m=0}^{M-1} C_m^{\nu}
-
\frac{\lambda}{1-\alpha}
(L_T^{\nu}-\eta_\lambda^\ast)_+
\right],
\]
up to the constant term $-\lambda\eta_\lambda^\ast$. With the threshold fixed,
the objective is an additive expected-value criterion with a terminal shortfall
penalty and therefore has the usual dynamic-programming interpretation.
We interpret the computed policy $\nu_\lambda^\ast$ as the induced
time-consistent strategy for the fixed-threshold objective associated with
$(\alpha,\lambda,\eta_\lambda^\ast)$;
see, for example,
\cite{PA2020a,strub2019,dang2026multi}.
\end{remark}

For a fixed confidence level $\alpha$, let $\nu_\lambda^\ast$ denote an
optimizer of \eqref{eq:household_scalarized_objective} for a given
$\lambda>0$. The household reward-risk frontier is the set
\begin{equation}
\label{eq:household_reward_risk_frontier}
\mathcal S(\alpha)
:=
\left\{
\left(
\mathcal Q_\alpha(\nu_\lambda^\ast),
\mathcal R(\nu_\lambda^\ast)
\right)
:\lambda>0
\right\}.
\end{equation}
{As $\lambda>0$ varies, the corresponding optimizers of
\eqref{eq:household_scalarized_objective} generate this set. We report the
frontier with $\mathcal Q_\alpha(\nu_\lambda^\ast)$ on the horizontal axis
and $\mathcal R(\nu_\lambda^\ast)$ on the vertical axis.}

\section{Neural network formulation}
\label{sec:NNs}
To solve the household reward--risk stochastic control problem numerically, we
adopt a neural-network policy-approximation approach in the spirit of the
rapidly developing literature on direct NN-based control methods
\cite{buehler2019deep, li2019data, reppen2023deep, reppen2023deepMF, van2024global, orozco2026money}.
These methods avoid classical dynamic-programming discretizations by
parameterizing the control itself and training it end-to-end. They are often
described as ``global-in-time'' machine-learning approaches to stochastic control
\cite{hu2024recent}, in contrast to stacked-NN schemes that use a separate
network at each rebalancing step \cite{tsang2020deep, han2016deep}.

\subsection{Approximation of admissible controls}
\label{ssc:NN_household_controls}
{The central idea of our NN formulation is to approximate an admissible household control policy $\nu\in\mathcal A$ by parameterizing its two feedback components with separate feedforward, fully connected neural networks:
a withdrawal network and a rebalancing network.
Specifically, given trainable network parameters
$\boldsymbol{\theta}_q$ and $\boldsymbol{\theta}_p$, define the
NN-parameterized withdrawal and rebalancing controls by}
\begin{equation}
\label{eq:NN_household_control}
\begin{aligned}
{
\widehat q_m(\cdot;\boldsymbol{\theta}_q)
}
&{:=
\widehat q\!\left(
W_{m^-},\,t_m,\,H_{m^-};\,\boldsymbol{\theta}_q
\right)
\simeq q_m(W_{m^-},t_m,H_{m^-}),}
\\[4pt]
{
\widehat{\boldsymbol p}_m(\cdot;\boldsymbol{\theta}_p)
}
&{:=
\widehat{\boldsymbol p}\!\left(
W_{m^+},\,t_m,\,H_{m^-};\,\boldsymbol{\theta}_p
\right)
\simeq \boldsymbol p_m(W_{m^+},t_m,H_{m^-}).}
\end{aligned}
\end{equation}
The NN-parameterized household control policy is the sequence of paired controls
\begin{equation}
\label{eq:NN_household_policy}
{
\widehat\nu
=
\left\{
\bigl(
\widehat q_m(\cdot;\boldsymbol{\theta}_q),
\widehat{\boldsymbol p}_m(\cdot;\boldsymbol{\theta}_p)
\bigr)
\right\}_{m=0}^{M-1},}
\end{equation}
where  $\widehat q_m(\cdot;\boldsymbol{\theta}_q)$
and
$\widehat{\boldsymbol p}_m(\cdot;\boldsymbol{\theta}_p)$
are defined in \eqref{eq:NN_household_control}.
The NN-parameterized controls in \eqref{eq:NN_household_policy} are applied according to the admissibility conventions in Section~\ref{sc:hh_modeling}: the withdrawal
control is applied only when $H_{m^-}\in\mathcal H_R$ and $W_{m^-}>0$, while
the rebalancing control is evaluated only when $H_{m^-}\in\mathcal H_R$ and
$W_{m^+}>0$. Outside these conditions, the corresponding network output is not
used; the household-payment convention, passive rebalancing update, termination
convention, or insolvency convention applies instead.

{Let
\[
\widehat{\mathcal A}
:=
\big\{
\widehat\nu\in\mathcal A
~|~
\widehat\nu \text{ has the form \eqref{eq:NN_household_policy}}
\big\}
\]
denote the class of admissible NN-parameterized household control policies.
We now approximate $\nu$ in the household objective
$V(x_0,h_0,t_0^-;\nu,\eta)$ by restricting attention to the NN-parameterized
policy class $\widehat{\mathcal A}$. Specifically, for fixed
$\widehat\nu\in\widehat{\mathcal A}$ and $\eta\in\mathbb R$, define the
NN-induced household objective by
}
\begin{equation}
\label{eq:VNN_household}
\begin{aligned}
V_{NN}(x_0,h_0,t_0^-;\widehat\nu,\eta)
:=
\mathbb E_{x_0,h_0}^{\widehat\nu}\Bigg[
\sum_{m=0}^{M-1} C_m^{\widehat\nu}
-\lambda\left(
\eta+\frac{1}{1-\alpha}(L_T^{\widehat\nu}-\eta)_+
\right)
\Bigg].
\end{aligned}
\end{equation}
{Here, the expectation in \eqref{eq:VNN_household} is taken under the
NN-induced system evolution obtained from
\eqref{eq:hh_control_problem_constraints}, with the controls
$q_m(\cdot)$ and $\boldsymbol p_m(\cdot)$ replaced by their NN
parameterizations $\widehat q_m(\cdot)$ and
$\widehat{\boldsymbol p}_m(\cdot)$ in
\eqref{eq:NN_household_control}.}
The household value function $V(x_0,h_0,t_0^-)$, defined in \eqref{eq:Vhh_value_function}, is approximated by restricting
the policy search to $\widehat{\mathcal A}$:
\begin{equation}
\label{eq:VNN_household_value}
V(x_0,h_0,t_0^-)
\simeq
V_{NN}(x_0,h_0,t_0^-)
:=
\sup_{\widehat\nu\in\widehat{\mathcal A}}\ \sup_{\eta\in\mathbb R}
V_{NN}(x_0,h_0,t_0^-;\widehat\nu,\eta).
\end{equation}

\subsection{Network architecture for controls}
\label{ssc:NN_architecture}
{The network architecture is designed so that admissibility is enforced by
construction. We use one feedforward network for the withdrawal control and one
feedforward network for the rebalancing control, with time, wealth, and
household state used as inputs. The output layers are chosen to map
unconstrained network outputs into the admissible sets
\eqref{eq:hh_Zq} and \eqref{eq:hh_Zp}.}

\paragraph{Withdrawal control.}
Let $z_q\in\mathbb R$ denote the pre-activation output of the withdrawal
network's final layer. For
{$H_{m^-}\in\mathcal H_R$ and $W_{m^-}>0$}, the state-dependent lower and upper withdrawal bounds are $\underline q(H_{m^-})$ and
$\overline q(H_{m^-})$, as defined in \eqref{eq:hh_state_bounds}. We enforce
the admissibility condition \eqref{eq:hh_Zq} using the scaled sigmoid
transformation
\begin{equation}
\label{eq:hatq_theta}
\widehat q\!\left(
W_{m^-},t_m,H_{m^-};\boldsymbol{\theta}_q
\right)
=
\underline q(H_{m^-})
+
\max\!\left(
\min\{\overline q(H_{m^-}),W_{m^-}\}
-\underline q(H_{m^-}),\,0
\right)
\frac{1}{1+e^{-z_q}}.
\end{equation}
Since the sigmoid function maps into $(0,1)$, this ensures that
$
\widehat q_m(\cdot)
\in
\mathcal Z_q(W_{m^-},t_m,H_{m^-})$
whenever {$H_{m^-}\in\mathcal H_R$ and $W_{m^-}>0$}. If
{$H_{m^-}\notin\mathcal H_R$ or $W_{m^-}\le0$}, the withdrawal
network output is not used.

\paragraph{Rebalancing control.}
{The rebalancing network outputs four logits
$(z_1,z_2,z_3,z_4)\in\mathbb R^4$. Applying a four-way softmax gives}
\begin{equation}
\label{eq:hatp_theta}
{
\widehat p_m^{\, (i)}
=
\frac{e^{z_i}}{\sum_{\ell=1}^{4}e^{z_\ell}},
\qquad i=1,\ldots,4.
}
\end{equation}
{This guarantees
$\widehat p_m^{\, (i)}\in[0,1]$ and
$\sum_{i=1}^{4}\widehat p_m^{\, (i)}=1$. The first three components define
the explicitly parameterized rebalancing control,
$\widehat{\boldsymbol p}_m(\cdot)
=\bigl(\widehat p_m^{\, (1)},\widehat p_m^{\, (2)},
\widehat p_m^{\, (3)}\bigr)$, corresponding to domestic stock, domestic bond,
and foreign stock, while $\widehat p_m^{\, (4)}$ is the residual foreign-bond
allocation.}

{For $m=0,\ldots,M-1$, the softmax transformation enforces the simplex
constraint in $\mathcal Z_p(W_{m^+},t_m)$ automatically whenever
$H_{m^-}\in\mathcal H_R$ and $W_{m^+}>0$.
If $H_{m^-}\notin\mathcal H_R$ or $W_{m^+}\le0$, the rebalancing network
output is not used. If the contract becomes insolvent after the household
payment, the insolvency convention applies before any rebalancing is carried
out. In state $01$, the passive rebalancing rule applies; in state $00$, the
contract is absorbing. No terminal rebalancing control is introduced at
$t_M$.}

{With these output transformations in place, an NN-parameterized policy
$\widehat\nu$ is fully determined by its trainable network parameters
$(\boldsymbol{\theta}_q,\boldsymbol{\theta}_p)$. Accordingly, we write
$V_{NN}(\cdot;\widehat\nu,\eta)$ as
$V_{NN}(\cdot;\boldsymbol{\theta}_q,\boldsymbol{\theta}_p,\eta)$ to
make this dependence explicit. The optimization problem
\eqref{eq:VNN_household_value} then becomes the unconstrained optimization
problem}
\begin{equation}
\label{eq:VNN_theta}
{
V_{NN}(x_0,h_0,t_0^-)
=
\sup_{\boldsymbol{\theta}_q,\boldsymbol{\theta}_p,\eta}
V_{NN}
(x_0,h_0,t_0^-;\boldsymbol{\theta}_q,\boldsymbol{\theta}_p,\eta).
}
\end{equation}
{Here, the supremum is taken over all trainable weights and biases in the
withdrawal and rebalancing networks, together with the scalar CVaR threshold
$\eta$. We denote the optimal trainable parameters and CVaR threshold by
$\boldsymbol{\theta}_q^\ast$, $\boldsymbol{\theta}_p^\ast$, and $\eta^\ast$,
respectively. }

{We emphasize that, although the original household control problem
\eqref{eq:Vhh_value_function} is constrained, the NN-parameterized training
problem in \eqref{eq:VNN_theta} is unconstrained in the trainable variables. The
withdrawal and rebalancing admissibility constraints are enforced through the
output transformations in \eqref{eq:hatq_theta}--\eqref{eq:hatp_theta}. This
permits the use of standard gradient-based optimizers such as Adam.}

\subsection{Empirical objective}
\label{ssc:approx_NN}
{To evaluate the NN-induced objective in \eqref{eq:VNN_theta}, we approximate
its expectation using a finite training sample of $N$ independent realizations
of the exogenous asset-return and household-state processes. These realizations
are indexed by $n=1,\ldots,N$, and all sample-dependent quantities carry the
superscript ``$(n)$''. For example,  $\{X_t^{(n)}\}_{0\le t\le T}$ is the sample-$n$ realization of the controlled index-position process, $\{H_{m^-}^{(n)}\}_{m=0}^{M}$ is the corresponding household-state realization, and $W_T^{(n)}$ is the terminal
contract wealth. Unless otherwise noted, the dependence of controlled pathwise quantities, such as $X_t^{(n)}$ and $W_T^{(n)}$, on the current
network parameters $(\boldsymbol{\theta}_q,\boldsymbol{\theta}_p)$ is suppressed
to keep notation light.

For any
$(\boldsymbol{\theta}_q,\boldsymbol{\theta}_p)$, let
$C_m^{(n)}$ denote the period-$m$ household payment generated 
along the $n$-th training path under the current network parameters
$(\boldsymbol{\theta}_q,\boldsymbol{\theta}_p)$.
The corresponding sample-wise household terminal shortfall is
\begin{equation}
\label{eq:NN_sample_shortfall}
L_T^{(n)}
=
\mathbf 1_{\{H_T^{(n)}=11\}}(w_{11}^\ast-W_T^{(n)})_+
+
\mathbf 1_{\{H_T^{(n)}=10\}}(w_{10}^\ast-W_T^{(n)})_+
+
\mathbf 1_{\{H_T^{(n)}=01\}}(w_{01}^\ast-W_T^{(n)})_+.
\end{equation}
where $H_T^{(n)}:=H_{M^-}^{(n)}$.}
The associated NN-induced objective
$V_{NN} (x_0,h_0,t_0^-;\boldsymbol{\theta}_q,\boldsymbol{\theta}_p,\eta)$
is approximated by the empirical objective
\begin{equation}
\label{eq:hh_VNN_sim}
\begin{aligned}
V_{NN}
(x_0,h_0,t_0^-;\boldsymbol{\theta}_q,\boldsymbol{\theta}_p,\eta)
&\approx
\widehat V_{NN}
(x_0,h_0,t_0^-;\boldsymbol{\theta}_q,\boldsymbol{\theta}_p,\eta)
\\
&:=
\frac{1}{N}\sum_{n=1}^{N}
\sum_{m=0}^{M-1} C_m^{(n)}
-\lambda\left[
\eta
+
\frac{1}{(1-\alpha)N}
\sum_{n=1}^{N}
\bigl(L_T^{(n)}-\eta\bigr)_+
\right].
\end{aligned}
\end{equation}
{We train the NN policy by maximizing the empirical objective
\eqref{eq:hh_VNN_sim} over the network parameters and the CVaR threshold. Thus,}
\begin{equation}
\label{eq:loss}
(\boldsymbol{\theta}_q^\ast,\boldsymbol{\theta}_p^\ast,\eta^\ast)
:=
\arg\max_{\boldsymbol{\theta}_q,\boldsymbol{\theta}_p,\eta}
\widehat V_{NN}
(x_0,h_0,t_0^-;\boldsymbol{\theta}_q,\boldsymbol{\theta}_p,\eta).
\end{equation}
Equivalently, one may minimize the empirical loss
\[
\mathcal L(\boldsymbol{\theta}_q,\boldsymbol{\theta}_p,\eta)
:=
-\widehat V_{NN}
(x_0,h_0,t_0^-;\boldsymbol{\theta}_q,\boldsymbol{\theta}_p,\eta).
\]
The resulting learned withdrawal and rebalancing controls are the sequences
\[
\widehat q^\ast
:=
\left\{
\widehat q_m^\ast(\cdot)
\right\}_{m=0}^{M-1},
\qquad
\widehat{\boldsymbol p}^{\,\ast}
:=
\left\{
\widehat{\boldsymbol p}_m^{\,\ast}(\cdot)
\right\}_{m=0}^{M-1},
\]
where, for each $m=0,\ldots,M-1$,
\[
\widehat q_m^\ast(\cdot)
:=
\widehat q_m(\cdot;\boldsymbol{\theta}_q^\ast),
\qquad
\widehat{\boldsymbol p}_m^{\,\ast}(\cdot)
:=
\widehat{\boldsymbol p}_m(\cdot;\boldsymbol{\theta}_p^\ast).
\]
Thus, the learned NN household control policy is
\[
\widehat\nu^\ast
=
\left(
\widehat q^\ast,\widehat{\boldsymbol p}^{\,\ast}
\right)
=
\left\{
\bigl(
\widehat q_m^\ast(\cdot),
\widehat{\boldsymbol p}_m^{\,\ast}(\cdot)
\bigr)
\right\}_{m=0}^{M-1}.
\]
The network architecture, training hyperparameters, sequential
transfer-learning protocol, and out-of-sample evaluation design used in the
numerical experiments are reported in
Appendix~\ref{app:NN_training}.

\section{Spouse-continuation load and book-level diversification}
\label{sc:spouse_continuation_load}
This section quantifies the spouse-continuation cost induced by an optimal
household control policy. We first consider the representative-contract
expected-cost load and tail risk, then book-level diversification, and finally
use the excess upper-tail cost after diversification to construct the
risk-loaded continuation cost and payment-scale load. These are reporting
quantities and do not enter the household optimization.

Let $\nu^\ast$ denote an optimal household control policy for the reward--risk
problem \eqref{eq:household_scalarized_objective}. Under $\nu^\ast$, define the
total payments in the retiree-alive states and in the spouse-only phase
 respectively by
\begin{equation}
\label{eq:cont_payment_components}
Y_r^{\nu^\ast}
:=
\sum_{m=0}^{M-1}
C_m^{\nu^\ast}\,
\mathbf 1_{\{H_{m^-}\in\mathcal H_R\}},
\qquad
Z_c^{\nu^\ast}
:=
\sum_{m=0}^{M-1}
C_m^{\nu^\ast}\,
\mathbf 1_{\{H_{m^-}=01\}}.
\end{equation}
Here, $C_m^{\nu^\ast}$ denotes the household payment at $t_m$ generated by the
unified payment rule \eqref{eq:hh_household_cashflow} under $\nu^\ast$.
All payment quantities are expressed in real dollars, and no additional discount factor is applied.

For the remainder of this section, we suppress the policy superscript and write
$Y_r:=Y_r^{\nu^\ast}$ and $Z_c:=Z_c^{\nu^\ast}$. We refer to $Z_c$ as the
representative-contract spouse-continuation cost. Here, ``cost'' refers to cumulative spouse-continuation payments
over the planning horizon. Under the withdrawal and capped spouse-continuation payment rules, and using
$\overline q_{11}\ge\overline q_1$, the finite horizon gives
\begin{equation}
\label{eq:cont_payment_bounds}
0\le Y_r\le M\overline q_{11},
\qquad\qquad
0\le Z_c\le M\overline q_1.
\end{equation}

\subsection{Representative-contract expected-cost load and tail risk}
\label{ssc:cont_actuarial_load}
For brevity, write
$\mathbb E^{\nu^\ast}[\cdot]
:=\mathbb E_{x_0,h_0}^{\nu^\ast}[\cdot]$
for expectation under the fixed policy and initial state.
The expected-cost spouse-continuation load, expressed as the share of expected total household payments attributable to the spouse-continuation cost, is
\begin{equation}
\label{eq:cont_load_ev}
f_c(\nu^\ast)
:=
\frac{
\mathbb E^{\nu^\ast}[Z_c]
}{
\mathbb E^{\nu^\ast}[Y_r]
+
\mathbb E^{\nu^\ast}[Z_c]
}.
\end{equation}
Let $\alpha_c\in(0,1)$ denote an upper-tail probability. For an integrable
nonnegative cost $Z$ generated under $\nu^\ast$, define
\begin{equation}
\label{eq:cont_cvar_def}
\mathrm{CVaR}_{\alpha_c}^{+}(Z)
:=
\inf_{\eta\in\mathbb R}
\left\{
\eta+
\frac{1}{\alpha_c}
\mathbb E^{\nu^\ast}[(Z-\eta)_+]
\right\},
\end{equation}
and define the excess upper-tail cost above the mean by
\begin{equation}
\label{eq:cont_excess_cvar}
\Delta_{\alpha_c}^{+}(Z)
:=
\mathrm{CVaR}_{\alpha_c}^{+}(Z)
-
\mathbb E^{\nu^\ast}[Z].
\end{equation}
For example, $\alpha_c=0.05$ corresponds to the upper $5\%$ tail. In
particular, $\Delta_{\alpha_c}^{+}(Z_c)$ measures the amount by which the
representative-contract upper-tail spouse-continuation cost exceeds its
mean. We retain
$\mathrm{CVaR}_{\alpha_c}^{+}(Z_c)$ as a
representative-contract tail-cost benchmark.
The tail probability $\alpha_c$ is used for risk loading and need not coincide
with the confidence level in the household reward--risk objective.

\subsection{Book-level diversification}
\label{ssc:cont_book_risk}
Let $\mathcal G$ denote the sigma-field generated by the common market
environment, namely, the asset returns over the planning horizon to which all
contracts in the book are exposed. Consider a homogeneous book of
spouse-protected contracts following the same policy $\nu^\ast$, with household
mortality histories conditionally independent and identically distributed
given~$\mathcal G$. We use $J$ to denote the number of contracts over which
spouse-continuation costs are aggregated; it is distinct from the tontine-pool
size underlying the homogeneous large-pool gain rates. We first examine the
average per-contract cost for a finite book and then its large-book limit.

\subsubsection{Finite-book average per-contract cost}
Consider a book of $J$ spouse-protected contracts, and let $Z_{c,j}$ denote the
continuation cost of contract $j$, where $j=1,\ldots,J$. Define the finite-book
average per-contract spouse-continuation cost by
\begin{equation}
\label{eq:cont_book_average}
\overline Z_{c,J}
:=
\frac{1}{J}\sum_{j=1}^{J}Z_{c,j}.
\end{equation}
By homogeneity, linearity of expectation, and the positive homogeneity and
subadditivity of CVaR,
\begin{equation}
\label{eq:cont_book_subadditivity}
\mathbb E^{\nu^\ast}
\bigl[\overline Z_{c,J}\bigr]
=
\mathbb E^{\nu^\ast}
\bigl[Z_c\bigr],
\qquad
\mathrm{CVaR}_{\alpha_c}^{+}
\bigl(\overline Z_{c,J}\bigr)
\le
\mathrm{CVaR}_{\alpha_c}^{+}
\bigl(Z_c\bigr),
\qquad
\Delta_{\alpha_c}^{+}
\bigl(\overline Z_{c,J}\bigr)
\le
\Delta_{\alpha_c}^{+}
\bigl(Z_c\bigr).
\end{equation}

\subsubsection{Large-book limiting average per-contract cost}
By \eqref{eq:cont_payment_bounds}, $Z_c$ is bounded. Hence, under the
conditional-i.i.d.\ assumption, the conditional law of large numbers gives
\begin{equation}
\label{eq:cont_large_book_limit}
\overline Z_{c,J}
\longrightarrow
\overline Z_c
:=
\mathbb E^{\nu^\ast}
\bigl[Z_c\mid\mathcal G\bigr],
\qquad J\to\infty, \quad \text{almost surely}.
\end{equation}
We refer to $\overline Z_c$ as the large-book limiting average
per-contract spouse-continuation cost.
Equation~\eqref{eq:cont_large_book_limit} shows that idiosyncratic
household-mortality risk in the continuation cost is diversified away in the
large-book limit, whereas common-market risk remains.
Furthermore, the law of iterated expectations gives
\begin{equation}
\label{eq:cont_tower_property}
\mathbb E^{\nu^\ast}
\bigl[\overline Z_c\bigr]
=
\mathbb E^{\nu^\ast}
\left[
\mathbb E^{\nu^\ast}
\bigl[Z_c\mid\mathcal G\bigr]
\right]
=
\mathbb E^{\nu^\ast}
\bigl[Z_c\bigr].
\end{equation}
Thus, book-level diversification leaves the expected cost
per contract unchanged.

\vspace*{-0.25cm}
\paragraph{Finite-book convergence rate.}
Define the average conditional idiosyncratic variance by
\begin{equation}
\label{eq:cont_idiosyncratic_variance}
\sigma_{c,\mathrm{id}}^2(\nu^\ast)
:=
\mathbb E^{\nu^\ast}
\left[
\operatorname{Var}^{\nu^\ast}
\bigl(Z_c\mid\mathcal G\bigr)
\right]
\le
\frac{M^2\overline q_1^{\,2}}{4}
<
\infty,
\end{equation}
where the inequality follows from \eqref{eq:cont_payment_bounds}
and Popoviciu's inequality. In particular, the bound is independent of the book
size $J$.

The conditional i.i.d.\ assumption and the Cauchy-Schwarz inequality give
\begin{equation}
\label{eq:cont_book_mean_square_rate}
\mathbb E^{\nu^\ast}
\left[
\left(
\overline Z_{c,J}
-
\overline Z_c
\right)^2
\right]
=
\frac{\sigma_{c,\mathrm{id}}^2(\nu^\ast)}{J},
\qquad
\mathbb E^{\nu^\ast}
\left[
\left|
\overline Z_{c,J}
-
\overline Z_c
\right|
\right]
\le
\frac{\sigma_{c,\mathrm{id}}(\nu^\ast)}{\sqrt J}.
\end{equation}
The variational representation \eqref{eq:cont_cvar_def} implies that upper-tail
CVaR is $1/\alpha_c$-Lipschitz in $L^1$. Hence
\begin{equation}
\left|
\mathrm{CVaR}_{\alpha_c}^{+}
\bigl(\overline Z_{c,J}\bigr)
-
\mathrm{CVaR}_{\alpha_c}^{+}
\bigl(\overline Z_c\bigr)
\right|
\le
\frac{\sigma_{c,\mathrm{id}}(\nu^\ast)}
{\alpha_c\sqrt J},
\qquad
\left|
\Delta_{\alpha_c}^{+}
\bigl(\overline Z_{c,J}\bigr)
-
\Delta_{\alpha_c}^{+}
\bigl(\overline Z_c\bigr)
\right|
\le
\frac{\sigma_{c,\mathrm{id}}(\nu^\ast)}
{\alpha_c\sqrt J},
\label{eq:cont_book_excess_cvar_rate}
\end{equation}
because $\overline Z_{c,J}$ and $\overline Z_c$ have the same
mean by \eqref{eq:cont_book_subadditivity} and
\eqref{eq:cont_tower_property}. Thus,
$\Delta_{\alpha_c}^{+}
\bigl(\overline Z_{c,J}\bigr)$ converges to
$\Delta_{\alpha_c}^{+}
\bigl(\overline Z_c\bigr)$
with an $\mathcal O(J^{-1/2})$ error bound.

\begin{remark}[Square-root book-size approximation]
\label{rm:cont_sqrt_book_approximation}
If the common market environment is degenerate,
i.e.\ the shared asset returns are deterministic, then
$\overline Z_c=\mathbb E^{\nu^\ast}[Z_c]$ is deterministic and
$\Delta_{\alpha_c}^{+}(\overline Z_c)=0$. A normal approximation to the
finite-book average $\overline Z_{c,J}$ suggests that its excess upper-tail
cost decreases approximately as $J^{-1/2}$, motivating the heuristic
$\Delta_{\alpha_c}^{+}
\bigl(\overline Z_{c,J}\bigr)
\approx
\frac{\Delta_{\alpha_c}^{+}(Z_c)}{\sqrt J}$.

The bound \eqref{eq:cont_book_excess_cvar_rate} establishes only an
$\mathcal O(J^{-1/2})$ bound on the difference between $\Delta_{\alpha_c}^{+}(\overline Z_{c,J})$ and $\Delta_{\alpha_c}^{+}(\overline Z_c)$; it does not identify the coefficient in
the heuristic.
If instead the common market environment is non-degenerate,
$\Delta_{\alpha_c}^{+}(\overline Z_{c,J})$ converges to
$\Delta_{\alpha_c}^{+}(\overline Z_c)$, which is generally nonzero.
The numerical experiments therefore estimate both quantities directly rather
than impose the heuristic.
\end{remark}

Under the independent-mortality specification of
Section~\ref{sc:tontine_modeling}, retiree and spouse mortality are independent,
and household mortality is independent of the shared asset returns.
Consequently, $\overline Z_c$ has a probability-weighted representation.
Specifically, for $\beta=1,\ldots,M-1$, define
\begin{equation}
\label{eq:cont_entry_survival_weights}
\begin{aligned}
a_\beta
&:=
\mathbb P
\bigl(
H_{(\beta-1)^+}=11,\,
H_{\beta^-}=01
\bigr)
=
\mathbb P
\bigl(H_{(\beta-1)^+}=11\bigr)
\delta_{R,\beta-1}(1-\delta_{S,\beta-1}),
\\
s_{\beta,m}
&:=
\mathbb P
\bigl(
H_{m^-}=01
\mid
H_{(\beta-1)^+}=11,\,
H_{\beta^-}=01
\bigr)
=
\prod_{\ell=\beta}^{m-1}(1-\delta_{S,\ell}),
\qquad
m=\beta,\ldots,M-1,
\end{aligned}
\end{equation}
with an empty product equal to one.
Let $C_m^{(\beta)}$, $m=\beta,\ldots,M-1$, denote the
spouse-continuation payment at $t_m$ for a household that enters state $01$ at
$t_\beta^-$ and whose spouse remains alive through $t_m^-$. Then
\begin{equation}
\label{eq:cont_probability_weighted_cost}
\overline Z_c
=
\sum_{\beta=1}^{M-1}a_\beta
\sum_{m=\beta}^{M-1}s_{\beta,m}\,C_m^{(\beta)}.
\end{equation}
Equation~\eqref{eq:cont_probability_weighted_cost} averages over all possible
entry times and durations of the spouse-only phase while retaining the
uncertainty in the shared asset returns. Its derivation is given in
Appendix~\ref{app:cont_probability_weighted_derivation}.

For reporting purposes, when $\Delta_{\alpha_c}^{+}(Z_c)>0$, define the
large-book diversification ratio by
\begin{equation}
\label{eq:cont_diversification_ratio}
D_c(\nu^\ast)
:=
1-
\Delta_{\alpha_c}^{+}(\overline Z_c)/\Delta_{\alpha_c}^{+}(Z_c).
\end{equation}
Conditional Jensen's inequality gives
$\bigl(\mathbb E^{\nu^\ast}[Z_c\mid\mathcal G]-\eta\bigr)^+
\le
\mathbb E^{\nu^\ast}[(Z_c-\eta)^+\mid\mathcal G]$
for each $\eta\in\mathbb R$. Hence, by \eqref{eq:cont_cvar_def} and
\eqref{eq:cont_tower_property}, $D_c(\nu^\ast)\in[0,1]$ and measures the
fraction of the representative-contract excess upper-tail cost diversified
away in the large-book limit.

\subsection{Book-level risk-loaded cost and payment-scale load}
\label{ssc:cont_book_load}
For notational convenience, set
$\overline Z_{c,\infty}:=\overline Z_c$. For
$J\in\mathbb N\cup\{\infty\}$ and prudential-buffer coefficient
$\lambda_c\ge0$, define
\begin{align}
\Pi_{c,J}(\nu^\ast)
&:=
\mathbb E^{\nu^\ast}
\bigl[\overline Z_{c,J}\bigr]
+
\lambda_c\,
\Delta_{\alpha_c}^{+}
\bigl(\overline Z_{c,J}\bigr),
\label{eq:cont_cost_rl}
\\
f_{c,J}^{\mathrm{RL}}(\nu^\ast)
&:=
\frac{
\Pi_{c,J}(\nu^\ast)
}{
\mathbb E^{\nu^\ast}
[Y_r]
+
\Pi_{c,J}(\nu^\ast)
}.
\label{eq:cont_load_rl}
\end{align}
Here, $\Pi_{c,J}(\nu^\ast)$ is the risk-loaded average per-contract
continuation cost, including a prudential loading on the excess upper-tail
cost remaining after book-level diversification, while
$f_{c,J}^{\mathrm{RL}}(\nu^\ast)$ is the corresponding normalized
continuation-cost share under the fixed policy.  For the representative-contract case $J=1$,
write $\Pi_c(\nu^\ast):=\Pi_{c,1}(\nu^\ast)$ and
$f_c^{\mathrm{RL}}(\nu^\ast):=f_{c,1}^{\mathrm{RL}}(\nu^\ast)$.

When $\lambda_c=0$, \eqref{eq:cont_load_rl} reduces to the expected-cost load
$f_c(\nu^\ast)$ in \eqref{eq:cont_load_ev}. Since
$\mathbb E^{\nu^\ast}[\overline Z_{c,J}]
=\mathbb E^{\nu^\ast}[Z_c]$ for every
$J\in\mathbb N\cup\{\infty\}$, when $\lambda_c=1$,
\eqref{eq:cont_cost_rl} reduces to
$\Pi_{c,J}(\nu^\ast)
=\mathrm{CVaR}_{\alpha_c}^{+}
(\overline Z_{c,J})$.
Moreover, \eqref{eq:cont_book_excess_cvar_rate} implies
\[
\left|
\Pi_{c,J}(\nu^\ast)-\Pi_{c,\infty}(\nu^\ast)
\right|
\le
\frac{\lambda_c\,\sigma_{c,\mathrm{id}}(\nu^\ast)}
{\alpha_c\sqrt J},
\]
so the finite-book risk-loaded cost $\Pi_{c,J}(\nu^\ast)$ converges to its
large-book limit $\Pi_{c,\infty}(\nu^\ast)$ with an
$\mathcal O(J^{-1/2})$ error bound. If $\overline Z_{c,J}$ is deterministic,
its excess upper-tail cost vanishes and
$\Pi_{c,J}(\nu^\ast)=\mathbb E^{\nu^\ast}[Z_c]$.

\subsection{Simulation-based numerical method}
\label{ssc:cont_simulation_method}

In numerical work, the exact optimal household control policy $\nu^\ast$ is
replaced by the learned NN-parameterized household control policy
$\widehat\nu^\ast$ from Section~\ref{sec:NNs}. The learned policy is held fixed
during load estimation, and the notation above is used with expectations taken
under $\widehat\nu^\ast$.

Representative-contract quantities are estimated from out-of-sample
realizations of the joint asset-return and household-state processes. For each
selected finite book size $J$ and each Monte Carlo replication, one realization
of the shared asset returns is common to all contracts in the book, while $J$
conditionally independent household mortality histories are generated. The
resulting finite-book averages $\overline Z_{c,J}$ are used to estimate
$\mathrm{CVaR}_{\alpha_c}^{+}(\overline Z_{c,J})$,
$\Delta_{\alpha_c}^{+}(\overline Z_{c,J})$,
$\Pi_{c,J}(\widehat\nu^\ast)$, and
$f_{c,J}^{\mathrm{RL}}(\widehat\nu^\ast)$, and to assess convergence to the
large-book limit.

Separately, the large-book limiting variable $\overline Z_c$ is evaluated
using \eqref{eq:cont_probability_weighted_cost}, which integrates over
household mortality conditional on each realization of the shared asset
returns. These values estimate
$\mathrm{CVaR}_{\alpha_c}^{+}(\overline Z_c)$,
$D_c(\widehat\nu^\ast)$,
$\Pi_{c,\infty}(\widehat\nu^\ast)$, and
$f_{c,\infty}^{\mathrm{RL}}(\widehat\nu^\ast)$. As a consistency check,
the sample means of $Z_c$, $\overline Z_{c,J}$, and $\overline Z_c$ should
agree up to Monte Carlo error.

The complete numerical algorithm and empirical estimators are given in
Appendix~\ref{app:cont_load_algorithm}.

\section{Numerical experiments}
\label{sec:numerics}

{This section quantifies the economic and actuarial effects of spouse
protection in the tontine. After specifying the baseline household, mortality
inputs, financial data, and return-generation procedure, we examine household-state
incidence, the household reward--risk frontier, the representative-contract
spouse-continuation load, finite-book diversification, and the conditional
large-book limit. We conclude with their implications for product design. The main analysis distinguishes the continuation-cost tail
risk that is diversified across contracts from the common-market component
that remains at the book level. Supplementary analyses of the continuation cap
and learned household controls are reported in the appendices; Monte Carlo
evaluation uncertainty is reported in
Appendix~\ref{app:num_numerical_robustness}.}
\subsection{Baseline household scenario and mortality inputs}
\label{ssc:num_scenario}

{The baseline household consists of a male retiree aged $a_R=65$ and
a female spouse aged $a_S=60$ at inception. With terminal reference age $95$
for the younger protected life and annual decision times, the horizon is}
\[
T=M=\left\lceil 95-\min\{65,60\}\right\rceil=35.
\]
Unless otherwise noted, all monetary quantities are reported in
thousands of real Australian dollars. The household starts with initial wealth
$W_0=1000$ and, while solvent, is subject to the no-shorting and no-leverage
constraints in Section~\ref{sc:hh_modeling}. The annual tontine management fee
is set to $\varrho=0.11\%$, consistent with the fee structure of QSuper's
Lifetime Pension pool \cite{qsuper2025pds}.
Debt positions accrue at the domestic bond-index return plus a
borrowing spread of $2\%$ per annum.

{We calibrate the household withdrawal bands using the modest and
comfortable spending estimates in the December 2025 quarter of the ASFA
Retirement Standard \cite{ASFA2026RetirementStandard}, and set}
\[
\underline q_{11}=51.299,\qquad
\overline q_{11}=77.375,\qquad
\underline q_{1}=35.503,\qquad
\overline q_{1}=54.840.
\]
{The first pair applies to the two-person state $11$. The second is
the one-survivor band used for controlled withdrawals in state $10$ and for
the spouse-continuation rule in state $01$.}

The terminal targets represent capital reserves for spending beyond
$T$, not annual payments within the horizon. Since the younger protected life
reaches age $95$ at $T$, we use the ASFA over-85 homeowner spending figures
\cite{ASFA2026RetirementStandard}, rather than the age-65--84 figures used
for the withdrawal bands. We choose a fixed ten-year spending multiple as
the baseline, giving
\[
w_{11}^{\ast}=10\times72.093=720.93,\qquad
w_{10}^{\ast}=w_{01}^{\ast}=10\times52.235=522.35.
\]

{For deterministic mortality, we use the 2021 single-year Australian
male and female period life tables from the Human Mortality Database
\cite{HMD}. We hold these age-specific period schedules fixed over the
horizon, so that only age advances. With $t_m=m$, the retiree and spouse
one-year death probabilities are}
\[
\delta_{R,m-1}
=
q^{M}_{a_R+(m-1),\,2021},
\qquad
\delta_{S,m-1}
=
q^{F}_{a_S+(m-1),\,2021},
\qquad m=1,\ldots,M.
\]
{The homogeneous baseline pool uses the same mortality sequences for
every contract:}
\[
\delta_{R,m-1}^{\ell}\equiv \delta_{R,m-1},
\qquad
\delta_{S,m-1}^{\ell}\equiv \delta_{S,m-1}.
\]
{These sequences determine the household-state transitions
\eqref{eq:hh_transition_11}--\eqref{eq:hh_transition_01}, the contract-exit
probabilities \eqref{eq:hh_exit_prob}, and the state-dependent tontine gain
rates \eqref{eq:hh_tontine_gain_piecewise}.}
\subsection{Financial data and return generation}
\label{ssc:num_data}

{Having specified the household and mortality inputs, we now describe
the market data used to train and evaluate the learned policies. We use a
four-asset monthly return panel comprising Australian equities, Australian
government bonds, U.S.\ equities converted to AUD, and U.S.\ 30-day Treasury
bills converted to AUD. Following \cite{orozco2026money}, the four series are
aligned over the common sample period 1935:1--2022:12 and converted to real
AUD returns using Australian CPI. Detailed data construction and descriptive
statistics are provided in Appendix~\ref{app:financial_data}.}

{Future return paths are generated nonparametrically by applying the
stationary block bootstrap jointly to the four-dimensional monthly return
vector
\cite{politis1994stationary,politis2004automatic,patton2009correction,
dichtl2016testing}. We use an expected block length of 24 months, thereby
retaining serial dependence within each series and contemporaneous dependence
across assets. The resulting real AUD return paths are used for NN training and
out-of-sample policy evaluation. Network specifications and the training and evaluation samples are
reported in Appendix~\ref{app:NN_training}; Monte Carlo evaluation
uncertainty is assessed in Appendix~\ref{app:num_numerical_robustness}.}
\subsection{Household-state probabilities and spouse-only duration}
\label{ssc:num_household_state_probs}

We first assess the incidence and persistence of spouse continuation
under the baseline household and fixed-period mortality schedules in
Subsection~\ref{ssc:num_scenario}. Let
\[
\pi_h(m):=\mathbb P(H_{m^-}=h), \qquad h\in\{11,10,01,00\}, \qquad m=0,\ldots,M,
\]
denote the probability that the household is in state $h$ at time $t_m^-$. The initial condition is
$\pi_{11}(0)=1$ and $\pi_{10}(0)=\pi_{01}(0)=\pi_{00}(0)=0$.
The analytic household-state probabilities are obtained directly from the household-state transition law \eqref{eq:hh_transition_11}--\eqref{eq:hh_transition_01}. Under the deterministic one-year death probabilities $\delta_{R,m-1}$ and $\delta_{S,m-1}$,
\begin{align*}
\pi_{11}(m)
&=
\pi_{11}(m-1)(1-\delta_{R,m-1})(1-\delta_{S,m-1}),
\\
\pi_{10}(m)
&=
\pi_{10}(m-1)(1-\delta_{R,m-1})
+
\pi_{11}(m-1)(1-\delta_{R,m-1})\delta_{S,m-1},
\\
\pi_{01}(m)
&=
\pi_{01}(m-1)(1-\delta_{S,m-1})
+
\pi_{11}(m-1)\delta_{R,m-1}(1-\delta_{S,m-1}),
\\
\pi_{00}(m)&=1-\pi_{11}(m)-\pi_{10}(m)-\pi_{01}(m).
\end{align*}Thus, the full household-state probability profile is available analytically, year by year, from the mortality inputs alone.
A particularly useful summary statistic is the probability that the household ever enters the spouse-only state. Since state $01$ can only be entered from state $11$, this probability is
\begin{equation}
\mathbb P\!\left(\exists\,m\in\{1,\ldots,M\}: H_{m^-}=01\right)
=
\sum_{j=0}^{M-1}
\left(
\prod_{\ell=0}^{j-1}(1-\delta_{R,\ell})(1-\delta_{S,\ell})
\right)
\delta_{R,j}(1-\delta_{S,j}).
\label{eq:num_prob_enter_01}
\end{equation}
For the baseline Australian household, this gives
\[
\mathbb P\!\left(\exists\,m\le M:\;H_{m^-}=01\right)\approx 0.7100.
\]
Hence, roughly $71\%$ of baseline households are observed in the
spouse-only state at least once by the terminal date.

\begin{figure}[!htb]
\centering
\begin{subfigure}[t]{0.48\textwidth}
\centering
\includegraphics[width=\textwidth]{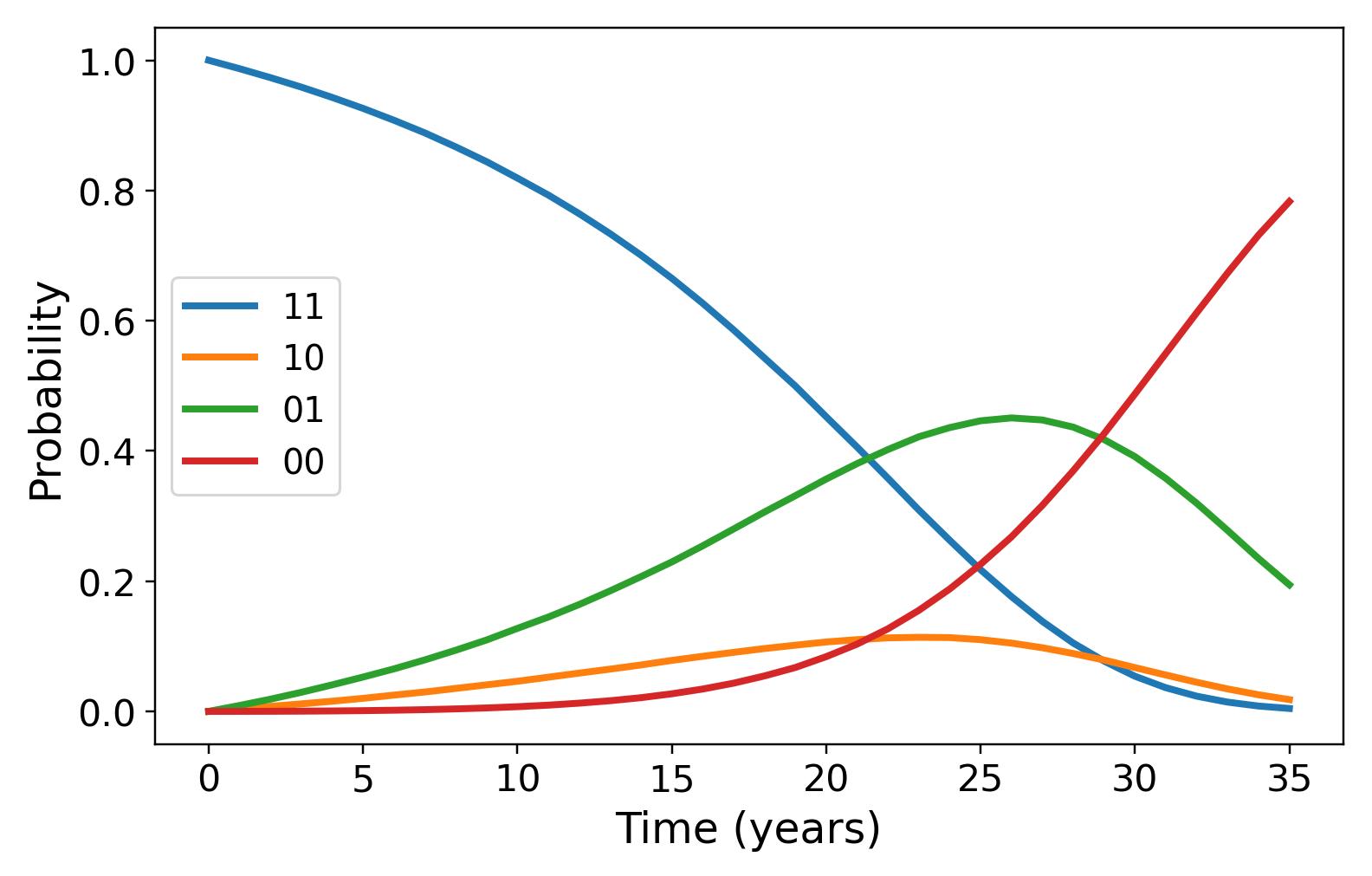}
\caption{Analytic household-state probabilities over time.}
\label{fig:hhs_probs}
\end{subfigure}
\hfill
\begin{subfigure}[t]{0.48\textwidth}
\centering
\includegraphics[width=\textwidth]{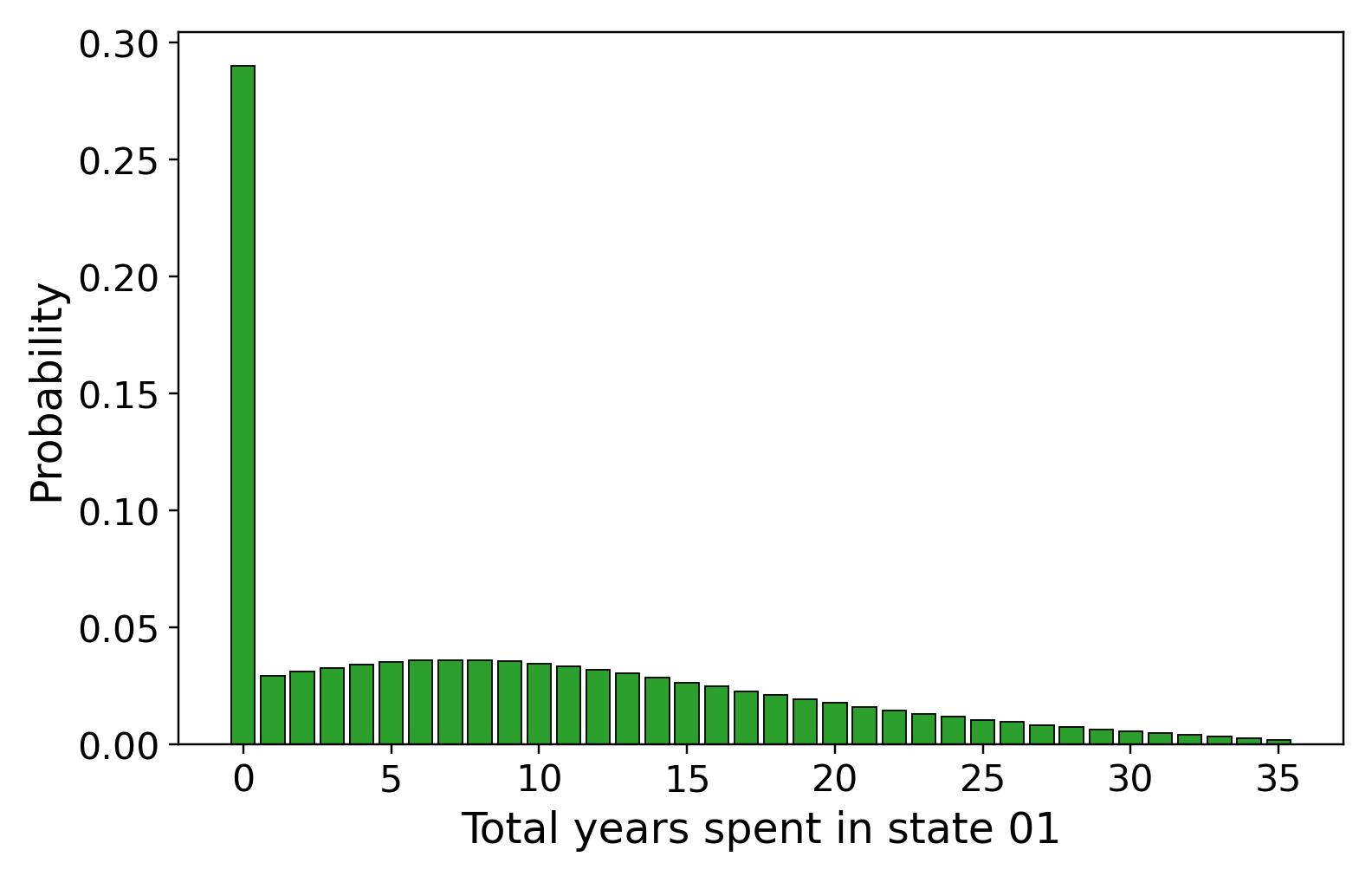}
\caption{Analytic distribution of spouse-only duration~$D_{01}$.}
\label{fig:duration_probabilities}
\end{subfigure}
\caption{Household-state probabilities and spouse-only duration for the baseline Australian household.}
\label{fig:household_state_probs}
\end{figure}

Figure~\ref{fig:household_state_probs}(\subref{fig:hhs_probs}) reports the
analytic household-state probabilities
$\pi_{11}(m),\pi_{10}(m),\pi_{01}(m),\pi_{00}(m)$ over the full horizon.
As expected, the probability of state $11$ declines monotonically, while the probability of state $00$ rises monotonically. More importantly, the spouse-only state $01$ is materially more likely than the retiree-only state $10$ over the economically relevant part of the horizon, reflecting both the younger spouse age in the baseline scenario and lower female mortality in the Australian life tables.

To quantify the length of the spouse-only phase, define
$\displaystyle D_{01}:=\sum_{m=1}^{M}\mathbf 1_{\{H_{m^-}=01\}}$,
an annual state-count measure that includes the terminal observation. Its distribution is obtained analytically by combining the probability of entering state $01$ at each date with the probability of remaining there for a given number of subsequent years before either the spouse dies or the horizon is reached. Figure~\ref{fig:household_state_probs}(\subref{fig:duration_probabilities}) reports the resulting distribution. The unconditional mean spouse-only duration is approximately $8.70$ years, and conditional on ever entering state $01$ it is approximately $12.25$ years. These state-duration statistics differ from the expected number of
spouse-only payment dates: payments stop at $M-1$, giving
$\sum_{m=1}^{M-1}\pi_{01}(m)\approx8.5058$, consistent with the
constant-payment diagnostic in Appendix~\ref{app:num_numerical_robustness}.

Overall, these results show that under realistic Australian mortality inputs, spouse continuation is both likely to arise and likely to persist for a substantial period. This provides direct motivation for modeling spouse protection explicitly in the household decumulation problem. The corresponding international comparison in Appendix~\ref{app:international_household_states} shows that this qualitative pattern is not specific to Australia. A separate Monte Carlo validation based on $10^7$ household paths, not reported here, closely matches the analytic formulas.

Because every contract enters the pool in state $11$, spouse protection
also weakens mortality pooling at the start of retirement. While both household
members are alive, pool exit requires both deaths within the period, so the
tontine gain is based on the joint-death probability
$\delta_{R,m-1}\delta_{S,m-1}$ rather than an individual death probability.
Spouse protection therefore delays forfeitures and attenuates mortality-credit
gains while the contract remains in state $11$.
\subsection{Household reward--risk frontier}
\label{ssc:num_reward_risk_frontier}

{Having established that spouse continuation is both likely and
persistent, we next examine the trade-off between expected cumulative real
household payments and state-contingent terminal-shortfall risk. For each
scalarization parameter $\lambda$ in a non-uniform grid over $[0.15,100]$, we
train an NN-parameterized household control policy
$\widehat\nu_\lambda^\ast$ using the NN parameterization in Section~\ref{sec:NNs} and the
sequential transfer-learning protocol in Appendix~\ref{app:NN_training}.}

For later reference, define the ever-insolvent probability under the learned
policy $\widehat\nu_\lambda^\ast$ by
\begin{equation}
\label{eq:num_ever_insolvent_prob}
p_{\mathrm{ins}}^{\mathrm{ever}}
\bigl(\widehat\nu_\lambda^\ast\bigr)
:=
\mathbb P_{x_0,h_0}^{\widehat\nu_\lambda^\ast}
\left(
\exists\,m\in\{0,\ldots,M-1\}:
H_{m^-}\neq 00
\ \text{and}\
W_{m^+}^{\widehat\nu_\lambda^\ast}\le 0
\right).
\end{equation}
Thus,
$p_{\mathrm{ins}}^{\mathrm{ever}}(\widehat\nu_\lambda^\ast)$ is the
probability that the contract becomes insolvent at least once while it remains
active during the planning horizon.

{From the computed frontier, we select the policies
$\widehat\nu_{0.3}^\ast$, $\widehat\nu_{1}^\ast$, and
$\widehat\nu_{15}^\ast$, which we call \emph{payment-seeking},
\emph{intermediate}, and \emph{conservative}, respectively. These policies
span materially different reward--risk positions and are carried through the
subsequent continuation-cost experiments. Table
\ref{tab:num_reward_risk_selected} reports their out-of-sample evaluation
statistics.}

\begin{table}[!htb]
\centering
\caption{Selected positions on the household reward--risk frontier and their
associated learned policies.}
\label{tab:num_reward_risk_selected}
\begin{tabular}{lrrrr}
\hline
\noalign{\vspace{1mm}}
\shortstack{Frontier\\position}
&
$\lambda$
&
$\mathcal R(\widehat\nu_\lambda^\ast)$
&
$\operatorname{CVaR}_{0.95}
 \bigl(L_T^{\widehat\nu_\lambda^\ast}\bigr)$
&
\shortstack{
$p_{\mathrm{ins}}^{\mathrm{ever}}
(\widehat\nu_\lambda^\ast)$\\
(\%)
}
\\
\noalign{\vspace{1mm}}
\hline
Payment-seeking & 0.3 & 1860.4 & 1218.0 & 26.7\\
Intermediate    & 1   & 1704.8 & 840.5  & 12.6\\
Conservative    & 15  & 1584.8 & 793.3  & 12.0\\
\hline
\end{tabular}
\end{table}

Table~\ref{tab:num_reward_risk_selected} shows that the payment-seeking
policy generates the highest expected household payments but also the largest
terminal-shortfall tail and insolvency probability.
Moving to the intermediate policy reduces
$\operatorname{CVaR}_{0.95}(L_T)$ by approximately $31.0\%$ and
$p_{\mathrm{ins}}^{\mathrm{ever}}$ by $14.1$ percentage points, while
$\mathcal R$ decreases by approximately $8.4\%$. Moving from the intermediate
to the conservative policy reduces terminal-shortfall CVaR by a further
$5.6\%$ and insolvency by only $0.6$ percentage points, while reducing
$\mathcal R$ by another $7.0\%$.
{The intermediate policy therefore lies near the visible knee of the
frontier and is used as the reference policy for the learned household
withdrawal and allocation controls presented in
Appendix~\ref{app:additional_control_surfaces}.}

\begin{figure}[hbt!]
\centering
\begin{minipage}[t]{0.5\textwidth}
\vspace{0pt}
\centering
\includegraphics[width=0.9\textwidth]
{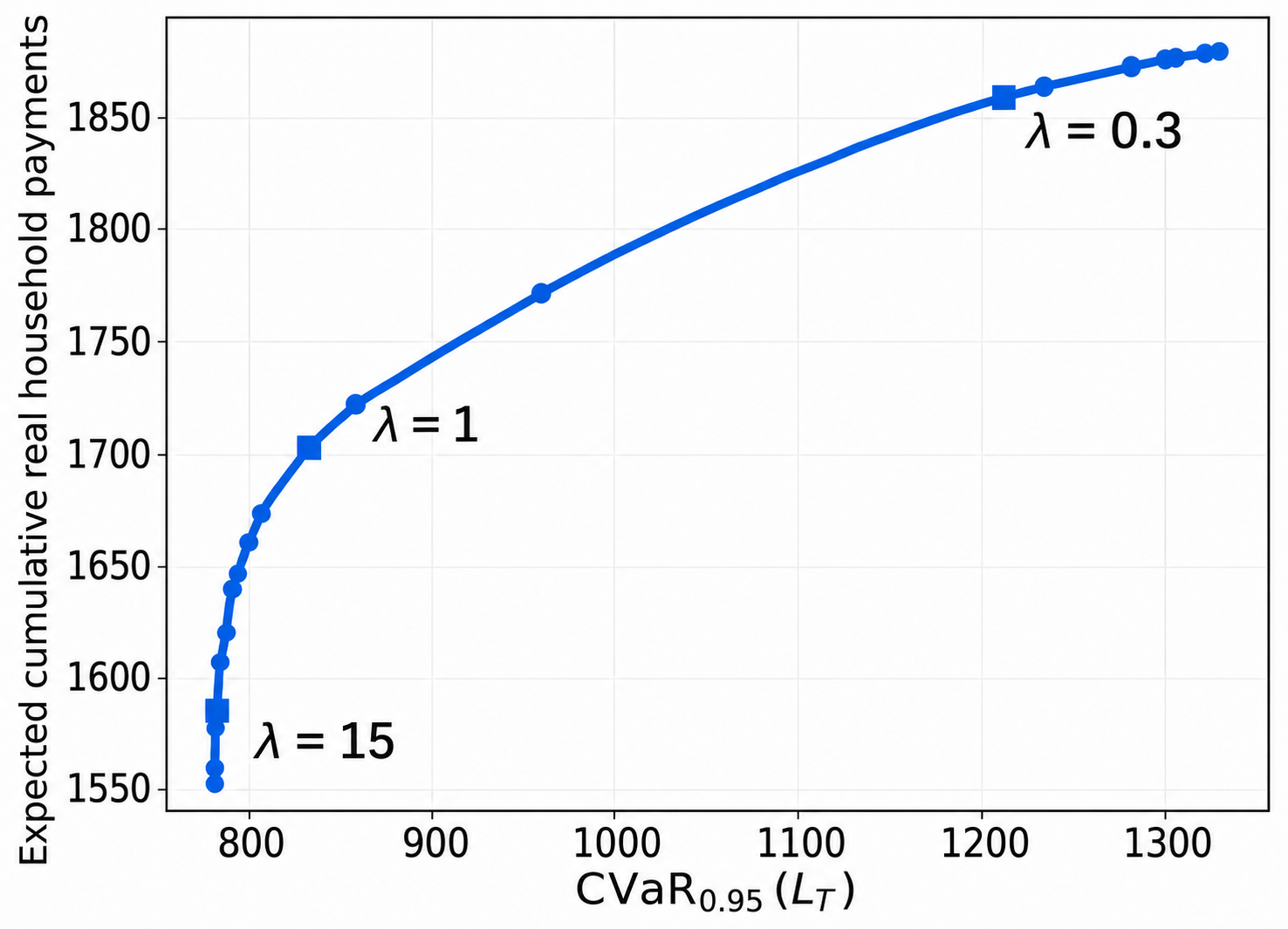}
\end{minipage}
\hfill
\begin{minipage}[t]{0.45\textwidth}
\vspace{0pt}
\raggedright
\begingroup
\setlength{\abovecaptionskip}{0pt}
\captionof{figure}{Household reward--risk frontier under the baseline
spouse-protected tontine specification. The square markers identify the three
representative positions in Table~\ref{tab:num_reward_risk_selected}.}
\label{fig:num_reward_risk_frontier}
\endgroup
\end{minipage}
\end{figure}

For each trained policy, the corresponding frontier point is
\[
\left(
\mathcal Q_{0.95}(\widehat\nu_\lambda^\ast),
\mathcal R(\widehat\nu_\lambda^\ast)
\right)
=
\left(
\operatorname{CVaR}_{0.95}
\bigl(L_T^{\widehat\nu_\lambda^\ast}\bigr),
\mathcal R(\widehat\nu_\lambda^\ast)
\right),
\]
consistent with \eqref{eq:household_reward_risk_frontier}. The resulting
frontier is shown in Figure~\ref{fig:num_reward_risk_frontier}.

{Figure~\ref{fig:num_reward_risk_frontier} shows the expected
reward--risk trade-off: increasing $\lambda$ lowers both cumulative household
payments and terminal-shortfall risk, with diminishing marginal risk reduction
beyond the intermediate position. The three selected policies therefore
provide distinct reward--risk environments for assessing whether greater
policy conservatism materially reduces the spouse-continuation obligation.}
\subsection{Representative-contract spouse-continuation load}
\label{ssc:num_spouse_continuation_load}

{We first evaluate the spouse-continuation obligation for one representative contract ($J=1$) under the three selected learned policies. Throughout this subsection, the learned household policy is held fixed and the baseline continuation rule is $B_m=\min\{q_m,\overline q_1\}$. The retiree-alive and spouse-only payment streams $Y_r^{\widehat\nu_\lambda^\ast}$ and $Z_c^{\widehat\nu_\lambda^\ast}$ are defined in \eqref{eq:cont_payment_components}. We set $\alpha_c=0.05$ and, unless otherwise stated, take the prudential-buffer coefficient to be $\lambda_c=1$.
Detailed capped-versus-uncapped sensitivity results under the same
fixed learned policies are reported in
Appendix~\ref{app:num_continuation_cap_sensitivity}.}

\begin{table}[!htb]
\centering
\caption{{Representative-contract spouse-continuation load under the selected learned policies. The last column reports the baseline risk-loaded payment-scale load $f_{c,1}^{\mathrm{RL}}$ at $\alpha_c=0.05$ and $\lambda_c=1$.}}
\label{tab:num_spouse_continuation_load}
\begin{tabular}{lrrrrrr}
\hline
\noalign{\vspace{1mm}}
\shortstack{Frontier\\position}
&
$\lambda$
&
$\mathbb E[Y_r]$
&
$\mathbb E[Z_c]$
&
$\Delta_{0.05}^{+}(Z_c)$
&
$f_c$ (\%)
&
\shortstack{$f_{c,1}^{\mathrm{RL}}$\\(\%)}
\\
\noalign{\vspace{1mm}}
\hline
Payment-seeking & 0.3 & 1422.2 & 438.1 & 1096.2 & 23.5 & 51.9\\
Intermediate    & 1   & 1264.6 & 440.1 & 1043.8 & 25.8 & 54.0\\
Conservative    & 15  & 1151.7 & 433.1 & 1027.5 & 27.3 & 55.9\\
\hline
\end{tabular}
\end{table}

{Table~\ref{tab:num_spouse_continuation_load} shows that expected
spouse-continuation payments remain in the narrow range $[433.1,440.1]$,
whereas expected retiree-alive payments fall by approximately $270.5$ from
the payment-seeking to the conservative policy. Consequently, $f_c$ rises from $23.5\%$ to $27.3\%$ even though the absolute expected continuation cost changes by less than $2\%$. Across these policies, the increase in $f_c$ is driven mainly by
the lower retiree-alive payment scale, not by higher expected continuation
cost.}
{The same table reports a large representative-contract excess
upper-tail cost, with $\Delta_{0.05}^{+}(Z_c)$ declining only modestly from
$1096.2$ to $1027.5$. Under the baseline capped rule, relatively stable
expected continuation costs therefore coexist with substantial
representative-contract tail risk. This motivates the finite-book analysis in the next subsection.}

\begin{figure}[!htb]
\centering
\begin{minipage}[t]{0.5\textwidth}
\raggedright
\vspace{10pt}
{Figure~\ref{fig:num_spouse_continuation_load} shows how the
risk-loaded payment-scale load varies with $\lambda_c\in[0,1]$.
At $\lambda_c=1$, Table~\ref{tab:num_spouse_continuation_load} reports
loads of $51.9\%$--$55.9\%$, substantially above the expected-cost loads.}
\end{minipage}
\hfill
\begin{minipage}[t]{0.45\textwidth}
\vspace{10pt}
\centering
\includegraphics[width=0.9\linewidth]{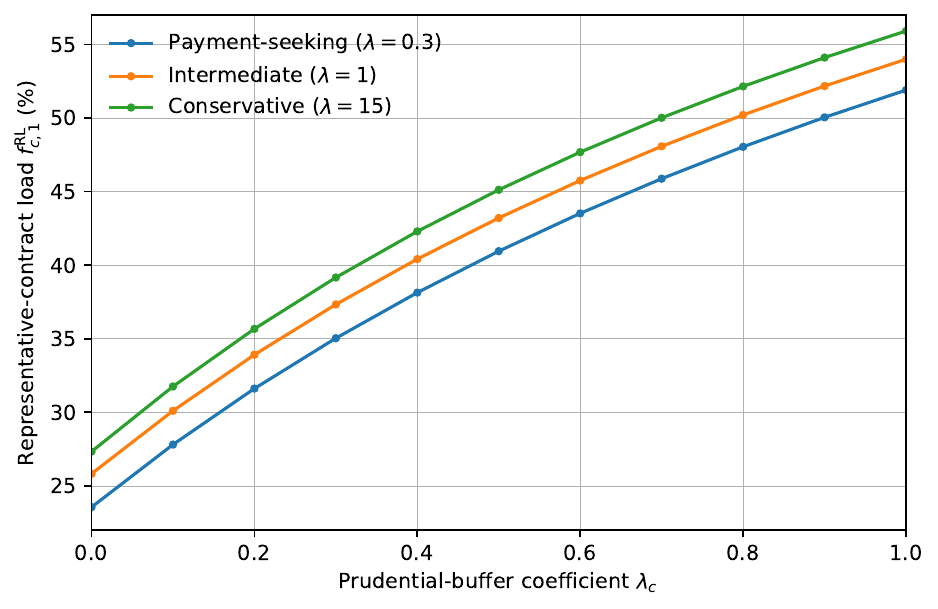}
\begingroup
\setlength{\abovecaptionskip}{4pt}
\captionof{figure}{{Representative-contract risk-loaded spouse-continuation load under the three learned policies.}}
\label{fig:num_spouse_continuation_load}
\endgroup
\end{minipage}
\end{figure}

\subsection{Finite-book spouse-continuation load}
\label{ssc:num_finite_book_spouse_continuation_load}

{We next evaluate finite-book spouse-continuation costs for
$J\in\{1,10,100,1000\}$ under capped continuation. In each of the
$K=256{,}000$ replications, all contracts in a book share the same
market-return path, while their household mortality histories are conditionally
independent given that path. Smaller books use the first $J$ histories from the same pool of
$1{,}000$ mortality draws for each market path; this shared sampling design is
described in Appendix~\ref{app:num_numerical_robustness}.}

\begin{table}[!htb]
\centering
\caption{Finite-book average per-contract spouse-continuation cost
under capped continuation. The excess upper-tail cost is evaluated at
$\alpha_c=0.05$, and $f_{c,J}^{\mathrm{RL}}$ is reported at
$\lambda_c=1$.}
\label{tab:num_finite_book_spouse_continuation_load}
\begin{tabular}{llrrr}
\hline
\noalign{\vspace{1mm}}
\shortstack{Frontier\\position}
&
$J$
&
$\mathbb E[\overline Z_{c,J}]$
&
$\Delta_{0.05}^{+}(\overline Z_{c,J})$
&
\shortstack{$f_{c,J}^{\mathrm{RL}}$\\(\%)}
\\
\noalign{\vspace{1mm}}
\hline
Payment-seeking & 1    & 438.1
& 1096.2 & 51.9\\
                          & 10   & 437.7
& 327.6  & 35.0\\
                          & 100  & 437.4
& 116.1  & 28.0\\
                          & 1000 & 437.4
& 54.8   & 25.7\\
\hline
Intermediate    & 1    & 440.1
& 1043.8 & 54.0\\
                          & 10   & 439.7
& 314.6  & 37.4\\
                          & 100  & 439.5
& 107.2  & 30.2\\
                          & 1000 & 439.4
& 46.9   & 27.8\\
\hline
Conservative    & 1    & 433.1
& 1027.5 & 55.9\\
                          & 10   & 432.7
& 307.3  & 39.1\\
                          & 100  & 432.5
& 103.3  & 31.7\\
                          & 1000 & 432.4
& 44.8   & 29.3\\
\hline
\end{tabular}
\end{table}

{Table~\ref{tab:num_finite_book_spouse_continuation_load} shows that,
for each policy, the estimated mean continuation cost varies by less than
$0.2\%$ over $J\in\{1,10,100,1000\}$, consistent with the mean identity in
\eqref{eq:cont_book_subadditivity}. The excess upper-tail
cost, by contrast, falls by about $70\%$ at $J=10$, $89\%$--$90\%$ at
$J=100$, and $95\%$--$96\%$ at $J=1000$, relative to the
representative-contract value. At $\lambda_c=1$, the risk-loaded load
consequently falls from $51.9\%$--$55.9\%$ for $J=1$ to
$25.7\%$--$29.3\%$ for $J=1000$. Book-level diversification therefore
reduces the prudential tail buffer, not the expected continuation cost.}
\subsection{Large-book limiting spouse-continuation load}
\label{ssc:num_large_book_spouse_continuation_load}

{We next evaluate the conditional large-book limit
$\overline Z_c=\mathbb E[Z_c\mid\mathcal G]$ in
\eqref{eq:cont_large_book_limit} under capped continuation. Conditional on
each common market-return path, household mortality is integrated out, so the
resulting distribution isolates the common-market component of
continuation-cost risk. The probability-weighted evaluation is specified in
Algorithm~\ref{alg:cont_load_large_book} in
Appendix~\ref{app:cont_load_algorithm}.}

\begin{table}[!htb]
\centering
\caption{Representative-contract excess upper-tail cost and
conditional large-book limiting average per-contract spouse-continuation cost
under capped continuation. The excess upper-tail costs and diversification
ratio are evaluated at $\alpha_c=0.05$, and
$f_{c,\infty}^{\mathrm{RL}}$ is reported at $\lambda_c=1$.}
\label{tab:num_large_book_spouse_continuation_load}
\begin{tabular}{lrrrrr}
\hline
\noalign{\vspace{1mm}}
\shortstack{Frontier\\position}
&
$\Delta_{0.05}^{+}(Z_c)$
&
$\mathbb E[\overline Z_c]$
&
$\Delta_{0.05}^{+}(\overline Z_c)$
&
\shortstack{$D_c$\\(\%)}
&
\shortstack{$f_{c,\infty}^{\mathrm{RL}}$\\(\%)}
\\
\noalign{\vspace{1mm}}
\hline
Payment-seeking
& 1096.2
& 437.4
& 29.0
& 97.35
& 24.7\\
Intermediate
& 1043.8
& 439.4
& 24.9
& 97.62
& 26.9\\
Conservative
& 1027.5
& 432.4
& 26.9
& 97.38
& 28.5\\
\hline
\end{tabular}
\end{table}

{The limiting means in
Table~\ref{tab:num_large_book_spouse_continuation_load} differ from the
representative-contract means in Table~\ref{tab:num_spouse_continuation_load}
by less than $0.2\%$, consistent with the tower-property identity
\eqref{eq:cont_tower_property}. Book-level
diversification therefore leaves the expected per-contract continuation cost
unchanged. By contrast, the limiting excess upper-tail cost is only
$24.9$--$29.0$, compared with $1027.5$--$1096.2$ for a representative
contract. Accordingly, $D_c=97.35\%$--$97.62\%$: diversification of
household mortality removes almost all of the representative-contract
excess upper-tail cost under capped continuation. The smaller but nonzero limiting
tail is the common-market component that remains even in an arbitrarily large
book.}

{At $\lambda_c=1$,
Table~\ref{tab:num_large_book_spouse_continuation_load} reports limiting
risk-loaded payment-scale loads of $24.7\%$--$28.5\%$, only about one
percentage point above the corresponding expected-cost loads in
Table~\ref{tab:num_spouse_continuation_load}. Moreover, the $J=1000$ loads in
Table~\ref{tab:num_finite_book_spouse_continuation_load} are only
$0.8$--$1.0$ percentage points above their large-book limits. Thus, large
books remove most of the prudential loading associated with household-specific
mortality risk, but not the underlying expected continuation cost or the
residual common-market tail. The corresponding capped-versus-uncapped
large-book comparison is reported in
Appendix~\ref{ssc:num_large_book_continuation_cap_sensitivity}.}

\begin{figure}[!htb]
\centering
\begin{subfigure}[t]{0.32\textwidth}
\centering
\includegraphics[width=\textwidth]
{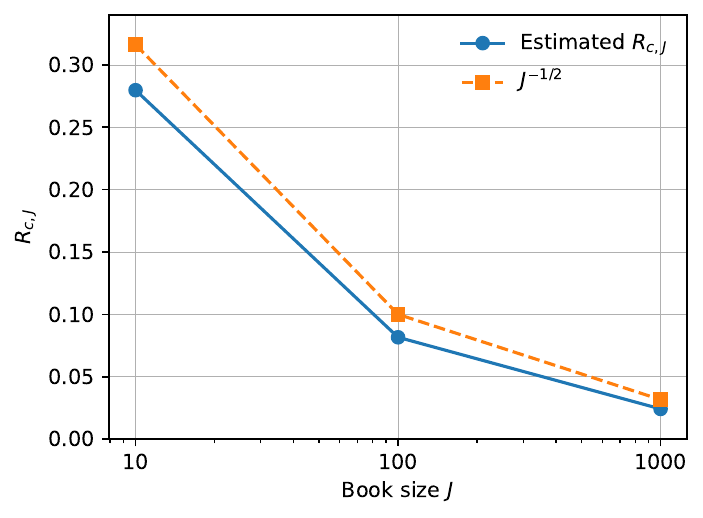}
\caption{$\lambda=0.3$.}
\label{fig:num_book_diversification_03}
\end{subfigure}
\hfill
\begin{subfigure}[t]{0.32\textwidth}
\centering
\includegraphics[width=\textwidth]
{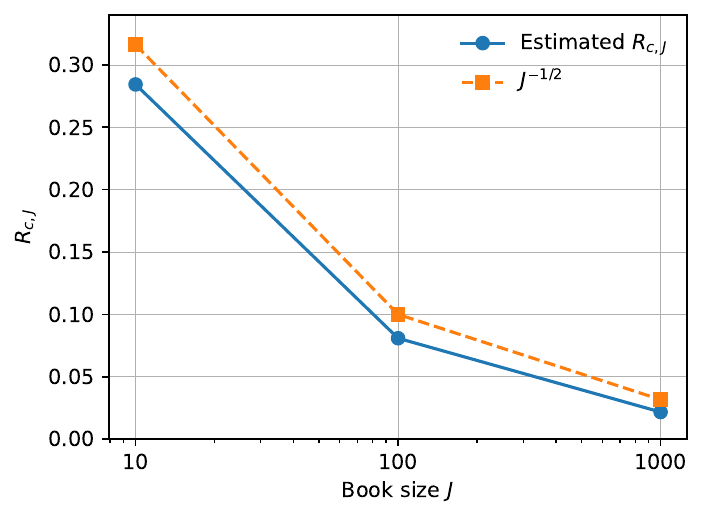}
\caption{$\lambda=1$.}
\label{fig:num_book_diversification_1}
\end{subfigure}
\hfill
\begin{subfigure}[t]{0.32\textwidth}
\centering
\includegraphics[width=\textwidth]
{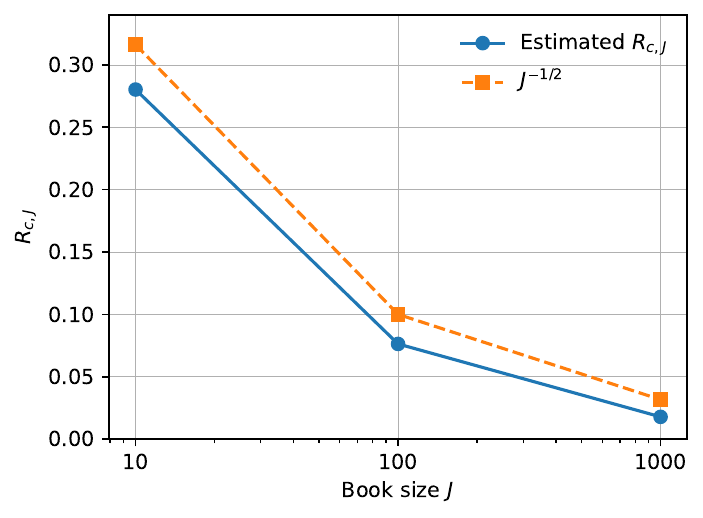}
\caption{$\lambda=15$.}
\label{fig:num_book_diversification_15}
\end{subfigure}
\caption{Finite-book convergence toward the conditional large-book
limiting excess upper-tail spouse-continuation cost under capped continuation.
The plotted statistic is $R_{c,J}$; the reference curve is $J^{-1/2}$.}
\label{fig:num_book_diversification}
\end{figure}

{To compare finite books with this limit, define, for $J>1$,}
\[
R_{c,J}
:=
\frac{
\Delta_{0.05}^{+}(\overline Z_{c,J})
-
\Delta_{0.05}^{+}(\overline Z_c)
}{
\Delta_{0.05}^{+}(Z_c)
-
\Delta_{0.05}^{+}(\overline Z_c)
},
\]
{whenever the denominator is positive. Figure
\ref{fig:num_book_diversification} compares the estimates with the
$J^{-1/2}$ benchmark from
Remark~\ref{rm:cont_sqrt_book_approximation}.}

{Across the three policies, $R_{c,J}$ is approximately
$0.280$--$0.284$ at $J=10$, $0.076$--$0.082$ at $J=100$, and
$0.018$--$0.024$ at $J=1000$. These values are broadly consistent with, but
generally below, the square-root benchmarks $0.316$, $0.100$, and $0.0316$.
The experiment supports the predicted rapid approach to the conditional
large-book limit; it does not establish an empirical convergence rate.}

\subsection{Implications for spouse-protected tontine design}
\label{ssc:num_design_implications}

Household risk control and continuation funding must be assessed
together. Tables~\ref{tab:num_reward_risk_selected} and
\ref{tab:num_spouse_continuation_load} show that greater policy conservatism
reduces terminal-shortfall risk but raises the expected-cost continuation
load, chiefly by lowering retiree-alive payments. Lower terminal-shortfall
risk therefore need not imply a lower continuation load.

Book size should affect the prudential buffer, not the expected
per-contract obligation. Using the representative-contract excess tail to set a prudential
buffer for a diversified book would fail to reflect the diversification
benefit shown in
Tables~\ref{tab:num_finite_book_spouse_continuation_load}
and \ref{tab:num_large_book_spouse_continuation_load}.
Scaling it toward zero ignores residual common-market risk.
The load should reflect the book's remaining tail risk, with expected cost
retained separately. This is a model-based funding assessment, not a market
price or regulatory capital requirement.

The continuation cap and book diversification are not substitutes.
The fixed-policy comparisons in
Appendix~\ref{app:num_continuation_cap_sensitivity} show that the cap reduces
expected cost and the residual common-market tail, neither of which disappears
with book size. This funding benefit must be weighed against the restriction
on the payment inherited by the spouse. The comparison quantifies the effect
of the rule; it does not establish an optimal cap or compare designs after
policy reoptimization.

\section{Conclusion}
\label{sec:conclusion}
This paper develops a tontine with spouse protection at the household-contract
level. Pool exit occurs at household extinction rather than at the retiree's
death, leading to household-state-dependent mortality credits and exact ex post
budget balance at the finite-pool level. The household chooses withdrawals and
portfolio rebalancing while the retiree is alive, whereas the spouse-only phase
follows a rule-based continuation payment and passive rebalancing convention.
Admissible global-in-time neural-network policies approximate the
solution of the resulting multidimensional reward--risk problem. The associated
policy-dependent continuation obligation is evaluated at both the
representative-contract and book levels, with the conditional large-book limit
isolating the common-market risk that remains after household-mortality
diversification.

The numerical results show that spouse continuation is frequent and
persistent, with a substantial expected cost that varies little across the
selected learned policies under capped continuation. 
In the baseline capped experiments, book-level diversification
reduces the representative-contract excess upper-tail cost by approximately
$97\%$ in the conditional large-book limit across the selected learned
policies. Expected per-contract continuation cost is unchanged, and a nonzero
common-market tail remains. The fixed-policy capped-versus-uncapped comparisons further show that the
continuation rule affects both the expected obligation and the residual
market-sensitive tail.
The cross-country mortality comparison indicates that the prevalence and
persistence of the spouse-only phase are not specific to Australia.

The framework therefore links household-extinction pooling,
endogenous decumulation, and the funding of spouse continuation within a
single model. Its central implication is that continuation funding must
distinguish the expected obligation from a prudential tail buffer assessed
at book level: neither representative-contract risk nor its mechanical
scaling toward zero captures the residual exposure of a large book.
Future work could incorporate dependent spousal mortality,
heterogeneous household types, stochastic mortality, and separately optimized
alternative continuation designs.

\section*{Appendices}
\appendix
\section{Neural-network training and evaluation specifications}
\label{app:NN_training}
The reported policies use two separate global-in-time, fully connected
feedforward networks: one for the withdrawal control and one for the
rebalancing control. At each decision time, both networks receive three inputs:
time-to-go, standardized contract wealth, and the current household state.
Table~\ref{tab:NN_training_architecture} summarizes the architecture and the
principal training specifications used in the production run.

\begin{table}[!htb]
\centering
\caption{Neural-network architecture and training specifications.}
\label{tab:NN_training_architecture}
\begin{tabular}{lcc}
\hline
Specification & Withdrawal network & Rebalancing network\\
\hline
Inputs
& \multicolumn{2}{c}{Time-to-go, standardized wealth, household state}\\
Hidden layers & 2 & 2\\
Nodes per hidden layer & 10 & 10\\
Hidden-layer activation & Logistic sigmoid & Logistic sigmoid\\
Output dimension & 1 & 4\\
Output transformation
& Scaled sigmoid, \eqref{eq:hatq_theta}
& Four-way softmax, \eqref{eq:hatp_theta}\\
Bias terms & Yes & Yes\\
Batch/layer normalization & No & No\\
Dropout & No & No\\
\hline
Training realizations & \multicolumn{2}{c}{$256{,}000$}\\
Training horizon & \multicolumn{2}{c}{$T=M=35$}\\
Stationary-bootstrap mean block length
& \multicolumn{2}{c}{24 months}\\
Training mini-batch size
& \multicolumn{2}{c}{$1{,}000$}\\
Number of training iterations
& \multicolumn{2}{c}{$30{,}000$}\\
CVaR confidence level & \multicolumn{2}{c}{$\alpha=0.95$}\\
Initial CVaR threshold & \multicolumn{2}{c}{$\eta_0=100$}\\
Policy-network optimizer & \multicolumn{2}{c}{Adam}\\
Initial policy-network learning rate & \multicolumn{2}{c}{$0.04$}\\
Policy-network Adam coefficients
& \multicolumn{2}{c}{$(\beta_1,\beta_2)=(0.9,0.998)$}\\
Policy-network weight decay
& \multicolumn{2}{c}{$10^{-4}$}\\
CVaR-threshold optimizer & \multicolumn{2}{c}{Adam}\\
Initial CVaR-threshold learning rate & \multicolumn{2}{c}{$0.05$}\\
CVaR-threshold regularization & \multicolumn{2}{c}{None}\\
Learning-rate schedule & \multicolumn{2}{c}{Multistep}\\
Random seed & \multicolumn{2}{c}{$5$}\\
Sequential transfer learning & \multicolumn{2}{c}{Yes}\\
Running-best checkpointing & \multicolumn{2}{c}{Yes}\\
\hline
\end{tabular}
\end{table}

The scalarization parameters are trained sequentially over the ordered grid
\[
\Lambda
=
\{0.15,0.16,0.17,0.18,0.185,0.2,0.25,0.3,0.5,0.7,1,
1.5,2,2.5,3,5,7.5,10,15,20,50,100\}.
\]
As $\lambda$ increases, the objective places greater weight on
$\operatorname{CVaR}_{0.95}(L_T)$. The gradient contribution from this risk
term is driven primarily by realizations in the upper tail of the terminal
shortfall distribution and can therefore be sparse within a training
mini-batch. This makes the more risk-sensitive policies harder to train from an
independent random initialization.

Following the sequential transfer-learning approach of
\cite{chen2023benchmark} and the broader transfer-learning literature
\cite{tan2018survey}, the policy at $\lambda=0.15$ is initialized randomly
(a cold start). For each subsequent $\lambda$, the withdrawal-network
parameters, rebalancing-network parameters, and CVaR threshold are initialized
from the immediately preceding trained solution. The ordering of $\Lambda$ is
therefore part of the numerical protocol. In related decumulation experiments,
this continuation strategy was found to improve training stability and reduce
the incidence of convergence to inferior local solutions at more
risk-sensitive frontier points \cite{chen2023benchmark}.

During each training run, running-best checkpointing retains the parameter set
that attains the best observed empirical objective, rather than necessarily
using the final iterate. After training, the learned policies are held fixed
and evaluated on fresh joint asset-return and household-state realizations.
For the selected policies $\lambda=0.3$, $1$, and $15$, Monte Carlo
uncertainty is assessed using 100 independent evaluation batches of $2{,}560$
realizations each. Common batch realizations are used across the selected
policies and in the paired capped--uncapped continuation comparisons. The shared-path evaluation design and batch-based uncertainty
estimates are detailed in Appendix~\ref{app:num_numerical_robustness}.
\section{Derivation of \textbf{(\ref{eq:cont_probability_weighted_cost})}}
\label{app:cont_probability_weighted_derivation}
Recall that
$Z_c=\sum_{m=0}^{M-1}C_m\mathbf 1_{\{H_{m^-}=01\}}$ and
$\overline Z_c=\mathbb E^{\nu^\ast}[Z_c\mid\mathcal G]$.
Since state $01$ can be entered at most once, through a transition
$11\to01$,
\[
Z_c
=
\sum_{\beta=1}^{M-1}\sum_{m=\beta}^{M-1}
C_m^{(\beta)}
\mathbf 1_{\{
H_{(\beta-1)^+}=11,\,
H_{\beta^-}=01,\,
H_{m^-}=01
\}},
\]
where, on the event in the indicator, $C_m=C_m^{(\beta)}$, and
$C_m^{(\beta)}$ is $\mathcal G$-measurable. Hence,
\begin{align*}
\overline Z_c
&=
\sum_{\beta=1}^{M-1}\sum_{m=\beta}^{M-1}
C_m^{(\beta)}
\mathbb P\!\left(
H_{(\beta-1)^+}=11,\,
H_{\beta^-}=01,\,
H_{m^-}=01
\mid\mathcal G
\right)
\\
&=
\sum_{\beta=1}^{M-1}\sum_{m=\beta}^{M-1}
C_m^{(\beta)}
\mathbb P\!\left(
H_{(\beta-1)^+}=11,\,
H_{\beta^-}=01,\,
H_{m^-}=01
\right)
=
\sum_{\beta=1}^{M-1}\sum_{m=\beta}^{M-1}
C_m^{(\beta)}a_\beta s_{\beta,m}
=
\sum_{\beta=1}^{M-1}a_\beta
\sum_{m=\beta}^{M-1}s_{\beta,m}C_m^{(\beta)},
\end{align*}
where the second equality uses independence of household mortality and the
shared asset returns. This gives~
\eqref{eq:cont_probability_weighted_cost}.
\section{Simulation algorithm for spouse-continuation loads}
\label{app:cont_load_algorithm}

{The two algorithms below implement the finite-book and large-book
spouse-continuation calculations in
Section~\ref{sc:spouse_continuation_load}.
Algorithm~\ref{alg:cont_load} estimates the finite-book quantities for a
selected book size $J$, with $J=1$ corresponding to the
representative-contract case. Algorithm~\ref{alg:cont_load_large_book}
evaluates the large-book limiting quantities by integrating over household
mortality using \eqref{eq:cont_probability_weighted_cost}. In the algorithms below, the ``continuation rule'' refers to the choice between
capped and uncapped continuation: $B_m=\min\{q_m,\overline q_1\}$ under capped
continuation and $B_m=q_m$ under uncapped continuation.}

\subsection{Representative-contract or finite-book estimates}
Let $K$ be the number of Monte Carlo replications and let $J\ge1$ be
the book size. Within each replication, all $J$ contracts share the same asset
returns, while their household mortality histories are conditionally
independent.

\begingroup
\captionsetup{aboveskip=1pt,belowskip=1pt}
\noindent\rule{\linewidth}{0.75pt}
\captionof{algorithm}{Representative-contract or finite-book estimates}
\label{alg:cont_load}
\noindent\rule{\linewidth}{0.75pt}
\endgroup

\begingroup
\begin{algorithmic}[1]
\STATE choose the number of evaluation realizations $K$, book size $J$,
prudential-buffer coefficient $\lambda_c\ge0$, upper-tail probability
$\alpha_c\in(0,1)$, and continuation rule;
\STATE {initialize arrays
$\overline{\mathcal Y}_{J}\leftarrow[0,\ldots,0]$ and
$\overline{\mathcal Z}_{J}\leftarrow[0,\ldots,0]$, each of length $K$;}
\FOR{$k=1,\ldots,K$}
    \STATE {generate one realization
    $\{\varepsilon_{m+1}^{(k)}\}_{m=0}^{M-1}$ of the exogenous return drivers
    in \eqref{eq:hh_dynamics_generic};}
    \FOR{$j=1,\ldots,J$}
        \STATE initialize
        $X_{0^-}^{(k,j)}\leftarrow x_0$,
        $H_{0^-}^{(k,j)}\leftarrow11$,
        $B_{-1}^{(k,j)}\leftarrow0$,
        $Y_r^{(k,j)}\leftarrow0$, and
        $Z_c^{(k,j)}\leftarrow0$;
        \FOR{$m=0,\ldots,M-1$}
            \STATE \label{algline:cont_recursion_start}
            compute
            $\mathcal W_{m^-}^{(k,j)}$,
            $\widetilde W_{m^-}^{(k,j)}$, and
            $W_{m^-}^{(k,j)}$ using
            \eqref{eq:hh_mod_tontine_gain}--\eqref{eq:hh_Wm_minus};
            \IF{$H_{m^-}^{(k,j)}\in\mathcal H_R$}
                \IF{$W_{m^-}^{(k,j)}>0$}
                    \STATE set
                    $q_m^{(k,j)}
                    \leftarrow
                    \widehat q_m^\ast
                    (W_{m^-}^{(k,j)},t_m,H_{m^-}^{(k,j)})$;
                \ELSE
                    \STATE set
                    $q_m^{(k,j)}
                    \leftarrow
                    \underline q(H_{m^-}^{(k,j)})$;
                \ENDIF
                \STATE set
                $C_m^{(k,j)}\leftarrow q_m^{(k,j)}$;
                \STATE set
                $B_m^{(k,j)}
                \leftarrow
                \min\{q_m^{(k,j)},\overline q_1\}$
                under capped continuation, and
                $B_m^{(k,j)}\leftarrow q_m^{(k,j)}$
                under uncapped continuation;
                \STATE update
                $Y_r^{(k,j)}
                \leftarrow
                Y_r^{(k,j)}+C_m^{(k,j)}$
                and
                $W_{m^+}^{(k,j)}
                \leftarrow
                W_{m^-}^{(k,j)}-C_m^{(k,j)}$;
                \IF{$W_{m^+}^{(k,j)}>0$}
                    \STATE compute
                    $\widehat{\boldsymbol p}_m^{\,\ast}
                    (W_{m^+}^{(k,j)},t_m,H_{m^-}^{(k,j)})$
                    and update $X_{m^+}^{(k,j)}$ using
                    \eqref{eq:hh_Xmplus_active};
                \ELSE
                    \STATE update $X_{m^+}^{(k,j)}$ using
                    \eqref{eq:hh_insolvent_X};
                \ENDIF
            \ELSIF{$H_{m^-}^{(k,j)}=01$}
                \STATE set
                $B_m^{(k,j)}\leftarrow B_{m-1}^{(k,j)}$;
                \STATE set
                $C_m^{(k,j)}\leftarrow B_{m-1}^{(k,j)}$
                if $W_{m^-}^{(k,j)}>0$, and
                $C_m^{(k,j)}\leftarrow\underline q_1$ otherwise;
                \STATE update
                $Z_c^{(k,j)}
                \leftarrow
                Z_c^{(k,j)}+C_m^{(k,j)}$
                and
                $W_{m^+}^{(k,j)}
                \leftarrow
                W_{m^-}^{(k,j)}-C_m^{(k,j)}$;
                \IF{$\mathcal W_{m^-}^{(k,j)}>0$,
                    $W_{m^-}^{(k,j)}>0$, and
                    $W_{m^+}^{(k,j)}>0$}
                    \STATE update $X_{m^+}^{(k,j)}$ passively using
                    \eqref{eq:hh_X_prepay}--                    \eqref{eq:hh_spouse_passive_update};
                \ELSE
                    \STATE update $X_{m^+}^{(k,j)}$ using
                    \eqref{eq:hh_insolvent_X};
                \ENDIF
            \ELSE
                \STATE set
                $B_m^{(k,j)}\leftarrow0$,
                $C_m^{(k,j)}\leftarrow0$,
                $W_{m^+}^{(k,j)}\leftarrow0$, and
                $X_{m^+}^{(k,j)}\leftarrow(0,0,0,0)^\top$;
            \ENDIF
            \STATE \label{algline:cont_recursion_end}
            {propagate $X_{(m+1)^-}^{(k,j)}$ from
            $X_{m^+}^{(k,j)}$ using $\varepsilon_{m+1}^{(k)}$ via
            \eqref{eq:hh_dynamics_generic};}
            \STATE simulate $H_{(m+1)^-}^{(k,j)}$ from
            $H_{m^+}^{(k,j)}=H_{m^-}^{(k,j)}$ using
            \eqref{eq:hh_transition_11}--\eqref{eq:hh_transition_01};
        \ENDFOR
    \ENDFOR
    \STATE {set
    $\displaystyle
    \overline Y_{r,J}^{(k)}
    \leftarrow
    \frac{1}{J}\sum_{j=1}^{J}Y_r^{(k,j)}$
    and
    $\displaystyle
    \overline Z_{c,J}^{(k)}
    \leftarrow
    \frac{1}{J}\sum_{j=1}^{J}Z_c^{(k,j)}$;}
    \STATE {set
    $\overline{\mathcal Y}_{J}^{(k)}
    \leftarrow\overline Y_{r,J}^{(k)}$
    and
    $\overline{\mathcal Z}_{J}^{(k)}
    \leftarrow\overline Z_{c,J}^{(k)}$;}
\ENDFOR
\STATE {compute
$\widehat{\mathbb E}[Y_r]
\leftarrow
K^{-1}\sum_{k=1}^{K}\overline{\mathcal Y}_{J}^{(k)}$
and
$\widehat{\mathbb E}[Z_c]
\leftarrow
K^{-1}\sum_{k=1}^{K}\overline{\mathcal Z}_{J}^{(k)}$;}
\STATE {sort $\overline{\mathcal Z}_{J}$ in descending order as
$\overline Z_{c,J,(1)}
\ge\cdots\ge
\overline Z_{c,J,(K)}$;}
\STATE {compute
$\displaystyle
\widehat{\mathrm{CVaR}}_{\alpha_c}^{+}
(\overline Z_{c,J})
\leftarrow
(\lceil\alpha_cK\rceil)^{-1}\sum_{k=1}^{\lceil\alpha_cK\rceil}\overline Z_{c,J,(k)}$;}
\STATE {compute
$\widehat\Delta_{\alpha_c}^{+}
(\overline Z_{c,J})
\leftarrow
\widehat{\mathrm{CVaR}}_{\alpha_c}^{+}
(\overline Z_{c,J})
-
\widehat{\mathbb E}[Z_c]$;}
\STATE compute
$\displaystyle
\widehat f_c
\leftarrow
\frac{\widehat{\mathbb E}[Z_c]}
{\widehat{\mathbb E}[Y_r]+\widehat{\mathbb E}[Z_c]}$;
\STATE {compute
$\widehat\Pi_{c,J}
\leftarrow
\widehat{\mathbb E}[Z_c]
+
\lambda_c
\widehat\Delta_{\alpha_c}^{+}
(\overline Z_{c,J})$
and
$\displaystyle
\widehat f_{c,J}^{\mathrm{RL}}
\leftarrow
\frac{\widehat\Pi_{c,J}}
{\widehat{\mathbb E}[Y_r]+\widehat\Pi_{c,J}}$;}
\STATE {return
$\widehat{\mathbb E}[Y_r]$,
$\widehat{\mathbb E}[Z_c]$,
$\widehat{\mathrm{CVaR}}_{\alpha_c}^{+}(\overline Z_{c,J})$,
$\widehat\Delta_{\alpha_c}^{+}(\overline Z_{c,J})$,
$\widehat f_c$,
$\widehat\Pi_{c,J}$, and
$\widehat f_{c,J}^{\mathrm{RL}}$.}
\end{algorithmic}
\endgroup
\noindent\rule{\linewidth}{0.75pt}

\medskip
{When $J=1$,
$\overline Y_{r,1}^{(k)}=Y_r^{(k,1)}$ and
$\overline Z_{c,1}^{(k)}=Z_c^{(k,1)}$, giving the
representative-contract estimates; $J>1$ gives the corresponding finite-book
estimates.}
\subsection{Large-book limiting estimates}

{The large-book limiting average per-contract cost is evaluated using
\eqref{eq:cont_probability_weighted_cost}. For each possible entry time
$t_\beta^-$, household mortality is not simulated: state $11$ is imposed before
$t_\beta^-$, state $01$ is imposed from $t_\beta^-$ onward, and the weights
$s_{\beta,m}$ account for spouse survival through subsequent payment dates.}
\begingroup
\captionsetup{aboveskip=1pt,belowskip=1pt}
\noindent\rule{\linewidth}{0.75pt}
\captionof{algorithm}{Large-book limiting estimates}
\label{alg:cont_load_large_book}
\noindent\rule{\linewidth}{0.75pt}
\endgroup

\begingroup
\begin{algorithmic}[1]
\STATE {generate $K$ realizations
$\{\varepsilon_{m+1}^{(k)}\}_{m=0}^{M-1}$,
$k=1,\ldots,K$, of the exogenous return drivers in
\eqref{eq:hh_dynamics_generic};}
\STATE use the same $\alpha_c$, $\lambda_c$, and the continuation rule,
as in Algorithm~\ref{alg:cont_load} with $J=1$;
\\
input the resulting representative-contract estimates
$\widehat{\mathbb E}[Y_r]$ and
$\widehat\Delta_{\alpha_c}^{+}(Z_c)$;
\STATE input the weights
$\{a_\beta,s_{\beta,m}\}$ from
\eqref{eq:cont_entry_survival_weights};
\STATE {initialize
$\overline{\mathcal Z}\leftarrow[0,\ldots,0]$ of length $K$;}
\FOR{$k=1,\ldots,K$}
    \FOR{$\beta=1,\ldots,M-1$}
        \STATE {initialize
        $X_{0^-}^{(k,\beta)}\leftarrow x_0$,
        $B_{-1}^{(k,\beta)}\leftarrow0$,
        $Y_r^{(k,\beta)}\leftarrow0$, and
        $Z_c^{(k,\beta)}\leftarrow0$;}
        \STATE {impose
        $H_{m^-}^{(k,\beta)}=H_{m^+}^{(k,\beta)}=11$ for $m<\beta$ and
        $H_{m^-}^{(k,\beta)}=H_{m^+}^{(k,\beta)}=01$ for $m\ge\beta$;}
        \STATE {when computing the period-$\beta$ tontine gain, use
        $H_{(\beta-1)^+}^{(k,\beta)}=11$;}
        \FOR{$m=0,\ldots,M-1$}
            \STATE {apply
            lines~\ref{algline:cont_recursion_start}--            \ref{algline:cont_recursion_end} of
            Algorithm~\ref{alg:cont_load}, with $j$ replaced by $\beta$,
            using $\varepsilon_{m+1}^{(k)}$ and the imposed household states;}
        \ENDFOR
        \STATE {retain
        $C_m^{(k,\beta)}$, $m=\beta,\ldots,M-1$;}
    \ENDFOR
    \STATE {set
    $\displaystyle
    \overline Z_c^{(k)}
    \leftarrow
    \sum_{\beta=1}^{M-1}a_\beta
    \sum_{m=\beta}^{M-1}
    s_{\beta,m}C_m^{(k,\beta)}$
    and
    $\overline{\mathcal Z}^{(k)}
    \leftarrow
    \overline Z_c^{(k)}$;}
\ENDFOR
\STATE {compute
$\widehat{\mathbb E}[\overline Z_c]
\leftarrow
K^{-1}\sum_{k=1}^{K}\overline{\mathcal Z}^{(k)}$;}
\STATE {sort $\overline{\mathcal Z}$ in descending order as
$\overline Z_{c,(1)}
\ge\cdots\ge
\overline Z_{c,(K)}$;}
\STATE {compute
$\displaystyle
\widehat{\mathrm{CVaR}}_{\alpha_c}^{+}(\overline Z_c)
\leftarrow
(\lceil\alpha_c K\rceil)^{-1}\sum_{k=1}^{\lceil\alpha_c K\rceil}\overline Z_{c,(k)}$;}
\STATE {compute
$\widehat\Delta_{\alpha_c}^{+}(\overline Z_c)
\leftarrow
\widehat{\mathrm{CVaR}}_{\alpha_c}^{+}(\overline Z_c)
-
\widehat{\mathbb E}[\overline Z_c]$;}
\STATE {if
$\widehat\Delta_{\alpha_c}^{+}(Z_c)>0$, compute
$\displaystyle
\widehat D_c
\leftarrow
1-
\frac{
\widehat\Delta_{\alpha_c}^{+}(\overline Z_c)
}{
\widehat\Delta_{\alpha_c}^{+}(Z_c)
}$;}
\STATE {compute
$\widehat\Pi_{c,\infty}
\leftarrow
\widehat{\mathbb E}[\overline Z_c]
+
\lambda_c
\widehat\Delta_{\alpha_c}^{+}(\overline Z_c)$;}
\STATE {compute
$\displaystyle
\widehat f_{c,\infty}^{\mathrm{RL}}
\leftarrow
\frac{\widehat\Pi_{c,\infty}}
{\widehat{\mathbb E}[Y_r]+\widehat\Pi_{c,\infty}}$;}
\STATE {return
$\widehat{\mathbb E}[\overline Z_c]$,
$\widehat{\mathrm{CVaR}}_{\alpha_c}^{+}(\overline Z_c)$,
$\widehat\Delta_{\alpha_c}^{+}(\overline Z_c)$,
$\widehat D_c$ when defined,
$\widehat\Pi_{c,\infty}$, and
$\widehat f_{c,\infty}^{\mathrm{RL}}$.}
\end{algorithmic}
\endgroup
\noindent\rule{\linewidth}{0.75pt}

\medskip
\section{International household-state sensitivity}
\label{app:international_household_states}
To assess whether the importance of spouse continuation is specific to the
Australian mortality calibration, we repeat the analytic household-state
calculation using the 2021 male and female period life tables from the Human
Mortality Database for Australia, Canada, the United States, Japan, and the
Netherlands \cite{HMD}. In every country, the household consists at inception
of a male retiree aged 65 and a female spouse aged 60 and is followed for 35
years, until the spouse reaches age 95. We hold the 2021 age-specific mortality
schedule fixed over the horizon and retain the baseline assumption that the two
household lifetimes are independent. Durations use the same annual state count $D_{01}$, including the
terminal observation, as in Subsection~\ref{ssc:num_household_state_probs}.

\begin{table}[!htbp]
\centering
\caption{Household-state probabilities and spouse-only duration under selected
2021 HMD period life tables. The household consists of a male retiree aged 65
and a female spouse aged 60 and is followed over a 35-year horizon.}
\label{tab:international_household_states}
\begin{tabular}{lccc}
\hline
\noalign{\vspace{2mm}}
Country
& \shortstack{Probability spouse-only\\phase occurs (\%)}
& \shortstack{Mean spouse-only\\duration (years)}
& \shortstack{Conditional mean duration\\given entry (years)}\\
\noalign{\vspace{2mm}}
\hline
Australia   & 71.0 &  8.70 & 12.25\\
Canada      & 70.2 &  9.03 & 12.85\\
United States & 68.1 & 9.09 & 13.34\\
Japan       & 77.0 & 10.23 & 13.28\\
Netherlands & 71.0 &  8.84 & 12.44\\
\hline
\end{tabular}
\end{table}

Table~\ref{tab:international_household_states} shows that the
probability of entering the spouse-only phase ranges from approximately
$68.1\%$ in the United States to $77.0\%$ in Japan. Its unconditional mean
duration ranges from $8.70$ to $10.23$ years, while the conditional mean given
entry lies between $12.25$ and $13.34$ years. Canada and the Netherlands are
close to the Australian baseline; Japan has the highest entry probability and
unconditional duration, while the United States has the lowest entry
probability but the longest conditional duration. Thus, the conclusion that
spouse continuation is both likely and long-lasting is robust across the
selected mortality environments. This comparison isolates mortality effects;
it does not recalibrate financial returns, spending bands, or product rules for
each country.
\section{Financial data construction and return characteristics}
\label{app:financial_data}

{The four-asset investment universe consists of Australian equities,
Australian government bonds, U.S.\ equities, and U.S.\ 30-day Treasury bills.
The underlying monthly data are aligned over the common sample period
1935:1--2022:12 and expressed in real Australian dollars.}

{The Australian equity series is a long-horizon capitalization-weighted
total-return index constructed from Bloomberg All Ordinaries data together
with historical dividend-yield information from
\cite{Mathews2019AustEquities} and the Reserve Bank of Australia. The
Australian bond series is a long-run government-bond total-return index formed
by combining the historical government-bond return series of
\cite{jorda2019rate} and \cite{chen2022documentation} with a recent Bloomberg
Australian government-bond total-return index.}

\begin{table}[!htbp]
\centering
\caption{Summary statistics of monthly real returns, 1935:1--2022:12.
The first three return measures are annualized.
$\mathrm{VaR}_{0.05}$ and $\mathrm{CVaR}_{0.05}$ refer to the lower
$5\%$ tail of monthly returns; returns are reported as decimals.}
\label{tab:return_stats_summary}
\begin{tabular}{lccccc}
\hline
\noalign{\vspace{2mm}}
Asset
& \shortstack{Mean\\(ann.)}
& \shortstack{Geometric\\mean (ann.)}
& \shortstack{Volatility\\(ann.)}
& $\mathrm{VaR}_{0.05}$
& $\mathrm{CVaR}_{0.05}$\\
\hline
U.S.\ 30-day Treasury bill
& 0.003 & -0.001 & 0.095 & -0.041 & -0.056\\
U.S.\ equity index
& 0.082 & 0.071 & 0.165 & -0.064 & -0.096\\
Australian 10-year government bond
& 0.012 & 0.012 & 0.036 & -0.015 & -0.025\\
Australian equity index
& 0.069 & 0.059 & 0.151 & -0.065 & -0.100\\
\hline
\end{tabular}
\end{table}

\begin{table}[!htbp]
\centering
\caption{Correlation matrix of monthly real AUD returns,
1935:1--2022:12.}
\label{tab:return_stats_corr}
\begin{tabular}{lcccc}
\hline
\noalign{\vspace{2mm}}
&
\shortstack{U.S.\ 30-day\\Treasury bill}
&
\shortstack{U.S.\\equity}
&
\shortstack{Australian\\bond}
&
\shortstack{Australian\\equity}\\
\hline
U.S.\ 30-day Treasury bill & 1.00 & 0.34 & 0.17 & -0.22\\
U.S.\ equity               & 0.34 & 1.00 & 0.10 &  0.33\\
Australian bond            & 0.17 & 0.10 & 1.00 &  0.11\\
Australian equity          & -0.22 & 0.33 & 0.11 & 1.00\\
\hline
\end{tabular}
\end{table}

{The U.S.\ asset block consists of a broad equity total-return series and
a 30-day Treasury-bill total-return series. Both are converted from USD to AUD
using an end-of-month AUD/USD exchange-rate series compiled from Federal
Reserve archives and Reserve Bank of Australia data. All four nominal return
series are converted to real AUD returns using Australian CPI. Further details
on data construction, splicing, currency conversion, and preprocessing are
given in \cite{orozco2026money}.}

Tables~\ref{tab:return_stats_summary} and~\ref{tab:return_stats_corr} report
the main return characteristics and dependence structure of the resulting
monthly real AUD panel.

{U.S.\ equity has the highest historical average real return in the
sample and is only moderately correlated with Australian equity. The
full-sample correlation is approximately $0.33$, increasing to $0.48$ over
1990:1--2022:12. By contrast, after conversion to real AUD, the U.S.\ Treasury
bill has a near-zero average return and substantial exchange-rate-induced
volatility. The lower historical volatility of Australian government bonds
helps explain their defensive role in the learned allocations reported in
Appendix~\ref{app:additional_control_surfaces}.}
\section{Learned household withdrawal and allocation controls}
\label{app:additional_control_surfaces}
\label{ssc:num_learned_controls}

{This appendix presents the learned withdrawal and four-asset
allocation controls under the intermediate policy $\widehat\nu_{1}^\ast$
selected in Subsection~\ref{ssc:num_reward_risk_frontier}. Controls are
defined only in the retiree-alive states $\mathcal H_R=\{11,10\}$;
state $01$ follows the continuation-payment and passive-rebalancing rules.
We examine withdrawal behavior first, followed by the allocation pattern
across funding levels and household states.}

\begin{figure}[!htb]
\centering
\begin{subfigure}[t]{0.49\textwidth}
\centering
\includegraphics[width=\textwidth]
{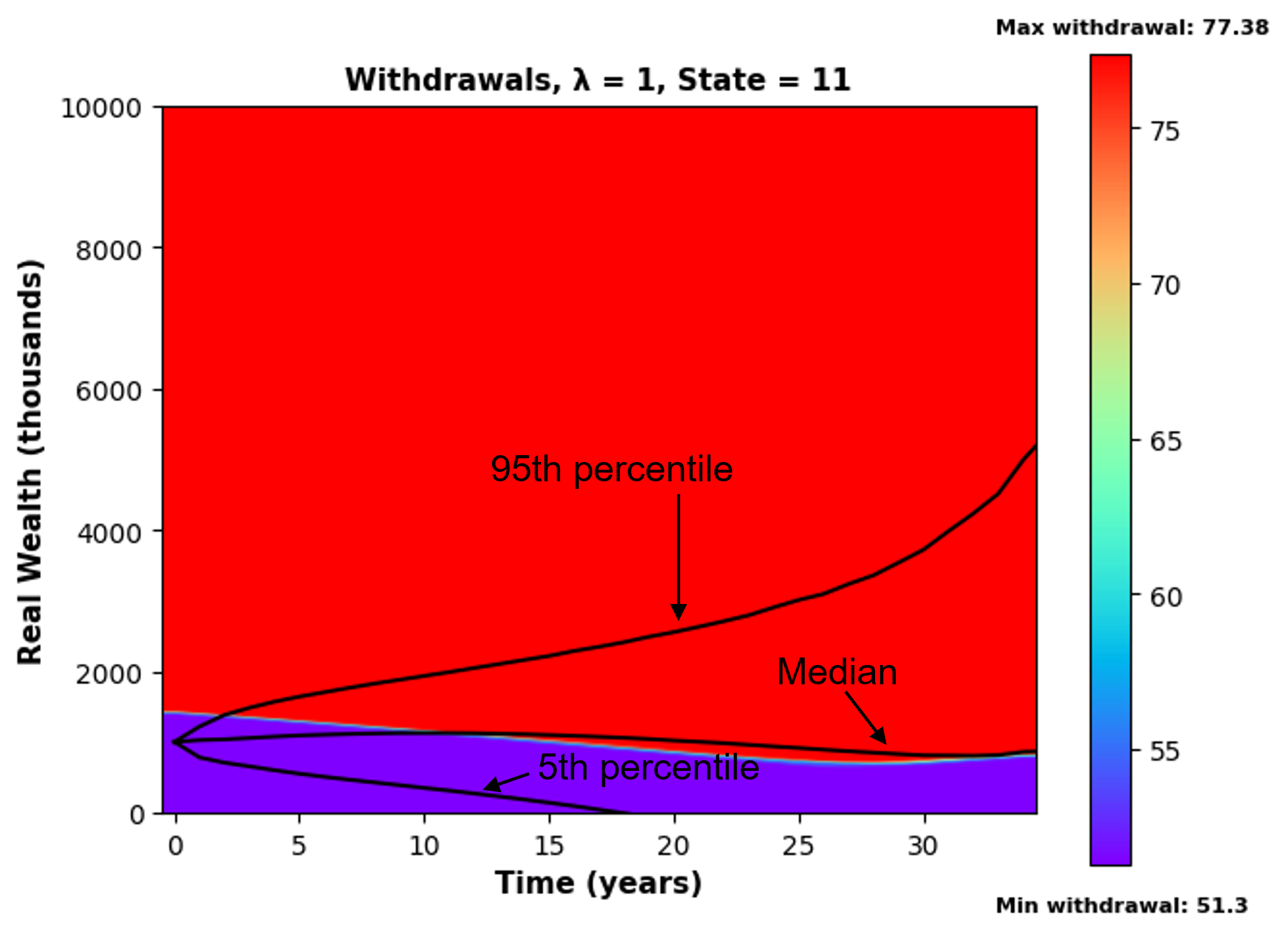}
\caption{Household state $11$.}
\label{fig:num_withdrawal_controls_lambda1_11}
\end{subfigure}
\hfill
\begin{subfigure}[t]{0.49\textwidth}
\centering
\includegraphics[width=\textwidth]
{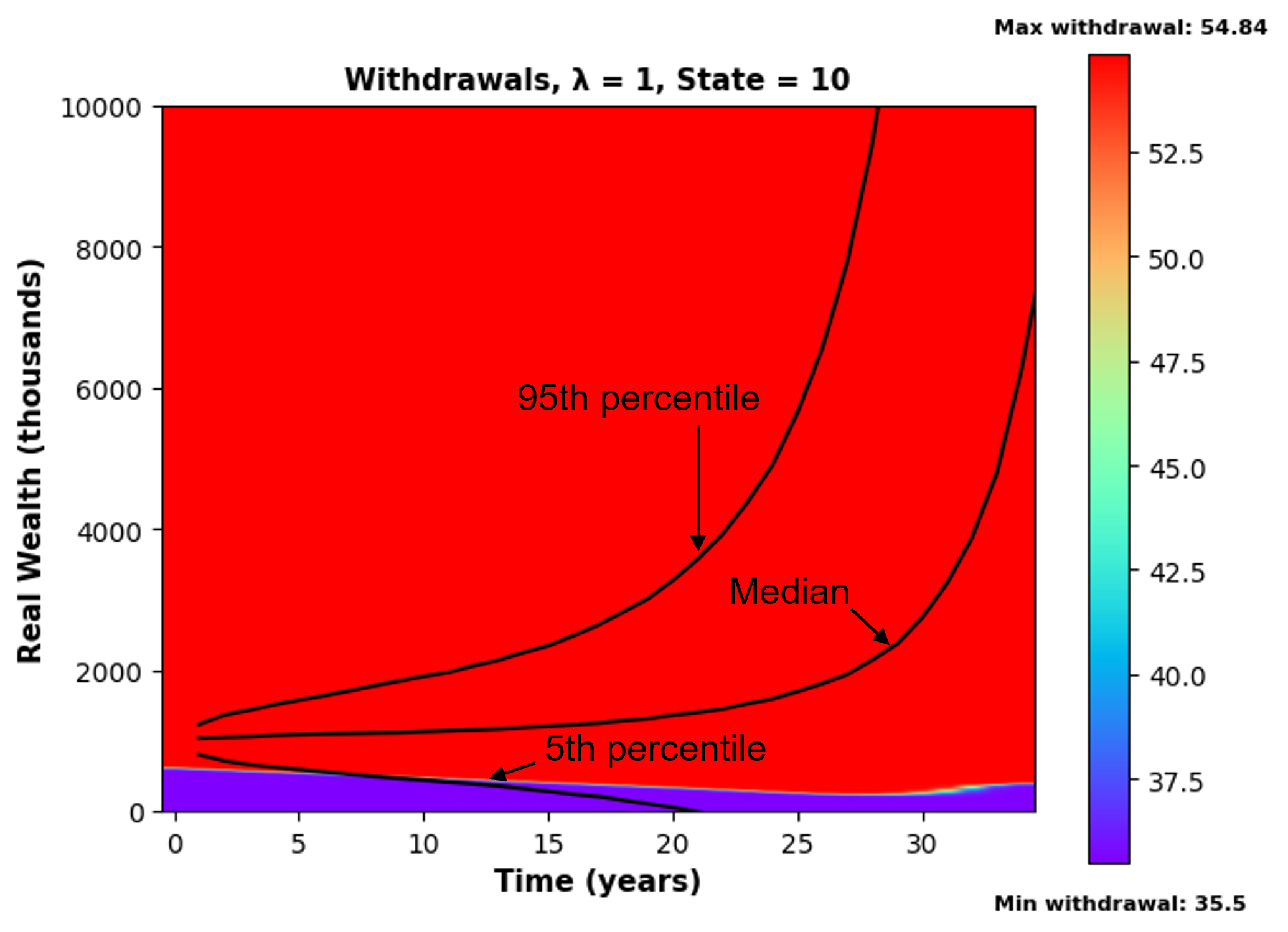}
\caption{Household state $10$.}
\label{fig:num_withdrawal_controls_lambda1_10}
\end{subfigure}
\caption{Learned withdrawal controls under the intermediate policy
$\widehat\nu_{1}^\ast$. Colors report annual controlled withdrawals as
functions of time and pre-payment wealth.}
\label{fig:num_withdrawal_controls_lambda1}
\end{figure}

{Figure~\ref{fig:num_withdrawal_controls_lambda1} shows that the
learned withdrawals are close to the relevant lower or upper band over most
of the wealth--time domain. The wealth threshold for the maximum withdrawal
is generally higher in state $11$ than in state $10$, leaving a larger
minimum-withdrawal region when both members are alive. This is consistent
with the larger two-person payment band and reserve target, together with
weaker mortality-credit support under the joint-death pool-exit rule.}

\begin{figure}[!htb]
\centering
\begin{subfigure}[t]{0.49\textwidth}
\centering
\includegraphics[width=\textwidth]
{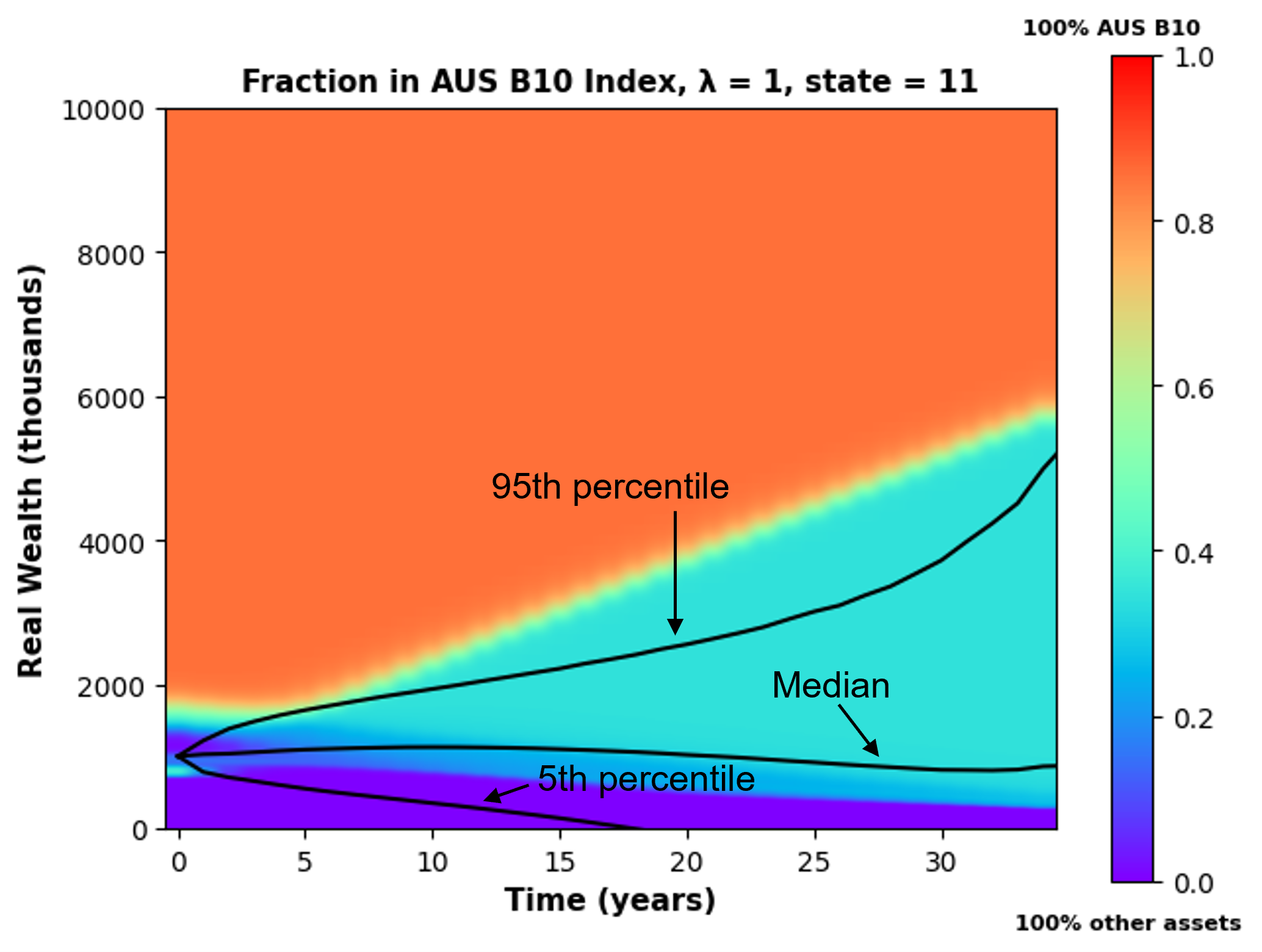}
\caption{Australian bond, state $11$.}
\label{fig:num_aus_bond_controls_lambda1_11}
\end{subfigure}
\hfill
\begin{subfigure}[t]{0.49\textwidth}
\centering
\includegraphics[width=\textwidth]
{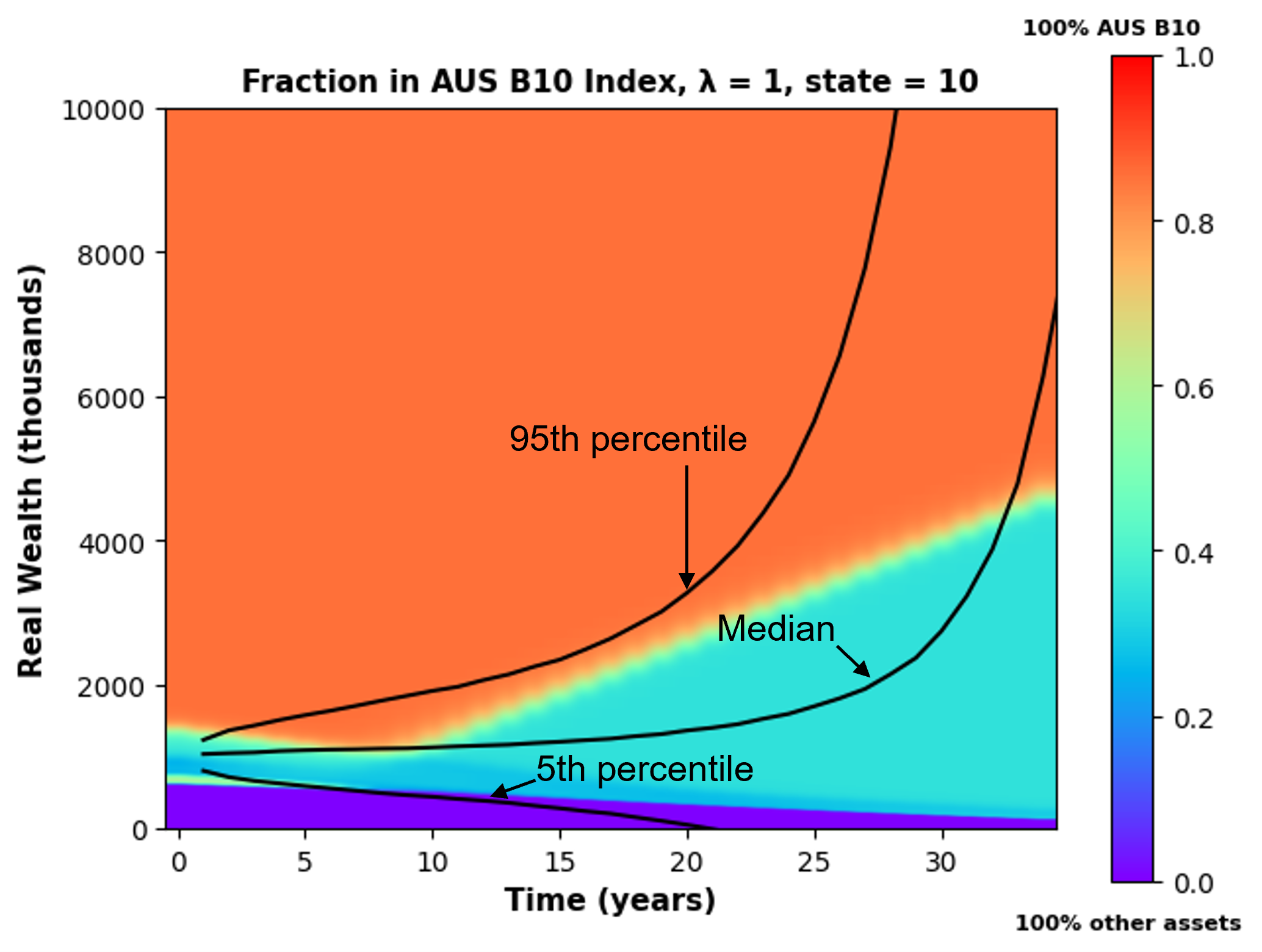}
\caption{Australian bond, state $10$.}
\label{fig:num_aus_bond_controls_lambda1_10}
\end{subfigure}

\vspace{0.6em}
\begin{subfigure}[t]{0.49\textwidth}
\centering
\includegraphics[width=\textwidth]
{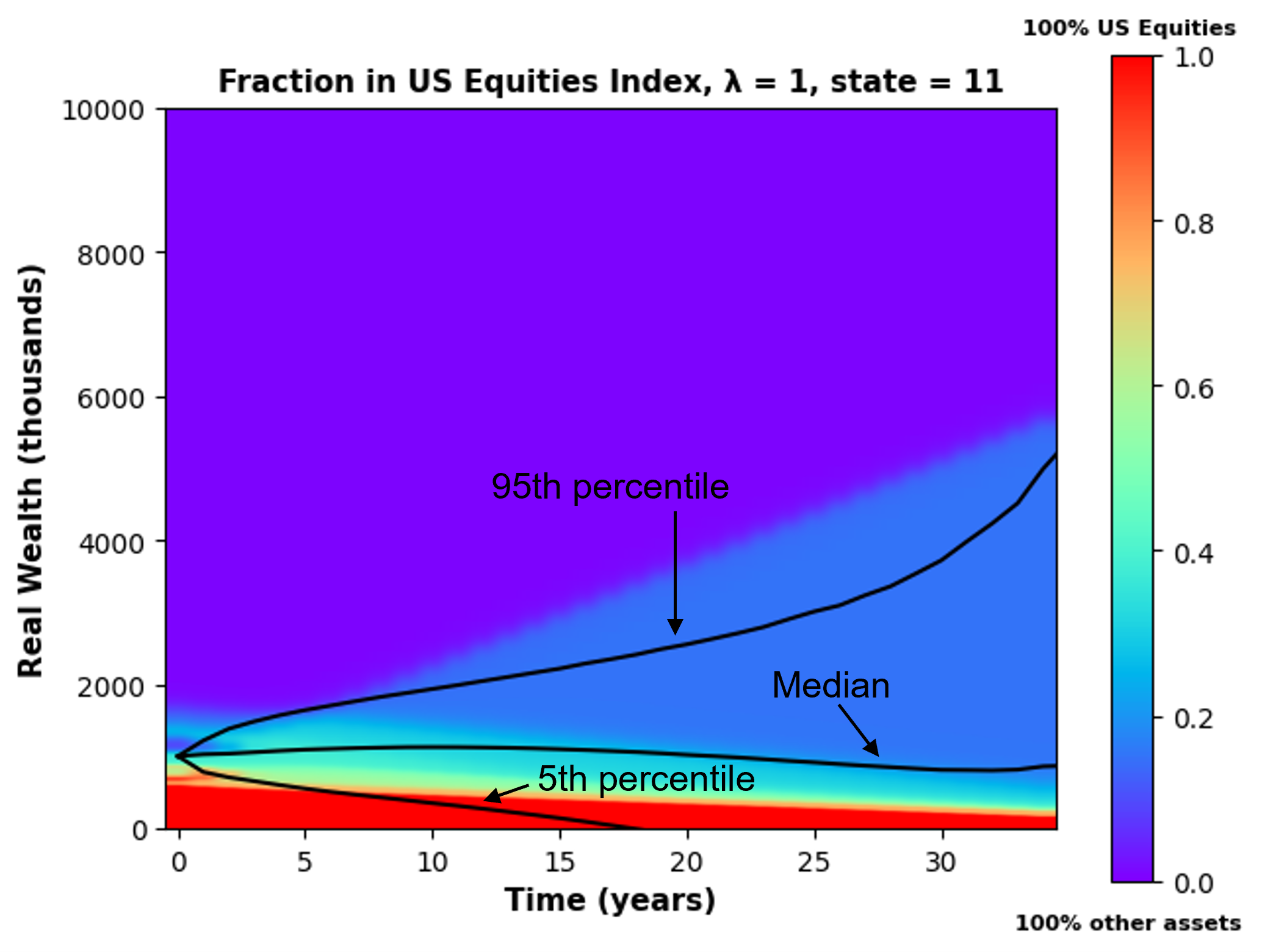}
\caption{U.S. equity, state $11$.}
\label{fig:num_us_equity_controls_lambda1_11}
\end{subfigure}
\hfill
\begin{subfigure}[t]{0.49\textwidth}
\centering
\includegraphics[width=\textwidth]
{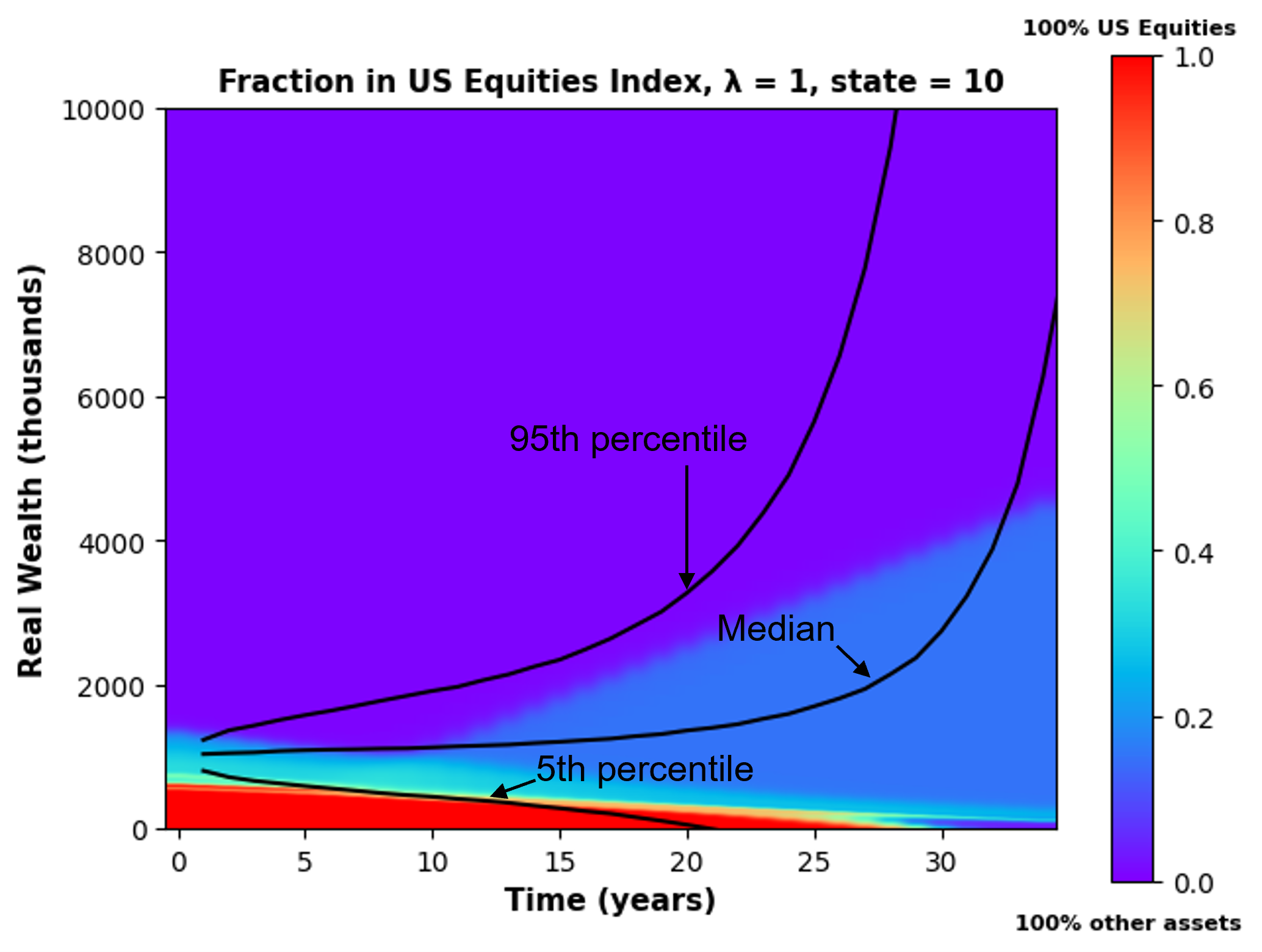}
\caption{U.S. equity, state $10$.}
\label{fig:num_us_equity_controls_lambda1_10}
\end{subfigure}
\caption{Learned Australian government-bond allocations (top row)
and U.S. equity allocations (bottom row) under the intermediate policy
$\widehat\nu_{1}^\ast$. Colors denote fractions of post-payment wealth.}
\label{fig:num_allocation_controls_lambda1}
\end{figure}

\begin{figure}[!htb]
\centering
\begin{subfigure}[t]{0.49\textwidth}
\centering
\includegraphics[width=\textwidth]
{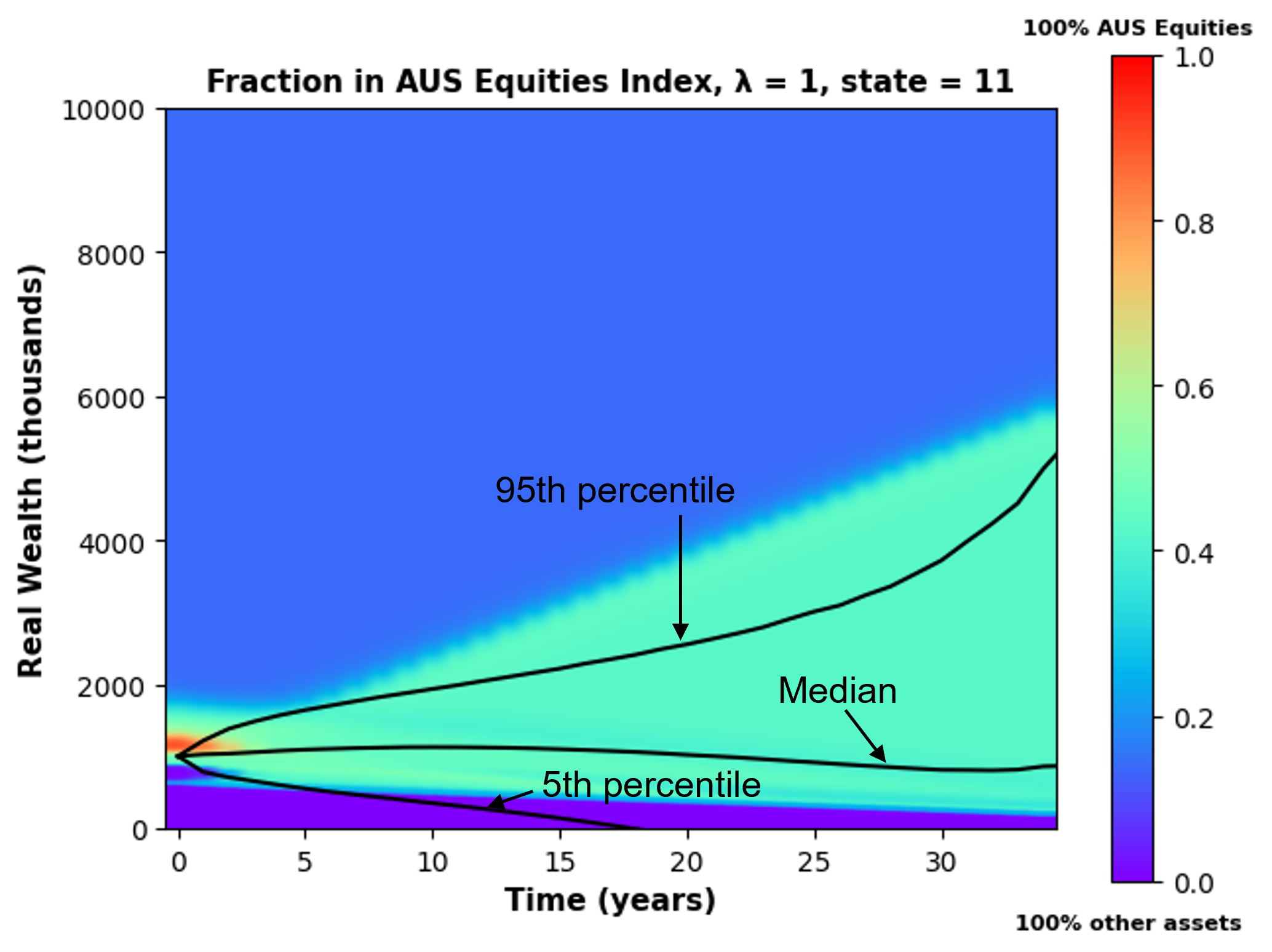}
\caption{Australian equity, state $11$.}
\label{fig:app_aus_equity_controls_lambda1_11}
\end{subfigure}
\hfill
\begin{subfigure}[t]{0.49\textwidth}
\centering
\includegraphics[width=\textwidth]
{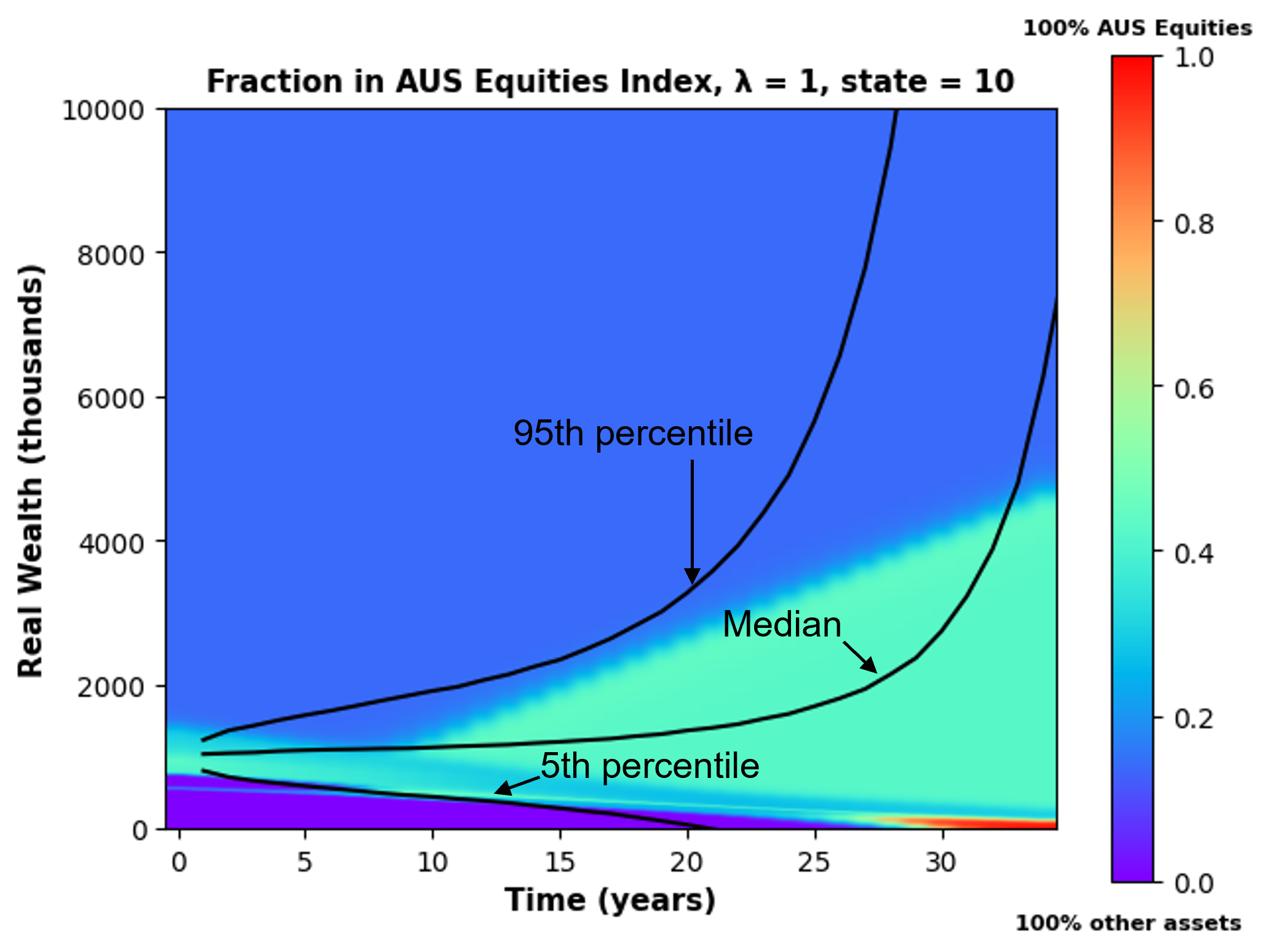}
\caption{Australian equity, state $10$.}
\label{fig:app_aus_equity_controls_lambda1_10}
\end{subfigure}

\vspace{0.6em}
\begin{subfigure}[t]{0.49\textwidth}
\centering
\includegraphics[width=\textwidth]
{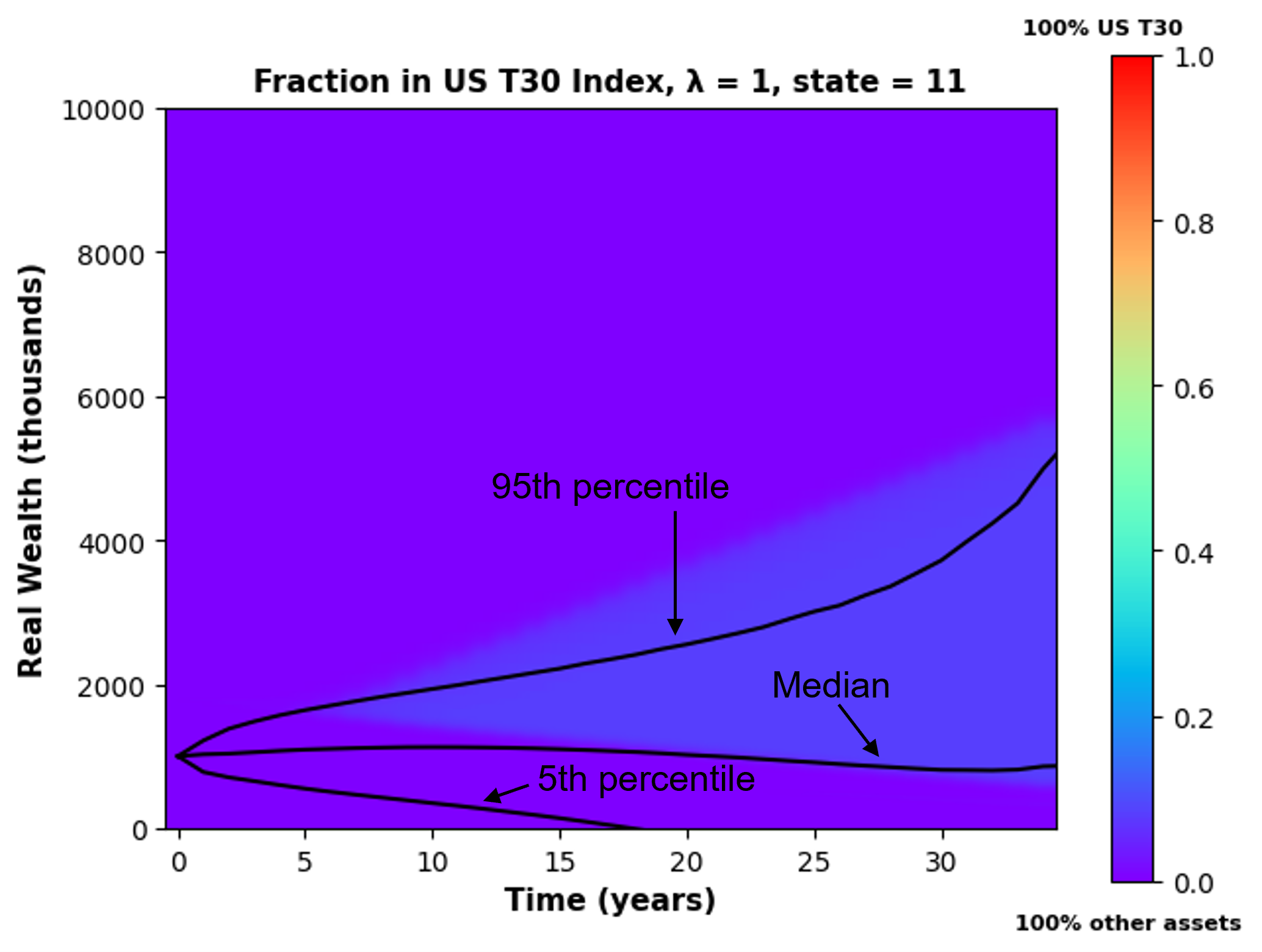}
\caption{U.S. Treasury bill, state $11$.}
\label{fig:app_us_tbill_controls_lambda1_11}
\end{subfigure}
\hfill
\begin{subfigure}[t]{0.49\textwidth}
\centering
\includegraphics[width=\textwidth]
{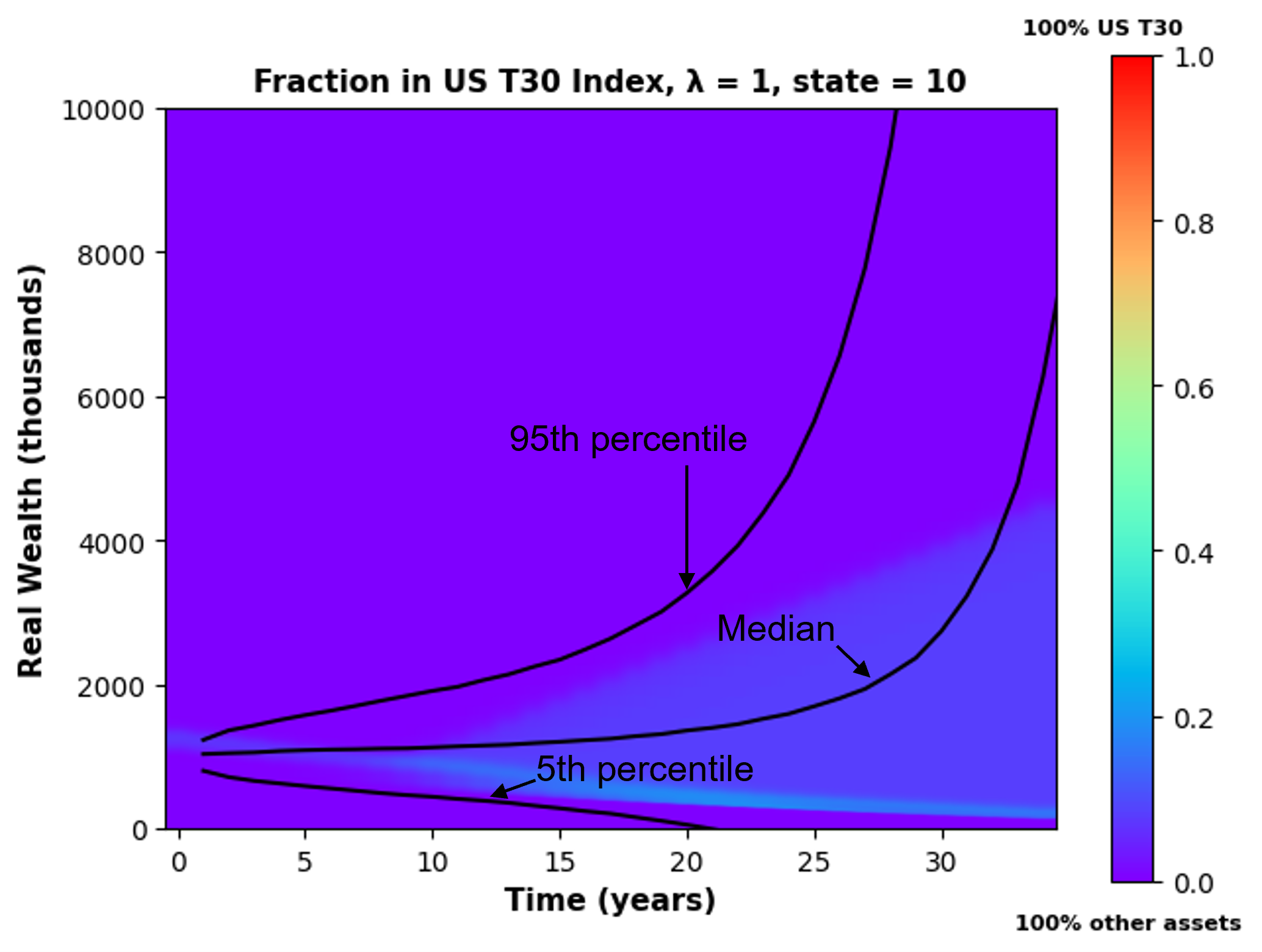}
\caption{U.S. Treasury bill, state $10$.}
\label{fig:app_us_tbill_controls_lambda1_10}
\end{subfigure}
\caption{Learned Australian equity allocations (top row) and
U.S. Treasury-bill allocations (bottom row) under the intermediate policy
$\widehat\nu_{1}^\ast$. Colors denote fractions of post-payment wealth.}
\label{fig:app_additional_allocation_controls}
\end{figure}

{Figures~\ref{fig:num_allocation_controls_lambda1} and
\ref{fig:app_additional_allocation_controls} together report the complete
four-asset allocation under the same learned policy, with paired panels
for states $11$ and $10$.}

{The allocation surfaces exhibit a growth-to-preservation pattern.
U.S. equity exposure is largest in poorly funded regions, particularly early
in the horizon (Figure~\ref{fig:num_allocation_controls_lambda1}), while
Australian equity is used mainly at low and intermediate funding levels
(Figure~\ref{fig:app_additional_allocation_controls}). Once wealth is
sufficiently high, the Australian government-bond allocation rises sharply;
the U.S. Treasury-bill component remains small over most of the plotted
domain. The learned policy therefore shifts from growth seeking when
underfunded to domestic-bond wealth preservation when the terminal reserve
target is well funded.}

{The shift toward Australian government bonds generally occurs at a
higher wealth level in state $11$
(Figure~\ref{fig:num_allocation_controls_lambda1}). This mirrors the
higher withdrawal threshold in
Figure~\ref{fig:num_withdrawal_controls_lambda1} and is consistent with
the greater spending and reserve requirements and weaker mortality-credit
support while both household members remain alive.}
\section{Continuation-cap sensitivity}
\label{app:num_continuation_cap_sensitivity}

{This appendix compares capped and uncapped spouse continuation under
the three selected learned policies. Each policy is held fixed, with no
retraining under the uncapped rule. We first examine representative-contract
continuation costs and household downside risk, then consider the cap's effect
on the conditional large-book limit.}
\subsection{Representative-contract comparison}
\label{ssc:num_continuation_cap_sensitivity}

{For a representative contract, we evaluate each learned policy
under}
\[
B_m^{\mathrm{cap}}=\min\{q_m,\overline q_1\},
\qquad
B_m^{\mathrm{uncap}}=q_m.
\]
{The asset-return and household-mortality paths are identical in each
paired comparison, so only the continuation-base convention changes. For
$\tau_R$ defined in \eqref{eq:hh_tau_R}, define the
unconditional cap-binding probability and mean cap excess by}
\begin{align}
p_{\mathrm{cap}}
\bigl(\widehat\nu_\lambda^\ast\bigr)
&:=
\mathbb P_{x_0,h_0}^{\widehat\nu_\lambda^\ast}
\left(
q_{\tau_R-1}^{\widehat\nu_\lambda^\ast}
>
\overline q_1,\;
\tau_R<\infty
\right),
\label{eq:num_cap_binding_probability}
\\
e_{\mathrm{cap}}
\bigl(\widehat\nu_\lambda^\ast\bigr)
&:=
\mathbb E_{x_0,h_0}^{\widehat\nu_\lambda^\ast}
\left[
\left(
q_{\tau_R-1}^{\widehat\nu_\lambda^\ast}
-
\overline q_1
\right)_+
\mathbf 1_{\{\tau_R<\infty\}}
\right].
\label{eq:num_mean_cap_excess}
\end{align}

\begin{table}[!htb]
\centering
\caption{Representative-contract cap incidence, expected continuation
costs, and expected-cost loads for the three selected learned policies.}
\label{tab:num_continuation_cap_sensitivity}
\begin{tabular}{lrrrrrrr}
\hline
\noalign{\vspace{1mm}}
\shortstack{Frontier\\position}
&
$\lambda$
&
\shortstack{$p_{\mathrm{cap}}$\\(\%)}
&
$e_{\mathrm{cap}}$
&
\shortstack{$\mathbb E[Z_c]$\\capped}
&
\shortstack{$\mathbb E[Z_c]$\\uncapped}
&
\shortstack{$f_c$ capped\\(\%)}
&
\shortstack{$f_c$ uncapped\\(\%)}
\\
\noalign{\vspace{1mm}}
\hline
Payment-seeking & 0.3 & 50.3 & 10.9 & 438.1 & 555.6 & 23.5 & 28.1\\
Intermediate    & 1   & 38.7 & 8.4  & 440.1 & 528.0 & 25.8 & 29.5\\
Conservative    & 15  & 22.7 & 5.0  & 433.1 & 476.3 & 27.3 & 29.3\\
\hline
\end{tabular}
\end{table}

{Table~\ref{tab:num_continuation_cap_sensitivity} shows that the
cap-binding probability and unconditional mean cap excess decrease across
the selected policies as conservatism increases. Removing the cap raises both
expected continuation cost and the expected-cost load under every policy,
with the largest increases under the payment-seeking policy. Expected
continuation costs remain close across policies under capped continuation
but are substantially more dispersed under the uncapped rule. Thus, limiting
the inherited continuation base to the one-survivor upper band
$\overline q_1$ stabilizes the expected continuation obligation across the
selected policies.}

\begin{figure}[!htb]
\centering
\begin{subfigure}[t]{0.32\textwidth}
\centering
\includegraphics[width=\textwidth]
{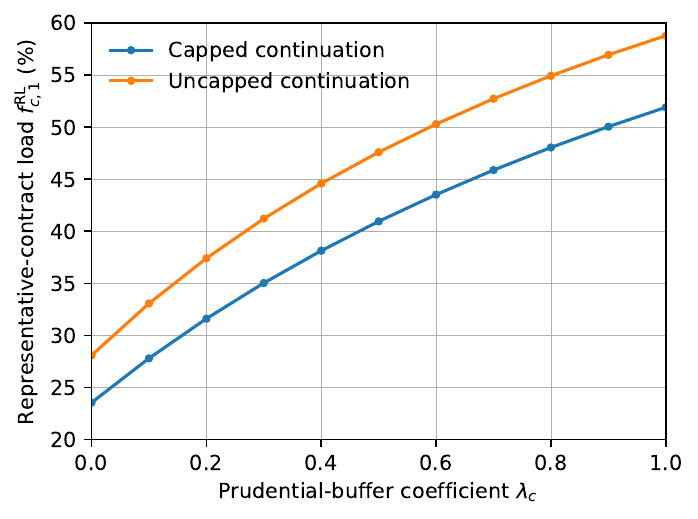}
\caption{$\lambda=0.3$.}
\label{fig:num_continuation_cap_sensitivity_03}
\end{subfigure}
\hfill
\begin{subfigure}[t]{0.32\textwidth}
\centering
\includegraphics[width=\textwidth]
{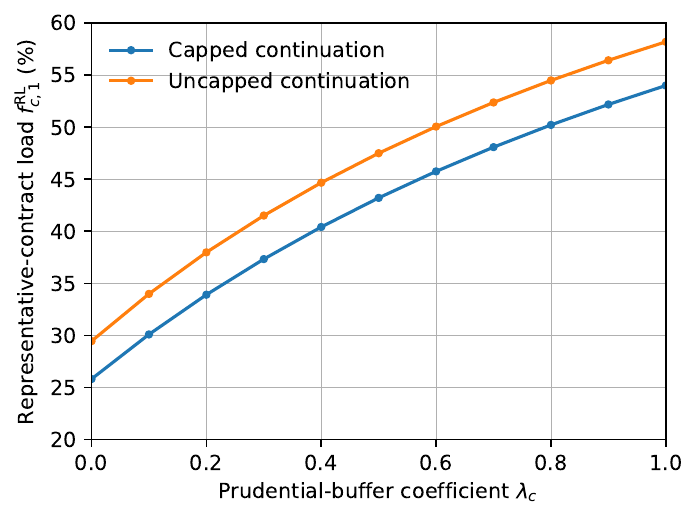}
\caption{$\lambda=1$.}
\label{fig:num_continuation_cap_sensitivity_1}
\end{subfigure}
\hfill
\begin{subfigure}[t]{0.32\textwidth}
\centering
\includegraphics[width=\textwidth]
{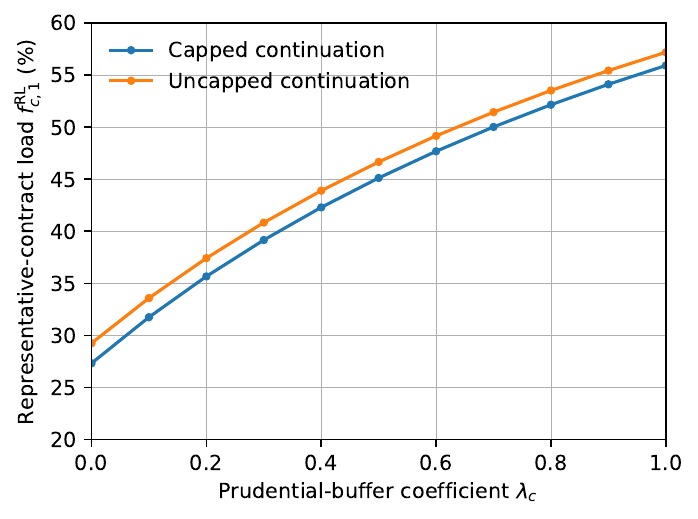}
\caption{$\lambda=15$.}
\label{fig:num_continuation_cap_sensitivity_15}
\end{subfigure}
\caption{Representative-contract risk-loaded spouse-continuation
loads under capped and uncapped continuation. The policy is held fixed,
$\alpha_c=0.05$, and the excess-CVaR loading is used.}
\label{fig:num_continuation_cap_sensitivity}
\end{figure}

{Figure~\ref{fig:num_continuation_cap_sensitivity} compares the
risk-loaded payment-scale loads as the prudential-buffer coefficient
$\lambda_c$ varies. Unlike the expected-cost loads in
Table~\ref{tab:num_continuation_cap_sensitivity}, these loads incorporate
the continuation-cost excess tail. Their values at $\lambda_c=1$, together
with the reported changes in terminal-shortfall risk and insolvency, are
collected in Table~\ref{tab:num_continuation_cap_risk_diagnostics}.}

\begin{table}[!htb]
\centering
\caption{Representative-contract risk-loaded loads and reported
changes in household downside risk when the continuation cap is removed.
Loads are evaluated at $\alpha_c=0.05$ and $\lambda_c=1$. Changes are
uncapped minus capped; terminal-shortfall CVaR changes are in thousands of
real Australian dollars, and insolvency-probability changes are in percentage
points.}
\label{tab:num_continuation_cap_risk_diagnostics}
\begin{tabular}{lrrrr}
\hline
\noalign{\vspace{1mm}}
\shortstack{Frontier\\position}
&
\shortstack{$f_{c,1}^{\mathrm{RL}}$\\capped (\%)}
&
\shortstack{$f_{c,1}^{\mathrm{RL}}$\\uncapped (\%)}
&
\shortstack{Change in\\$\operatorname{CVaR}_{0.95}(L_T)$}
&
\shortstack{Change in\\$p_{\mathrm{ins}}^{\mathrm{ever}}$}
\\
\noalign{\vspace{1mm}}
\hline
Payment-seeking & 51.9 & 58.8 & 35.4 & 5.7\\
Intermediate    & 54.0 & 58.2 & 31.6 & 0.4\\
Conservative    & 55.9 & 57.2 & 1.1  & $<0.1$\\
\hline
\end{tabular}
\end{table}

{Table~\ref{tab:num_continuation_cap_risk_diagnostics} shows that
removing the cap raises the risk-loaded continuation load and worsens both
household downside-risk measures under the fixed learned policies. The
increases are largest under the payment-seeking policy and smallest under the
conservative policy. Together with
Table~\ref{tab:num_continuation_cap_sensitivity}, these results show that
the cap has its strongest effect near the high-payment end of the frontier,
where larger withdrawals can otherwise be inherited by the surviving spouse.}
\subsection{Conditional large-book comparison}
\label{ssc:num_large_book_continuation_cap_sensitivity}
\label{app:num_large_book_continuation_cap_sensitivity}

{The representative-contract comparison includes both
household-specific mortality risk and common-market risk. We now compare the
two continuation rules in the conditional large-book limit, where household
mortality is integrated out. This separates the cap's effect on expected
continuation cost from its effect on the residual common-market tail. For
each rule, $D_c$ compares the limiting excess tail with the
representative-contract excess tail under that same rule.}

\begin{table}[!htb]
\centering
\caption{Conditional large-book continuation-cap sensitivity.
The upper-tail probability is $\alpha_c=0.05$;
$f_{c,\infty}^{\mathrm{RL}}$ is reported at $\lambda_c=1$.}
\label{tab:num_large_book_continuation_cap_sensitivity}
\begin{tabular}{llrrrrr}
\hline
\noalign{\vspace{1mm}}
\shortstack{Frontier\\position}
&
Rule
&
$\mathbb E[\overline Z_c]$
&
$\mathrm{CVaR}_{0.05}^{+}(\overline Z_c)$
&
$\Delta_{0.05}^{+}(\overline Z_c)$
&
\shortstack{$D_c$\\(\%)}
&
\shortstack{$f_{c,\infty}^{\mathrm{RL}}$\\(\%)}
\\
\noalign{\vspace{1mm}}
\hline
Payment-seeking & Capped   & 437.4 & 466.5 & 29.0  & 97.35 & 24.7\\
                          & Uncapped & 554.5 & 658.1 & 103.6 & 92.96 & 31.6\\
\hline
Intermediate    & Capped   & 439.4 & 464.3 & 24.9  & 97.62 & 26.9\\
                          & Uncapped & 526.9 & 641.2 & 114.4 & 90.72 & 33.6\\
\hline
Conservative    & Capped   & 432.4 & 459.4 & 26.9  & 97.38 & 28.5\\
                          & Uncapped & 475.6 & 604.0 & 128.4 & 87.91 & 34.4\\
\hline
\end{tabular}
\end{table}

{Table~\ref{tab:num_large_book_continuation_cap_sensitivity} shows
that removing the cap raises both the limiting expected continuation cost and
the residual excess tail under every policy. Across the three policies, the
residual excess tail is $24.9$--$29.0$ under capped continuation, compared
with $103.6$--$128.4$ under uncapped continuation. The lower uncapped
diversification ratios indicate that a larger share of the corresponding
representative-contract excess tail survives in the large-book limit.
Thus, the cap reduces not only the expected continuation-payment level but
also the market-sensitive tail that remains after household-specific mortality
risk has diversified.}

\begin{figure}[!htb]
\centering
\begin{subfigure}[t]{0.32\textwidth}
\centering
\includegraphics[width=\textwidth]
{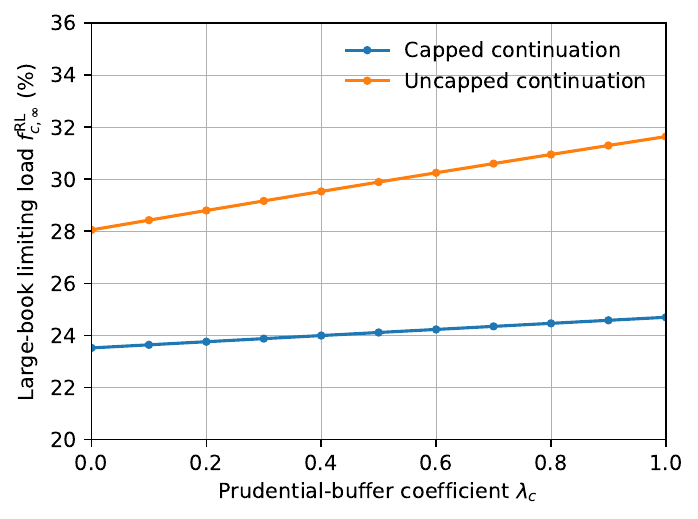}
\caption{$\lambda=0.3$.}
\label{fig:num_large_book_continuation_cap_03}
\end{subfigure}
\hfill
\begin{subfigure}[t]{0.32\textwidth}
\centering
\includegraphics[width=\textwidth]
{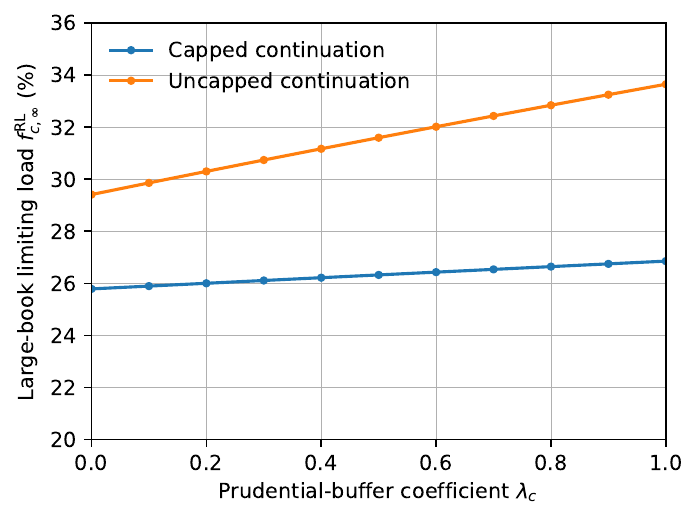}
\caption{$\lambda=1$.}
\label{fig:num_large_book_continuation_cap_1}
\end{subfigure}
\hfill
\begin{subfigure}[t]{0.32\textwidth}
\centering
\includegraphics[width=\textwidth]
{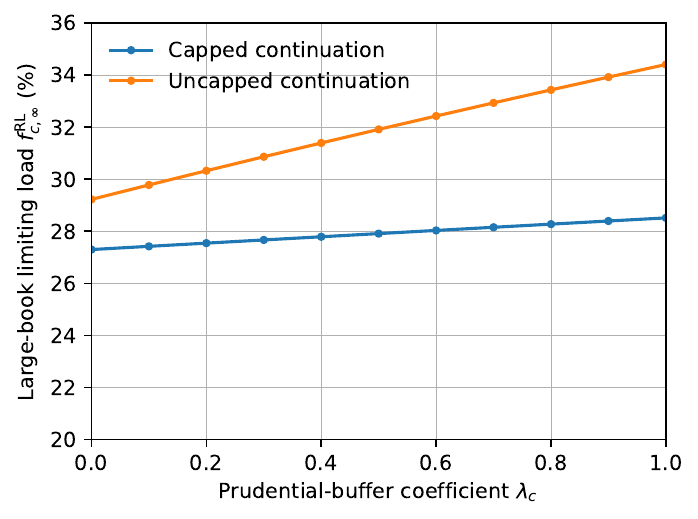}
\caption{$\lambda=15$.}
\label{fig:num_large_book_continuation_cap_15}
\end{subfigure}
\caption{Conditional large-book risk-loaded spouse-continuation loads
under capped and uncapped continuation. The upper-tail probability is
$\alpha_c=0.05$.}
\label{fig:num_large_book_continuation_cap}
\end{figure}

{Figure~\ref{fig:num_large_book_continuation_cap} shows how the
capped and uncapped large-book loads depend on $\lambda_c$. At
$\lambda_c=1$, Table~\ref{tab:num_large_book_continuation_cap_sensitivity}
reports capped loads of $24.7\%$--$28.5\%$ and uncapped loads of
$31.6\%$--$34.4\%$. The capped loads remain close to the corresponding
expected-cost loads in Table~\ref{tab:num_continuation_cap_sensitivity},
whereas uncapped continuation carries a larger residual tail buffer. This
difference cannot be captured by scaling the representative-contract excess
tail mechanically by $J^{-1/2}$, because the rule-dependent common-market
component does not diversify away.}
\section{Numerical robustness}
\label{app:num_numerical_robustness}

{The learned policies are held fixed and evaluated on
$K=256{,}000$ common market-return paths, organized into $B=100$
non-overlapping batches of $2{,}560$ paths. The same market paths are retained
across the three selected policies, book sizes, and paired capped--uncapped
comparisons. This analysis therefore measures Monte Carlo evaluation
uncertainty conditional on the learned policies; it does not include
variability from retraining the neural networks.}

{For each market path, $1{,}000$ independent household-mortality
histories are generated under the fixed 2021 Australian age-specific period
schedules. The first $1$, $10$, $100$, and $1{,}000$ histories define the
nested books with $J=1$, $10$, $100$, and $1{,}000$, respectively. Contract
outcomes are averaged within each book before any book-level statistic is
computed. In the conditional large-book calculation, household mortality is
instead integrated out using Algorithm~\ref{alg:cont_load_large_book} and
\eqref{eq:cont_probability_weighted_cost}. Setting each spouse-only payment
to one gives $8.5058442$ expected payment dates through $M-1$, the same
quantity $\sum_{m=1}^{M-1}\pi_{01}(m)$ identified in
Subsection~\ref{ssc:num_household_state_probs}. 
Debt positions accrue at the domestic bond-index return plus a borrowing spread of $2\%$ per annum.}

{The principal mean and CVaR estimates are calculated from the
complete $K$-vector. In particular, with $\alpha_c=0.05$, upper-tail CVaR is
the arithmetic mean of the largest $12{,}800$ observations. Batches are used only to estimate Monte Carlo uncertainty. If
$\widehat S_b$ denotes the statistic calculated within batch $b$, the
batch-based standard-error estimate is}
\[
\widehat{\operatorname{se}}(\widehat S)
=
\left[
\frac{1}{B(B-1)}
\sum_{b=1}^{B}
\left(\widehat S_b-\frac1B\sum_{r=1}^{B}\widehat S_r\right)^2
\right]^{1/2}.
\]
{For ratios and capped--uncapped comparisons, the relevant statistic
or difference is formed within each paired common-path batch before applying
this formula. The full-sample point estimate is not replaced by the average of
the batch estimates.}

\begin{table}[!htb]
\centering
\setlength{\tabcolsep}{3.2pt}
\caption{Representative-contract evaluation estimates and Monte Carlo
standard errors for the selected learned policies. Entries are reported as
estimate (standard error). Probability and load standard errors
are in percentage points. The continuation quantities use capped
continuation, $\alpha_c=0.05$, and $\lambda_c=1$.}
\label{tab:num_numerical_robustness}
\begin{tabular}{lccccccc}
\hline
\noalign{\vspace{2mm}}
\shortstack{Frontier\\position}
& $\lambda$
& $\mathcal R$
& $\operatorname{CVaR}_{0.95}(L_T)$
& \shortstack{$p_{\mathrm{ins}}^{\mathrm{ever}}$\\(\%)}
& $\mathbb E[Z_c]$
& \shortstack{$f_c$\\(\%)}
& \shortstack{$f_{c,1}^{\mathrm{RL}}$ (\%)\\$\lambda_c=1$}
\\
\noalign{\vspace{1mm}}
\hline
Payment-seeking
& 0.3
& 1860.4 (0.7)
& 1218.0 (5.1)
& 26.71 (0.08)
& 438.1 (0.9)
& 23.55 (0.05)
& 51.90 (0.04)
\\
Intermediate
& 1
& 1704.8 (0.7)
& 840.5 (9.9)
& 12.59 (0.07)
& 440.1 (0.8)
& 25.82 (0.05)
& 53.99 (0.04)
\\
Conservative
& 15
& 1584.8 (0.6)
& 793.3 (9.5)
& 11.95 (0.07)
& 433.1 (0.8)
& 27.33 (0.05)
& 55.91 (0.04)
\\
\hline
\end{tabular}
\end{table}

{For the capped finite-book results, the Monte Carlo standard error of
$\Delta_{0.05}^{+}(\overline Z_{c,J})$ is at most $0.75$ for $J=10$,
$0.22$ for $J=100$, and $0.11$ for $J=1000$; the corresponding standard
errors of $f_{c,J}^{\mathrm{RL}}$ are at most $0.03$, $0.01$, and $0.01$
percentage points. The standard errors of the large-book diversification
ratios are below $0.01$ percentage points. All estimated expected continuation
costs for $J=10,100,1000$ and the conditional large-book limit agree with the
representative-contract mean within paired $95\%$ Monte Carlo intervals. The
estimated tail costs also satisfy, for every selected policy and continuation
rule,}
\[
\mathrm{CVaR}_{0.05}^{+}(\overline Z_c)
\le
\mathrm{CVaR}_{0.05}^{+}(\overline Z_{c,J})
\le
\mathrm{CVaR}_{0.05}^{+}(Z_c).
\]
{The numerical conclusions in Section~\ref{sec:numerics} are
therefore not driven by evaluation-sample variation at the reported simulation
size.}

\section*{Acknowledgments}
 This work was supported by computational resources provided by The University of Queensland via the Friday supercomputer.
 
\setlength{\bibsep}{0pt plus 0.3ex}

\end{document}